\documentclass[aps, prc, tightenlines, letterpaper, preprintnumbers, floatfix, longbibliography, nofootinbib,superscriptaddress,10pt]{revtex4-2}

\usepackage[utf8]{inputenc}
\usepackage{amsmath}
\usepackage{graphicx}
\usepackage{dsfont}
\usepackage{physics}
\usepackage{placeins}
\usepackage{amssymb}
\usepackage{mathtools}
\usepackage{bbm}
\usepackage{tabularx}
\usepackage{array}
\usepackage[colorlinks=true, allcolors=blue]{hyperref}
\usepackage{multirow}
\usepackage[T1]{fontenc}
\usepackage{bm}
\usepackage{xspace}
\usepackage[dvipsnames]{xcolor}
\usepackage[normalem]{ulem}
\usepackage{diagbox}
\usepackage{soul}
\usepackage{nccmath}
\usepackage{lipsum}
\usepackage{cancel}
\usepackage{bm}
\usepackage{tabularx}
\usepackage{algorithm2e}
\usepackage{tikz}
\usetikzlibrary{quantikz2}

\newcolumntype{C}[1]{>{\centering\let\newline\\\arraybackslash\hspace{0pt}}m{#1}}
\newcolumntype{Y}{>{\centering\arraybackslash}X}

\usepackage{orcidlink}

\usepackage{stackengine}
\stackMath
\newcommand\tenq[2][1]{
 \def\useanchorwidth{T}%
  \ifnum#1>1%
    \stackunder[0pt]{\tenq[\numexpr#1-1\relax]{#2}}{\scriptscriptstyle\sim}%
  \else%
    \stackunder[1pt]{#2}{\scriptscriptstyle\sim}%
  \fi%
}

\definecolor{lavenderindigo}{rgb}{0.58, 0.34, 0.92}
\definecolor{colorslice}{HTML}{000580}

\newcommand{\infiL}{{\mathcal{I}_L}}

\usetikzlibrary{circuits.logic.US}
\tikzset{
    gateT/.style={
        draw,
        not gate US,
        inner sep=5pt
    },
    gateO/.style={
        draw,
        circle,
        minimum width=1.67em,
        inner sep=2pt,
        append after command={
            \pgfextra {
                \fill (\tikzlastnode) circle[radius=1pt];
            }
        }
    }
}

\begin{document}

\begin{figure}
\vskip -1.cm
\leftline{\includegraphics[width=0.15\textwidth]{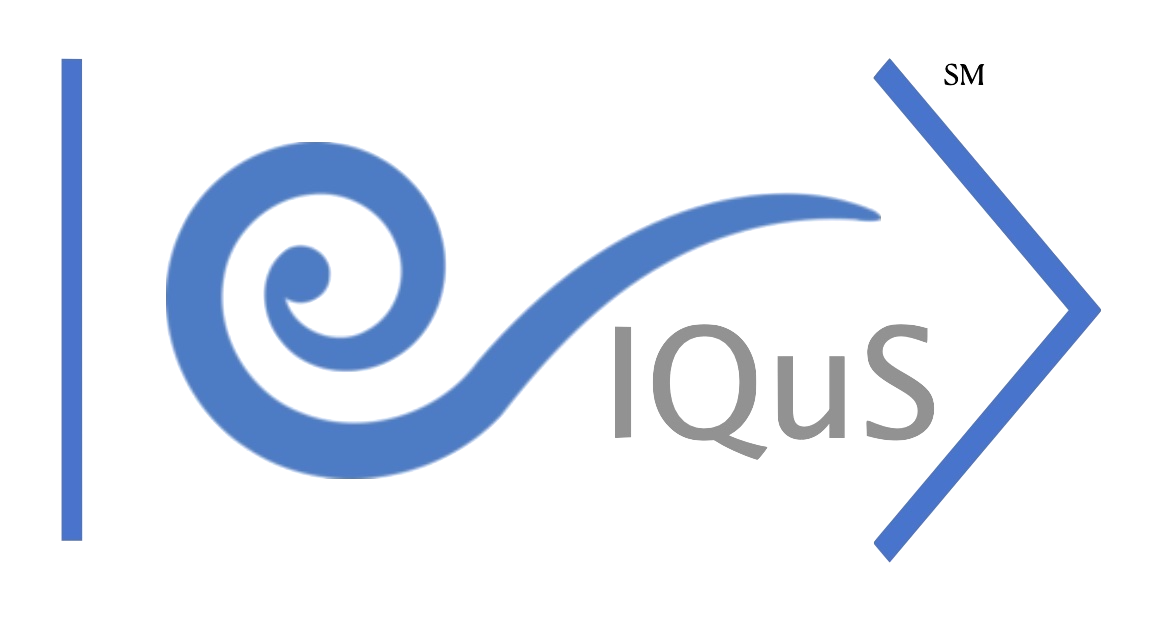}}
\end{figure}

\title{A Framework for Quantum Simulations of Energy-Loss and Hadronization in Non-Abelian Gauge Theories: SU(2) Lattice Gauge Theory in 1+1D}
\author{Zhiyao Li\,\orcidlink{0000-0002-7614-8496}}
\email{zhiyaol@uw.edu}
\affiliation{InQubator for Quantum Simulation (IQuS), Department of Physics, University of Washington, Seattle, WA 98195, USA}

\author{Marc Illa\,\orcidlink{0000-0003-3570-2849}}
\email{marc.illasubina@pnnl.gov}
\affiliation{InQubator for Quantum Simulation (IQuS), Department of Physics, University of Washington, Seattle, WA 98195, USA}
\affiliation{Physical Sciences Division, Pacific Northwest National Laboratory, Richland, WA 99354, USA}

\author{Martin J. Savage\,\orcidlink{0000-0001-6502-7106}}
\email{mjs5@uw.edu}
\affiliation{InQubator for Quantum Simulation (IQuS), Department of Physics, University of Washington, Seattle, WA 98195, USA}

\preprint{IQuS@UW-21-115}
\date{\today}

\begin{abstract}
\noindent
Simulations of energy loss and hadronization are essential for understanding  a range of  phenomena in non-equilibrium strongly-interacting matter.
We establish a framework for performing such simulations on a quantum computer and apply it to a heavy quark moving across a modest-sized 1+1D SU(2) lattice of light quarks.
Conceptual advances with regard to simulations of non-Abelian versus Abelian theories are developed, allowing for the evolution of the energy in light quarks, of their local non-Abelian charge densities, and of their multi-partite entanglement to be computed.
The non-trivial action of non-Abelian charge operators on arbitrary states suggests mapping the heavy quarks to qubits alongside the light quarks, and limits the heavy-quark motion to discrete steps among spatial lattice sites.
Further, the color entanglement among the heavy quarks and light quarks is most simply implemented using hadronic operators, and Domain Decomposition is shown to be effective in quantum state preparation.
Scalable quantum circuits that account for the heterogeneity of non-Abelian charge sectors across the lattice are used to prepare the interacting ground-state wavefunction in the presence of heavy quarks. 
The discrete motion of heavy quarks between adjacent spatial sites is implemented using fermionic SWAP operations.
Quantum simulations of the dynamics of a system on $L=3$ spatial sites are performed using IBM’s {\tt ibm\_pittsburgh} quantum computer using 18 qubits, for which the circuits for state preparation, motion, and one second-order Trotter step of time evolution have a two-qubit depth of 398 after transpilation.
A suite of error mitigation techniques  are used to extract the observables from the simulations, providing results that are in good agreement with classical simulations.
The framework presented here generalizes straightforwardly to other non-Abelian groups, including SU(3) for quantum chromodynamics.
\end{abstract}
\maketitle

\newpage

\tableofcontents

\newpage

\section{Introduction}
\label{sec:Intro}
\noindent
Extensive experimental campaigns in high-energy heavy-ion collisions are designed to reveal the structure and dynamics of matter under extreme conditions.  
The remarkable discovery of a near-ideal, low-viscosity quantum liquid of quarks and gluons produced in such collisions~\cite{PHENIX:2004vcz,BRAHMS:2004adc,PHOBOS:2004zne,STAR:2005gfr} has driven a program to measure and interpret its properties.
An important line of investigation is to quantify properties of transport in the liquid phase, where the identification of signatures and probes present a challenge.
One class of probes  is the production, motion and measurement of heavy quarks, including the production rates of quarkonia,  and processes such as dissociation and recombination.
Produced in hard collisions at early times, heavy quarks  witness the entire evolution of the exotic matter and its transition to hadronic final states. 
Therefore, the measured relative yields of heavy-quark systems reveal key attributes of the medium.
An excellent review of the state-of-the-art in this area can be found in Ref.~\cite{He_2023}.
An important element in extracting properties of this exotic matter is understanding how heavy quarks move and interact within the evolving system, including how they lose and gain energy as they move.
From a theoretical standpoint, first-principles simulations of these collisions present a monumental challenge.  
This challenge is not limited to evolution through exotic matter.   
Even simulations of high-energy probes impacting cold dense matter, such as a nucleus, or nuclear matter, suffer the same difficulties.  
While processes occurring over short-distances can be reliably determined in perturbation theory, the subsequent evolution (cascade) to longer length scales involves fragmentation and hadronization, both of which are intrinsically non-perturbative. 
Tuned to experimental data, modeling such processes is impressively successful, e.g., Geant4~\cite{GEANT4:2002zbu,Muskinja:2020ots,basaglia2024geant4gamechangerhigh,Pythia8},
but remains incomplete in describing fundamentally quantum observables, e.g., Refs.~\cite{Bauer:2019qxa,Klco:2021lap,Bauer:2022hpo,Gray:2022fou,Bauer:2023qgm,DiMeglio:2023nsa}.
While classical Euclidean-space
simulations of quantum chromodynamics (QCD) encounter a sign problem in summing complex phases in the time-evolution,
real-time evolution, starting from a specified initial state, is known to be efficient for quantum computers.

Constrained by available noisy intermediate-scale quantum (NISQ)~\cite{Preskill:2018jim} era quantum hardware, present-day quantum simulations of quantum field theories are focused on modest-sized, low-dimensional systems~\cite{Martinez:2016yna,
Klco:2018kyo,
Kokail:2018eiw,
Lu:2018pjk,
Klco:2019evd,
Mil:2019pbt,
Yang:2020yer,
Ciavarella:2021nmj,
Bauer:2021gup,
Atas:2021ext,
ARahman:2021ktn,
Mazzola:2021hma,
deJong:2021wsd,
Gong:2021bcp,
Zhou:2021kdl,
Riechert:2021ink,
Ciavarella:2021lel,
Nguyen:2021hyk,
Su:2022glk,
Illa:2022jqb,
Mildenberger:2022jqr,
ARahman:2022tkr,
Asaduzzaman:2022bpi,
Farrell:2022wyt,
Atas:2022dqm,
Farrell:2022vyh,
Mueller:2022xbg,
Pomarico:2023png,
Charles:2023zbl,
Zhang:2023hzr,
Ciavarella:2023mfc,
Farrell:2023fgd,
Meth:2023wzd,
Borzenkova:2023xaf,
Schuster:2023klj,
Angelides:2023noe,
Farrell:2024fit,
Kavaki:2024ijd,
Davoudi:2024wyv,
Turro:2024pxu,
Ciavarella:2024fzw,
Gyawali:2024hrz,
Gonzalez-Cuadra:2024xul,
Zemlevskiy:2024vxt,
Crippa:2024hso,
Ciavarella:2024lsp,
Than:2024zaj,
Turro:2025sec,
Alexandrou:2025vaj,
Farrell:2025nkx,
Davoudi:2025rdv,
Schuhmacher:2025ehh,
Chernyshev:2025lil,
Chen:2025zeh,
Cobos:2025krn,
Saner:2025nrq,
Than:2025gso},
with the Schwinger model~\cite{Schwinger:1962tp} (quantum electrodynamics (QED) in 1+1D) receiving significant attention.
Like QCD, the Schwinger model is a confining gauge theory, and remains a useful test-bed for exploring new algorithms and workflows and gaining improved understanding of infinite-volume and continuum limits.
Our present work is related to simulating transport, in a general sense, in non-equilibrium non-Abelian systems, 
and is on the path for robust simulations of jet physics and hadronization in QCD. 
There is a growing foundation for such simulations, e.g., 
Refs.~\cite{
Kuhn:2015zqa,
Kasper:2015cca,
Sala:2018dui,
Surace:2019dtp,
Magnifico:2019kyj,
Alexandru:2019nsa,
Li:2020uhl,
Bauer:2021gup,
Ji:2020kjk,
Barata:2021yri,
Li:2021zaw,
Yao:2022eqm,
Barata:2022wim,
Carena:2022kpg,
Florio:2023dke,
Barata:2023clv,
Florio:2024aix,
Alexandrou:2025vaj,
Barata:2025hgx}, 
along with progress in string breaking~\cite{
Hebenstreit:2013baa,
Lee:2023urk,
Belyansky:2023rgh,
Barata:2023jgd,
De:2024smi,
Surace:2024bht,
Ciavarella:2024lsp,
Gonzalez-Cuadra:2024xul,
grieninger2025st,
Borla:2025gfs,
Xu:2025abo,
Tian:2025mbv,
Cataldi:2025cyo}, 
and of the associated entanglement evolution~\cite{Chai:2023qpq,Papaefstathiou:2024zsu}.
Simulations of SU(2) and SU(3) theories including quarks and anti-quarks in 1+1D, with the gluon fields constrained by Gauss's law (either as dynamical fields or through long-range interactions), are now being performed using quantum computers~\cite{Klco:2019evd,
Ciavarella:2021nmj,
Atas:2021ext,
Farrell:2022wyt,
Farrell:2022vyh,
Atas:2022dqm,
Than:2024zaj,
Chernyshev:2025lil}.
With the Jordan-Wigner mapping of quarks, each spatial lattice site requires $2 n_f N_c$ qubits for $n_f$ flavors of light quarks with $N_c$ colors.
The inclusion of electroweak processes~\cite{Farrell:2022vyh,Chernyshev:2025lil} (using the low-energy effective field theory of the Standard Model) requires additional $4 n_g$ qubits encoding the charged-leptons and neutrinos, where $n_g$ is the number of generations included in the simulation.
Progress has also been made toward simulations using qudit systems, e.g., Refs.~\cite{Gustafson:2021qbt,
Meth:2023wzd,
Zache:2023cfj,
Gustafson:2023kvd,
Calajo:2024qrc,
Illa:2024kmf,
Gustafson:2024kym,
Crane:2024tlj,
Kurkcuoglu:2024cfv,
Gaz:2025ptu,
Jiang:2025ufg,
Huie:2025yzn,
Kurkcuoglu:2025gik,
Perez:2025cxl},
which have significant advantages in circuit depth and execution times.
Additionally, there have been a number of important quantum and classical algorithm developments for wavepacket preparation for scattering~\cite{Chai:2023qpq,
Farrell:2024fit,
Farrell:2024mgu,
Davoudi:2024wyv,
Zemlevskiy:2024vxt,
Turco:2025jot,
Farrell:2025nkx,
Chai:2025qhf,
Davoudi:2025rdv}, and for 
systematically-improvable truncation schemes~\cite{Li:2024lrl}.

Building on our previous work that established a framework for quantum simulations of energy-loss and fragmentation in systems of electrically-charged particles in the context of Abelian lattice gauge theory (LGT)~\cite{Farrell:2024mgu}, we now address energy loss in non-Abelian lattice gauge theories.\footnote{This should be considered to be a specific case of quantum thermodynamics of a non-equilibrium process~\cite{YungerHalpern:2019mvj,Davoudi2024a}, in the context of a non-Abelian lattice gauge theory.}
In particular, we establish a framework to simulate the energy loss and collisions of particles in an SU(2) non-Abelian lattice gauge theory in 1+1D using quantum computers.
Static heavy quark fields are mapped to qubits in order to simulate hadron motion and collisions on the lattice (as well as to establish regions of dense matter). 
To prepare ground-state  wavefunctions in the presence of a heavy quark, SC-ADAPT-VQE~\cite{Grimsley_2019,Farrell:2023fgd} is combined with Domain Decomposition 
(DDec), an algorithm used extensively in traditional lattice QCD calculations~\cite{10.1093/oso/9780198501787.001.0001,saad2003iterative,Luscher:2003qa,Heybrock2014LatticeQW,frommer2014DD,Ce:2016idq}, leveraging the heterogeneity of charge sectors across different regions on the lattice. 
The heavy quarks are translated across the lattice via discrete fermionic-SWAP operators, introducing additional finite lattice-spacing artifacts that vanish in the continuum limit.
The states are then time-evolved to allow the light quarks to respond dynamically to the heavy quark motion. 
Scalable quantum circuits~\cite{Farrell:2023fgd} are developed for state preparation, discrete translation, and time evolution, enabling measurements of the time dependence of local non-Abelian charge fluctuations, entanglement, and the energy in the light quarks and gluons.
Complete simulations (state preparation, translation, time evolution and measurement, with associated mitigation circuits) of a lattice with $L=3$ spatial sites are performed, requiring a total of 18 qubits, using IBM’s heron-processor quantum computers, with a transpiled CNOT depth of $\sim 400$. 
With multiple error-mitigation strategies applied, the results we obtain from the quantum simulations are in good agreement with corresponding classical simulations.

The simulations presented in this work were made possible by four conceptual advances.
One concerns the color entanglement between the heavy quark and the light degrees of freedom. 
Algorithms from traditional lattice QCD point toward working with color-singlet hadronic interpolating operators as an efficient way to handle color entanglement.
They provide a way to include the required symmetries of the wavefunction without qubit-wise manipulations.
A second development is appreciating the implications of mapping background static charges to the quantum register, and not including them through Gauss's law.
As the evolution operator for the light degrees of freedom does not commute with the heavy-quark fermionic SWAP operations, moving  heavy quarks to positions between lattice sites does not achieve what is intended, and as such, only translations between lattice sites are consistent.
Third, during the collision of two heavy quarks (when they are on adjacent sites), the SWAP operation interchanges them, and introduces an additional lattice artifact that can be addressed in two ways (or more).
Fourth, as the deviation of the wavefunction of the light degrees of freedom from the interacting vacuum state is localized around the heavy quark, it becomes an increasingly small part of total wavefunction infidelity in the variational process.
Therefore, partitioning the system in such a way that the heavy-quark modification is not sub-leading, DDec permits effective state preparation.

\section{The SU(2) Lattice with Heavy Quarks}
\label{sec:what}
\begin{figure}[ht!]
    \centering
\includegraphics[width=\linewidth]{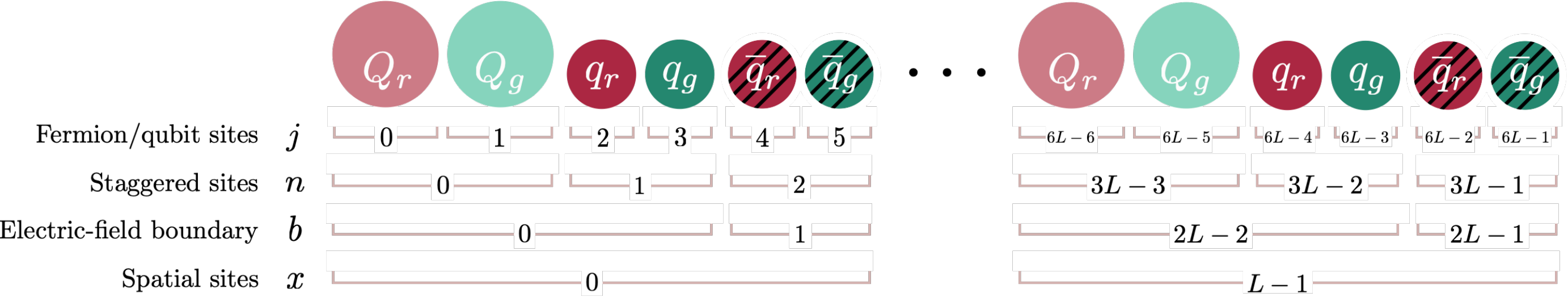}
\caption{
The lattice layout for heavy-quark, light-quark and light anti-quark fields on a lattice of spatial extent $L$.  
The spatial sites are labeled by $x$, which takes values from $x=0$ to $x=L-1$, the staggered sites are labeled by $n$ which takes values $n=0$ to $n=3L-1$, and the fermion (or qubit) sites are labeled by $j$ which takes values $n=0$ to $n=6L-1$.
The boundaries between quark and anti-quark sectors, denoted by $b$, correspond to the locations of chromo-electric field contributing to the Hamiltonian via the non-local charge operators. }
    \label{fig:layout_lattice}
\end{figure}

\noindent
The theory of focus in this work is the 1+1D SU(2) non-Abelian lattice gauge theory (LGT) with one flavor of light quark.
In the axial gauge with the Jordan-Wigner (JW) mapping of fermions to qubits, each spatial site of the lattice consists of non-dynamical heavy-quark fields, $Q_{r,g}$, light-quark fields, $q_{r,g}$, and light anti-quark fields, $\overline{q}_{r,g}$.
Heavy anti-quark fields are not included to reduce qubit overhead. 
The mapping of this system to qubits is shown in Fig.~\ref{fig:layout_lattice} for a lattice of spatial extent $L$.

The Hamiltonian describing these systems has three distinct contributions: 
kinetic energy, mass energy and chromo-electric interaction energy.
The kinetic energy operator  maps to spin operators acting on the spins associated with the 
light quark and anti-quark occupations.
In axial gauge with open boundary conditions (OBCs) it becomes,
\begin{equation}
\hat H_k = 
-\frac{1}{2}
\sum_{n\in q}\sum_{c=0,1}
\left(
\hat \sigma^+_{2n+c} \hat Z_{2n+1+c} \hat \sigma^-_{2n+2+c} 
 + \rm{h.c.} \right) - \frac{1}{2}
\sum_{n \in \overline{q}}\sum_{c=0,1}
\left( 
\hat \sigma^+_{2n+c} \hat Z_{2n+1+c} \hat Z_{2n+2+c} \hat Z_{2n+3+c} \hat \sigma^-_{2n+4+c} + \rm{h.c.} \right)
\ ,
\label{eq:kinL}
\end{equation}
where $n$ is the staggered index, $c$ the color index, and $\hat{\sigma}^{\pm} = (\hat{X} \pm i \hat{Y})/2$. 
The first term corresponds to hopping between quarks and anti-quarks
within one spatial site, while the second term corresponds to hopping between adjacent sites.
For the heavy quark to remain static (localized), effectively ``integrating it out'' of the low-energy effective lattice theory,
its kinetic term is not included. 

The mass-energy contribution to the Hamiltonian receives contributions from the $Q$, $q$ and $\overline{q}$.
For arbitrary heavy-quark and light-quark masses,  
the mass term becomes 
\begin{equation}
\hat H_m =
m_Q \sum\limits_{n \in Q}
\frac{1}{2}\left(\hat I + \hat Z\right)_n
 +\ 
m_q \sum\limits_{n \in q}
\frac{1}{2}\left(\hat I + \hat Z\right)_n
 +\ 
m_q \sum\limits_{n \in \overline{q}}
\frac{1}{2}\left(\hat I - \hat Z\right)_n
\ \rightarrow \ 
\frac{m_q}{2}
\sum\limits_{n \in q, \overline{q}}\sum_{c=0,1}
(-1)^{\eta_n}\  \hat Z_{2n+c}
\ \ ,
\label{eq:massterm}
\end{equation}
where $\eta_q=0$ and $\eta_{\overline{q}} = 1$.
We have set $m_Q=0$ for simplicity of calculation, 
and removed contributions from identity operators.

Contributions to the Hamiltonian
from the chromo-electric fields are determined by combinations of the non-Abelian SU(2) charge operators~\cite{Atas:2021ext}, 
\begin{equation}
\hat{\mathcal{Q}}_n^{(1)} = 
\frac{1}{2}
\left(
\hat \sigma^+_{2n}\hat \sigma^-_{2n+1}+
\hat \sigma^-_{2n}\hat \sigma^+_{2n+1}
\right)
\ ,\ 
\hat{\mathcal{Q}}_n^{(2)} =  \frac{-i}{2}  
\left(
\hat \sigma^+_{2n}\hat \sigma^-_{2n+1}-
\hat \sigma^-_{2n}\hat \sigma^+_{2n+1}
\right)
\ ,\ 
\hat{\mathcal{Q}}_n^{(3)} =  \frac{1}{4}  \left(  \hat Z_{2n}-\hat Z_{2n+1} \right)
\ \ ,
\label{eq:Qadefs}
\end{equation}
that act on pairs of fermion sites.
With the boundary condition that the chromo-electric field for $x<0$ vanishes,
the field at a given quark-anti-quark boundary, $E_b^{(a)}$, 
(denoted by $b$, depicted in Fig.~\ref{fig:layout_lattice})
is proportional to the summed charges from $n=0$ to the boundary,
\begin{equation}
    E_b^{(a)}  \propto \sum_{n=0}^{b+1+\lfloor b/2 \rfloor } \hat{\mathcal{Q}}_n^{(a)}
    \ \ ,
\end{equation}
where $\lfloor y \rfloor$ denotes the floor of $y$.
With OBCs, and the requirement that the chromo-electric field outside of the lattice ($x<0$ and $x>L-1$) has to vanish, the summation is restricted to the electric-field boundaries in the interior of the lattice, $b \in \{0, 1, \ldots, 2L-2\}$.
Color neutrality of the lattice is enforced by including a large energy penalty term in the Hamiltonian for non-zero $E_b^{(a)}$ at $b=2L-1$~\cite{Farrell:2022wyt}.
The chromo-electric field energy contribution to the Hamiltonian becomes
\begin{equation}
\hat H_g = 
\frac{g^2}{2}
\sum\limits_{\substack{a=1,2,3}}
\sum\limits_{b=0}^{2L-2}
\left(E_b^{(a)}\right)^2
\ \ ,\ \ 
\hat H_\lambda =
\frac{\lambda^2}{2}
\sum\limits_{a=1,2,3}
\left(E_{b=2L-1}^{(a)}\right)^2
\ \ ,
\end{equation}
where the second contribution is the quadratic penalty for non color-singlet configurations of the complete lattice, and $\lambda\rightarrow\infty$ is, in principle, taken.
The complete Hamiltonian is given by
\begin{equation}
    \hat H = \hat H_k + \hat H_m + \hat H_g + \hat H_\lambda
    \ .
    \label{eq:fullHami}
\end{equation}

\section{Conceptual Challenges and Advances}
\label{sec:Concepts}
\noindent
In order to study the energy loss of a heavy quark moving in the vacuum or through a region of finite density on a lattice, conceptual and algorithmic developments are required.
One challenge is to efficiently include heavy quarks into simulations, and understand the role of entanglement in color space across the lattice.   
A second is to develop quantum circuits to transport the heavy quark between adjacent spatial sites, and a third is to develop techniques to prepare the interacting ground-state in the presence of one or more heavy quarks.

\subsection{Color Entanglement}
\label{subsec:ColorEnt}
\noindent
In the case of U(1) gauge theory, heavy charges can be consistently included by discontinuities in Gauss's law and in values of the gauge field at each spatial location.
This is implemented by mapping the field to qubits (the occupation basis), where each basis state is an eigenstate of the electric charge operator.\footnote{When acted on by the electric charge operator, an electron remains an electron, and a positron remains a positron.}
In 1+1D QED, external charges are screened and charge-charge correlation functions fall off exponentially outside of a few confinement lengths.
The situation is more complex in the case of non-Abelian theories, even in 1+1D.
Specifically, because the non-Abelian charge operators act non-trivially on the fields, specifically, mixing states in the occupation basis, external charges cannot be efficiently included via discontinuities in Gauss's law and in the chromoelectric fields (see related works using tensor networks~\cite{Kuhn:2015zqa,Sala:2018dui}).\footnote{This leads to the use of a quadratic penalty for the energy in the electric field beyond the end of the lattice to enforce global color neutrality.}
Hence, heavy quarks are mapped to qubits in the same way as the light quarks, but a kinetic term is not included in the Hamiltonian to prevent its delocalization from a given lattice site.

To improve our intuition regarding the additional complexities of introducing  heavy quarks into a lattice of non-Abelian fields, we start by recalling the strong-coupling (SC) structure\footnote{The SC limit ($g\to\infty$) of the ground state is equivalent to ground state in the absence of the kinetic term in the Hamiltonian with arbitrary $g$.} of a single spatial site of the lattice Schwinger model with Kogut-Susskind (staggered) discretization of the light fermion fields, and with one-heavy charged particle.
In the absence of heavy charges, $n_Q=0$, the (anti-ferromagnetic) trivial-vacuum ground state is
\begin{equation}
    |\ Q^-\  e^- \ Q^+\ \ e^+ \rangle \rightarrow 
    |\downarrow \downarrow\ \uparrow \uparrow\rangle
    \ =\ |1100\rangle
    \ .
\label{eq:QEDSCQe}
\end{equation}
The minimum energy configuration when a static
negative electric charge is present, with vanishing global electric charge, is 
achieved by screening from a positron, giving the ($n_Q=1)$ state
\begin{equation}
    |\ Q^-\  e^- \ Q^+\ \ e^+ \rangle \rightarrow 
    |\uparrow \downarrow\ \uparrow \downarrow\rangle
    \ =\ |0101\rangle
    \ .
\label{eq:QEDSCQeQ1}
\end{equation}
The total energy of this state is higher than that of the trivial vacuum because of the single unit of electric flux absorbed by the classical $Q^-$ charge.
The structure of this state is somewhat trivial in the sense that it is unentangled, and the charge of the positron additively cancels that of the static charge (to ensure global charge neutrality).
This state can be thought of as the presence of a heavy meson, with quantum numbers of $Q^-e^+$, on top of the trivial vacuum.  
As a short-cut for this discussion, we could have considered the action of the simplest meson interpolating operator, $\hat O\sim \overline{Q}e$, on the trivial vacuum state.
As a point for the discussion that
follows, 
because the number of heavy charges in the system is fixed, 
if we consider only the presence or absence of $Q^-$'s,  the $Q^+$ field can be removed, reducing the state space to 
$|\ Q^-\  e^- \ e^+ \rangle$ 
(reducing the qubit footprint in quantum simulations).

The situation is more complex for non-Abelian theories.
We start by considering
one spatial site ($L=1$)
of a single flavor of color SU(2) fermions in the fundamental representation,
with the Kogut-Susskind discretization, along with a heavy-quark field, $Q$, (without heavy anti-quarks).
The lattice assignment of fields on a single spatial site, in terms of the two colors, red ($r$) and green ($g$), is $|\ Q_r\  Q_g\ q_r\  q_g\  \overline{q}_r\ \overline{q}_g\  \rangle$.
In the absence of a heavy quark, the SC vacuum state is
\begin{equation}
    |\ Q_r\  Q_g\ q_r\  q_g\  \overline{q}_r\ \overline{q}_g\  \rangle 
    \rightarrow 
    |\downarrow \downarrow \downarrow \downarrow\ \uparrow \uparrow\rangle
    \ =\ |111100\rangle
    \ .
\label{eq:SU2SCQ0}
\end{equation}
With a heavy-quark present, its non-Abelian charge can be screened in two ways.
The first is by the presence of light anti-quarks, corresponding to the presence of (color-singlet) heavy mesons (with interpolating operators of the form $\hat O \sim \sum_\alpha \overline{Q}{}^\alpha q_\alpha$). 
The second is by the presence of light quarks, corresponding to the presence of (color-singlet) heavy baryons (with interpolating operators of the form $\hat O \sim \sum_{\alpha\beta}  \epsilon_{\alpha\beta} \ \overline{Q}{}^\alpha \overline{q}{}^\beta$).\footnote{The Abelian theory does not have corresponding baryon operators.}
The latter is energetically favored (in the lattice theory) as there is no chromo-electric flux between the quark and anti-quark sectors.\footnote{As noted in Ref.~\cite{Atas:2021ext}, the baryon and meson masses converge in the continuum limit where lattice-spacing artifacts vanish.}
The SC ground-state wavefunction for this $n_Q=1$ system is of the form\footnote{The (entangled) wavefunction created from a meson interpolating operator is of the form
\begin{equation}
    |\ Q_r\ Q_g\ q_r\ q_g\ \overline{q}_r\ \overline{q}_g\  \rangle 
    \rightarrow 
\frac{1}{\sqrt{2}}\biggl[
|\uparrow \downarrow \downarrow \downarrow\ \downarrow \uparrow\rangle
\ +\ 
|\downarrow \uparrow \downarrow \downarrow\ \uparrow \downarrow\rangle
\biggl]
    \ =\ \frac{1}{\sqrt{2}}\biggl[
|011110\rangle \ +\ |101101\rangle
\biggr]
    \ .
\label{eq:SU2SCQ1mes}
\end{equation}
}
\begin{equation}
    |\ Q_r\  Q_g\ q_r\ q_g\  \overline{q}_r\ \overline{q}_g\  \rangle 
     \rightarrow 
\frac{1}{\sqrt{2}}\biggl[
|\uparrow \downarrow \downarrow \uparrow\ \uparrow \uparrow\rangle
\ -\ 
|\downarrow \uparrow \uparrow \downarrow\ \uparrow \uparrow\rangle
\biggr]
    \ =\ \frac{1}{\sqrt{2}}\biggl[
|011000\rangle \ -\ |100100\rangle
\biggr]
    \ ,
\label{eq:SU2SCQ1}
\end{equation}
where the colors of the heavy quark and the light quark
are entangled.
In the remainder of the manuscript, to set a convention, 
the ground state 
in the SC limit with $L$ spatial sites and $n_Q$ heavy quarks will be denoted by $|SC; L, n_Q\rangle$, while the true (interacting) ground state will be denoted by $|I; L, n_Q\rangle$.

\subsection{Implementing Heavy-Quark Motion}
\label{subsec:HQmove}
\noindent
For simulations of the Schwinger model, continuous spatial translations of the classical background charge can be implemented by a position-dependent reweighting of the discontinuities in the Gauss's law constraint~\cite{Florio:2023dke,Florio:2024aix,Farrell:2024mgu}.  
For the reasons mentioned above, specifically the action of the non-Abelian charge operator, spatial translations in non-Abelian theories cannot be efficiently used in this way, and instead are implemented by fermionic-SWAP (FSWAP) operators~\cite{Verstraete:2008qpa,Cervera-Lierta:2018bia,Algaba:2023enr} acting among the heavy-quark lattice sites.
Acting on a two-qubit system $\{ |00\rangle , |01\rangle , |10\rangle, |11\rangle \}$, the continuous FSWAP operation is
\begin{align}
\hat F(\theta) \ = \ \hat {\rm FSWAP}(\theta) & =\  
\left(
\begin{array}{cccc}
1&0&0&0\\
0 & e^{i\theta/2}\cos{\frac{\theta}{2}} & -ie^{i\theta/2}\sin{\frac{\theta}{2}}&0\\
0 & -ie^{i\theta/2}\sin{\frac{\theta}{2}} & e^{i\theta/2}\cos{\frac{\theta}{2}}&0\\
0&0&0&e^{i\theta}\\
\end{array}
\right)
\ =\ 
e^{-i 
\frac{\theta}{4} 
\left( \hat Y\otimes\hat Y + \hat X\otimes\hat X + \hat I\otimes \hat Z + \hat Z\otimes \hat I - 2 \hat I \right)
}
\ ,
\nonumber\\
&
 \xrightarrow[\theta\rightarrow\pi]{}
\left(
\begin{array}{cccc}
1&0&0&0\\
0& 0&1&0\\
0&1&0&0\\
0&0&0&-1\\
\end{array}
\right)
    \ .
    \label{eq:FSWAP}
\end{align}
As explained in App.~\ref{app:fswap}, the FSWAP gate can only be applied with $\theta=\pi$.
In the quantum simulations that follow, the FSWAP gates will be implemented for the red and green heavy-quark sites independently.
Their fermionic nature will be retained in the transformation, by including $\hat Z$ operators acting on the light and heavy quarks in between,
\begin{equation}
\hat F_{r,g} = 
e^{-i \frac{\pi}{4} 
\left( \hat X\otimes (- \hat Z)^{\otimes 5} \otimes\hat X 
+ \hat Y\otimes (- \hat Z)^{\otimes 5} \otimes\hat Y
+\hat I^{\otimes 6}\otimes \hat Z  
+\hat Z\otimes \hat I^{\otimes 6} 
- 2 \hat I^{\otimes 7}
\right) }
    \ .
    \label{eq:FSWAP6sites}
\end{equation}

Simulating a system involving multiple heavy quarks with one or more of them in motion requires modeling collisions. In 1+1D, this means characterizing the behavior when one $Q$ moves across another $Q$ when the motion is implemented by FSWAP gates.
\begin{figure}[ht!]
    \centering
\includegraphics[width=0.6\linewidth]{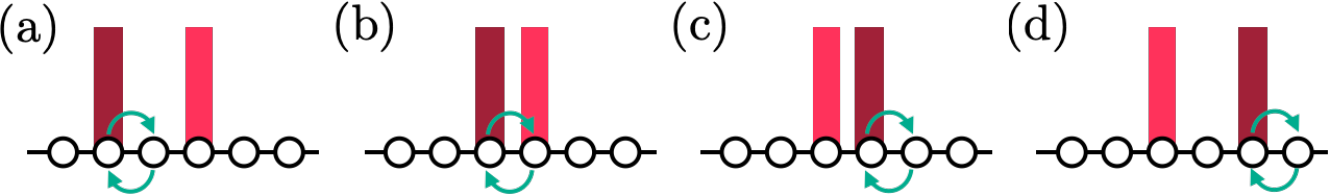}
\caption{A schematic of one heavy quark moving toward and passing a second heavy quark, implemented by FSWAP gates.
The circles denote spatial lattice sites, and the vertical colored bars denote heavy quarks.
Panels (a) through (d) show the discrete change of spatial site.  Panel (b) shows the 
action of the FSWAP gate when the heavy quarks are on adjacent site, in which the stationary heavy quark exchanges position with the moving heavy quark, displacing it by one lattice site.}
    \label{fig:QQswap}
\end{figure}
Consider the initial configuration shown in panel (a) of Fig.~\ref{fig:QQswap}. There are two heavy quarks, with one fixed in place and the other moving toward it, implemented by FSWAP gates.
As described in the caption of Fig.~\ref{fig:QQswap}, one important point to note is that when the heavy quarks are on adjacent spatial sites, they are exchanged by the action of 
FSWAP gates, displacing the stationary heavy quark by one site.
Subsequent motion involves displacing the moving heavy quark, but could also involve moving the stationary heavy quark back to its original location or not.
The difference between these possibilities is a lattice-spacing artifact that vanishes in the continuum limit.

\subsection{Ground-State Wavefunctions with Sources, and Domain Decomposition}
\label{subsec:GSDD}
\noindent
It is helpful to consider the structure of the ground-state wavefunction in the presence of a heavy quark.
Without the heavy quark, the ground-state wavefunction is invariant under discrete spatial translations, modified by effects from the boundaries.
In the case of OBCs, the boundaries modify the wavefunction by components that decrease exponentially away from the boundary, with a length scale set by the longest correlation length (the inverse mass of the lightest particle) in the theory.  
While local regions in the contained SU(2) charges fluctuate, the total charge of the lattice vanishes.
In the presence of a heavy quark, the light quarks are entangled with it in such a way to ensure global color neutrality.
Further, screening of the color charges means that far from the source the wavefunction approaches the ground state of the system without a heavy quark.   
A coherent sum of colored eigenstates arranges itself to generate these features.

In previous works, we developed scalable quantum circuits~\cite{Farrell:2023fgd,Farrell:2024fit} implemented by 
ADAPT-VQE~\cite{Grimsley_2019}, called SC-ADAPT-VQE (for a variant, see Ref.~\cite{Gustafson:2024bww}).
Classical computing was used to determine the operator sequence and their coefficients in creating a state approaching the ground state in modest-sized volumes, and then scaling to arbitrarily large systems on quantum computers in a controlled and converging way using the symmetries of the (lattice) system.
When wavepackets or heavy-quark sources are included, localized regions of the system are modified from the vacuum by operators with finite extent to approach the target wavefunction.

For the heavy-quark systems, it is the size of a heavy meson (or heavy baryon) that dictates the spatial extent of the region that needs to be modified away from the vacuum state.
This modified region is then combined with a region of the ground-state wavefunction without the heavy-quark source.
Both wavefunctions are classically optimized, and both suffer from boundary effects.
In such heterogeneous systems, DDec is often an efficient method to pursue.
We classically perform SC-ADAPT-VQE to determine the unitary operators to prepare the region with heavy-quarks, and combine it with a region of vacuum that is also prepared classically using SC-ADAPT-VQE.
This forms an initial wavefunction, which then undergoes another round of SC-ADAPT-VQE to optimize a sequence of unitary operators with support around the boundary.  
Specifically, operators that ``heal the wound'' caused by the boundary effects on both sides have a typical length-scale set by the confinement scale (or the lightest hadron).
For a system with multiple boundaries, as long as the boundaries are far from the heavy quark(s) and the edges, they can be healed with the same operators due to translational invariance.
This is made explicit in subsequent sections.

There are also practical reasons for using DDec in preparing these states.
For a large system with only one heavy quark, the volume of the wavefunction that is distorted away from the vacuum state is suppressed by the hadronic volume as opposed to the lattice volume. 
With increasing lattice volume, eventually this distortion becomes less than the tolerance of the optimizer, and the variational wavefunction becomes unresolvable between the interacting vacuum states with and without the heavy quark.
Therefore, a region around the heavy quark will necessarily need to be optimized separately from the rest of the lattice.

\section{Classical Simulation of SU(2) in 1+1D}
\label{sec:CSim}
\noindent
All the ingredients are in place to perform classical and quantum simulations of
heavy quarks moving through systems of light and heavy quarks in non-Abelian lattice gauge theories in 1+1D. 
It is helpful to consider the structure and dynamics of small systems in understanding how to perform simulations at scale. 
Small systems provide convenient environments to demonstrate the concepts and techniques discussed above, along with the subsequent implementation on quantum computers at scale.
As classical simulations are central to the framework of quantum simulations, we start by developing them for small-size systems, including for ``stitching them together'' in DDec, and for exploring their quantum complexity.
We also classically simulate time-evolution with discrete translation of the heavy quark, as discussed above.

\subsection{Structure of a Single Spatial Site}
\label{sec:L1}
\noindent
We begin with a somewhat detailed discussion of the simplest system of a heavy quark, a light quark and a light anti-quark on one spatial site, $L=1$, with basis states denoted by $ |\ Q_r\  Q_g\ q_r\  q_g\  \overline{q}_r\ \overline{q}_g\  \rangle $.
As shown in Fig.~\ref{fig:layout_lattice}, the staggered sites are labeled by $n=\{0,1,2\}$, with one internal electric-field boundary, $b=0$ after $n=1$.
As discussed in Sec.~\ref{sec:what}, the Hamiltonian describing the system is 
\begin{equation}
\hat H =
-\frac{1}{2}
\sum_{c=0}^{1}
\left( 
\hat \sigma^+_{2+c} \hat Z_{3+c} \hat \sigma^-_{4+c}  + {\rm h.c.} \right)
 +
\frac{m_q}{2}
\sum\limits_{n=1}^{2} \sum_{c=0}^{1}
(-1)^{\eta_n}  \hat Z_{2n+c}
+
\frac{g^2}{2}
\sum\limits_{a=1}^{3}
\left(\sum\limits_{n=0}^{1}\hat{\mathcal{Q}}_n^{(a)}\right)^2
 + 
\frac{\lambda^2}{2}
\sum\limits_{a=1}^{3}
\left(\sum\limits_{n=0}^{2}\hat{\mathcal{Q}}_n^{(a)}\right)^2
\ ,
\end{equation}
where $\eta_n$ is defined below Eq.~(\ref{eq:massterm}).

The SC ground state in the absence of heavy quarks is
\begin{equation}
    |SC; 1, 0\rangle = |\downarrow\downarrow\downarrow\downarrow\ \uparrow\uparrow\rangle 
     = |111100\rangle 
    \ ,
    \label{eq:SCGS10}
\end{equation}
and exact diagonalization of the Hamiltonian in the $2^6$ dimensional basis for $g=1$ and $m_q=0.1$ gives the interacting ground-state wavefunction,
\begin{equation}
|I; 1, 0\rangle = 
0.6663 |\downarrow\downarrow\downarrow\downarrow\ \uparrow\uparrow \rangle
\ +\ 
0.4383 |\downarrow\downarrow\downarrow\uparrow\ \uparrow\downarrow \rangle
\ -\ 
0.4383 |\downarrow\downarrow\uparrow\downarrow\ \downarrow\uparrow \rangle
\ +\ 
0.4143 |\downarrow\downarrow\uparrow\uparrow\ \downarrow\downarrow \rangle
\ \ ,
\label{eq:psiL1Q0}
\end{equation}
with an energy of $E_0=-0.6578$, and with spin expectation values $\langle \hat Z_j \rangle$ of $\{ -1 , -1 , -0.2723 , -0.2723 , +0.2723 , +0.2723  \}$.
In this state, the expectation values of the components of the Hamiltonian are
\begin{equation}
\langle \hat H_k \rangle = 
-0.9474
\ ,\ 
\langle \hat H_m \rangle  =
+0.14553
\ ,\ 
\langle \hat H_g \rangle =
+0.1440
\ ,\ 
\langle \hat H_\lambda\rangle = 0
\ \ .
\end{equation}

The operators required to create $|I; 1, 0\rangle$ from $|SC; 1, 0\rangle$ correspond to interpolating operators for color-singlet states in the spectrum, restricted to one-meson and one-baryon operators, defined to be
\begin{equation}
    \hat{O}_{M0} \equiv 
    \hat I^2 (\hat X \hat Z\hat Y - \hat Y \hat Z\hat X )   
    \hat I + \hat I^3  (\hat X \hat Z\hat Y - \hat Y \hat Z\hat X )
    \ ,\ 
    \hat O_{B0} \equiv   4i \hat I^2 (\hat\sigma^+\hat\sigma^+\hat\sigma^-\hat\sigma^- - {\rm h.c.}) 
     \ ,
     \label{eq:mesOpsL1}
\end{equation}
where the subscript $M$ denotes a meson operator, and $B$ denotes a baryon operator.
A variational wavefunction of the form
\begin{equation}
|\psi_{\rm var}\rangle = 
e^{-i \theta_2 \hat O_{B0}}\ 
e^{-i \theta_1  \hat{O}_{M0}}\ 
 |SC; 1, 0\rangle
\ ,\ \  
{\rm with}\ \ 
\theta_1 =  0.267215
\ ,\ \theta_2 = 0.05484
    \ ,
    \label{eq:L1nQ0VQE}
\end{equation}
recovers the wavefunction with an infidelity of ${\cal I} =  1 - |\langle\psi_{var}|I; 1, 0\rangle|^2 \sim 10^{-6}$.
The infidelity can be reduced to zero with this complete operator set.

In systems with one heavy quark, $n_Q=1$, the SC vacuum is given in Eq.~\eqref{eq:SU2SCQ1}, and the exact wavefunction for $g=1$ and $m_q=0.1$
is
\begin{equation}
|I; 1, 1\rangle = 
0.6120 |\downarrow\uparrow\uparrow\downarrow\  \uparrow\uparrow \rangle
-
0.6120 |\uparrow\downarrow\downarrow\uparrow\ \uparrow\uparrow \rangle
+
0.3540 |\downarrow\uparrow\uparrow\uparrow\  \uparrow\downarrow \rangle
-
0.3540 |\uparrow\downarrow\uparrow\uparrow\  \downarrow\uparrow \rangle
\ \ ,
\end{equation}
giving expectation values
of the components of the Hamiltonian,
\begin{equation}
\langle \hat H_k \rangle  = 
-0.4334
\ ,\ 
\langle \hat H_m \rangle = 
+0.1501
\ ,\ 
\langle \hat H_g \rangle = 
+0.0940
\ ,\ 
\langle \hat H_{\rm ext.}\rangle = 0
\ \ .
\end{equation}
The energy of this state is 
$E_1=-0.1892$, 
giving a hadron mass (minus the mass of the heavy quark) of $\Lambda_Q=E_1 - E_0 = 0.4685$, with spin alignments of $\langle \hat Z_j \rangle= \{  0 , 0 , 0.2507 , 0.2507 , 0.7492 , 0.7492\}$.
The vanishing heavy-quark expectation values reflect the color entanglement with the light quarks.
Without the possibility of adding a baryon-antibaryon pair to the SC vacuum, due to the color entanglement with the heavy quark, only mesons can contribute to the wavefunction.
A variational wavefunction of the form
\begin{equation}
|\psi_{\rm var}\rangle = 
e^{-i \theta_1  \hat{O}_{M0}}\ 
 |SC; 1, 1\rangle
\ ,\ \  
{\rm with}\ \ 
\theta_1 =  0.26224
    \ ,
    \label{eq:L1nQ1VQE}
\end{equation}
recovers the exact wavefunction.

\subsection{Hadronic Interpolating Operators for State Preparation}
\label{sec:L2}
\noindent
The $L=2$ system introduces a further layer of complexity, primarily in 
identification of the pool of variational operators.
Section~\ref{sec:what} contains the basic tools required to address the $L=2$ system,
and App.~\ref{app:L2} provides details of the exact diagonalization.
For notational clarity and specificity, site indices are added to the operators to indicate the spatial site that the first and last operator act on, i.e., $(x,y)$ in $\hat O_{Md}^{(x,y)}$ 
(if the operator acts only on one site, this is reduced to a single index, i.e., $\hat O_{M0}^{(x)}$).
The operators that we have identified for state preparation with sub-percent infidelity are:
\begin{equation}
\hat O_{M0}^{(0)}, \ 
\hat O_{M0}^{(1)}, \ 
\hat O_{M1}^{(0,1)}, \ 
\hat O_{B0}^{(0)}, \ 
\hat O_{B0}^{(1)}, \ 
\hat O_{B1}^{(0,1)} \ 
\ ,
\label{eq:OPxyL2}
\end{equation}
where
\begin{equation}
\hat{O}_{M1} \equiv (\hat Y \hat Z^3\hat X - \hat X \hat Z^3 \hat Y )  \hat I
+ \hat I   (\hat Y \hat Z^3\hat X - \hat X \hat Z^3 \hat Y )
\ ,\ 
    \hat O_{B1}  \equiv  -4i(\hat\sigma^+\hat\sigma^+ \hat{I}^2 \hat\sigma^-\hat\sigma^-
    - \rm{h.c.})
\ ,
\label{eq:OPxyL2def}
\end{equation}
where, as for the $L=1$ system, the $\hat X, \hat Y$ operators only act on the light quarks and anti-quarks.
These are both operators that ``straddle''  heavy $Q$ sites, i.e., the extra $\hat Z^2$ or $\hat I^2$ acts on  $Q_r, Q_g$ sites. 
A more detailed presentation of this operator pool can be found in Eq.~\eqref{eq:OPcomsL2}.
In general, for arbitrary systems, meson and baryon-antibaryon operators, or in fact any color-singlet configurations, with support across multiple sites, located at arbitrary positions on the lattice should be included.
Their relative contributions are determined by their (hadronic) form factors and the structure of the ground-state wavefunction.

In the variational process, as in previous works, the infidelity density, $\infiL$,  is chosen as the objective function,
\begin{equation}
\infiL = \frac{1}{L} 
\left( 1 - 
|\langle\psi_{\rm var}|\psi_{\rm target}\rangle|^2 \right)
\ ,
\label{eq:infidelitydens}
\end{equation} 
where $|\psi_{\rm target}\rangle$ is the target wavefunction (in this case  $|I; 2, n_Q\rangle$) and $|\psi_{\rm var}\rangle$ is the variational ansatz wavefunction that is being optimized to approach $|\psi_{\rm target}\rangle$, as discussed above in Sec.~\ref{sec:L1}.
This definition becomes increasingly useful as $\infiL\rightarrow 0$, as it gives comparable measures of the ``goodness of fit'' of the wavefunction per spatial site.

ADAPT-VQE is used, guided by physics input, to identify the optimal operator ordering and subsequent layer-by-layer parameter optimizations.  
To respect translational invariance and scalability in the $n_Q=0$ sector in the absence of OBCs, the meson operators are applied first to the SC vacuum in Eq.~\eqref{eq:psiSCL2Q0}, followed by the baryon operators, to give
\begin{equation}
\hat U^{n_Q=0}_{L=2} = 
e^{-i \theta_5 \hat O_{B1}^{(0,1)}}
e^{-i \theta_4 \hat O_{M1}^{(0,1)}}
e^{-i \theta_3 (\hat O_{B0}^{(0)}+\hat O_{B0}^{(1)})}
e^{-i \theta_2 (\hat O_{M0}^{(0)}+\hat O_{M0}^{(1)})}
e^{-i \theta_1 \hat O_{M1}^{(0,1)}}
\ ,
\label{eq:L2UNITQ0}
\end{equation}
with five variational parameters.
\begin{figure}[ht!]
    \centering
\includegraphics[width=0.48\linewidth]{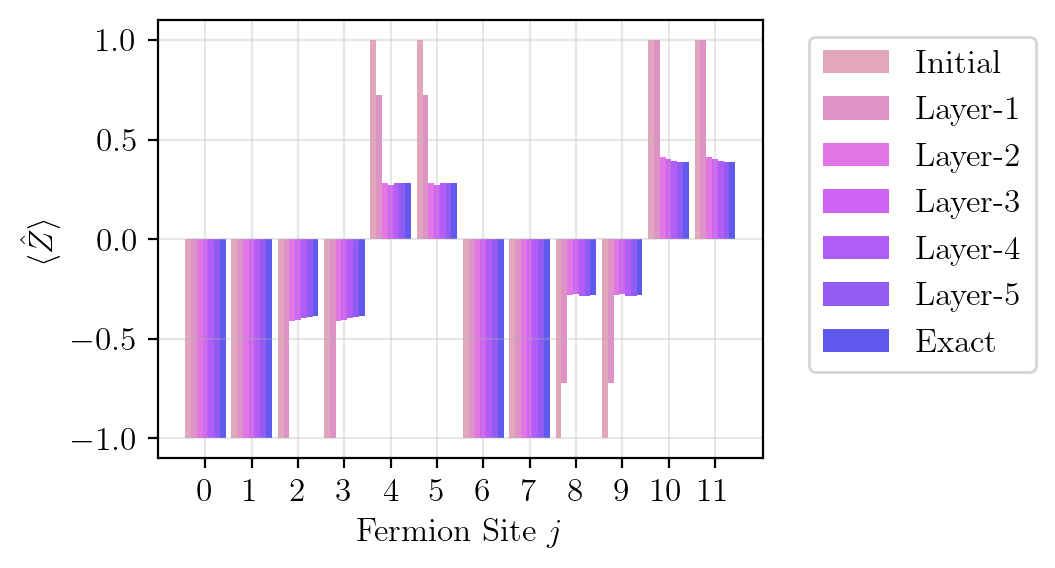}
\ \ 
\includegraphics[width=0.48\linewidth]{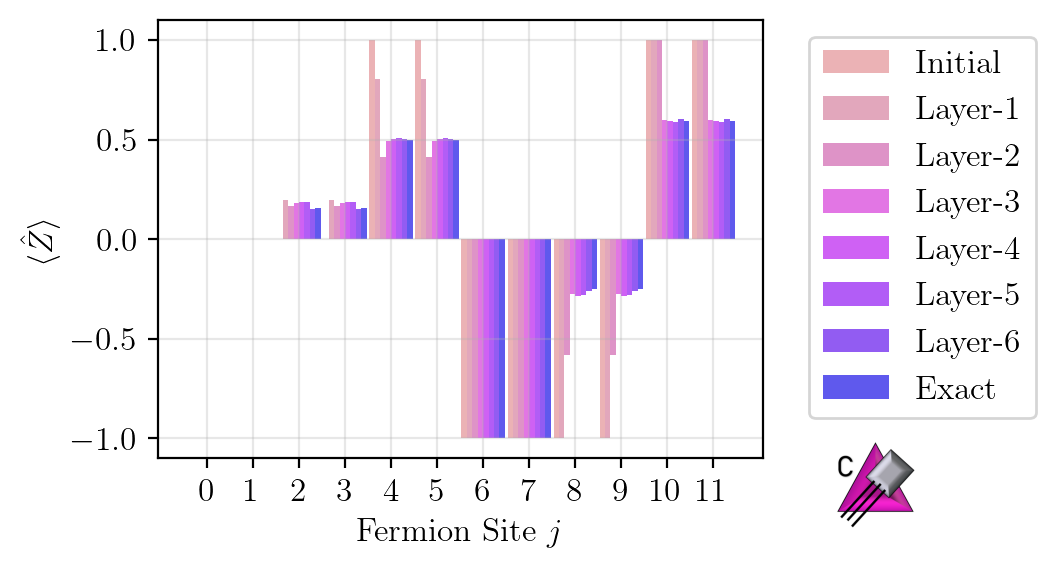}
\caption{The $\langle \hat Z_j \rangle$ as a function of the fermion site $j$, 
starting  from the SC vacuum, through the classically optimized layers, through to the exact values for $L=2$ with $m_q=0.1$ and $g=1.0$.  
The left panel shows the $\langle \hat Z_j \rangle$ in the $n_Q=0$ sector with up to five layers of optimization starting from $|SC; 2, 0\rangle$ toward $|I; 2, 0\rangle$.  
The right panel shows the analogous results in the $n_Q=1$ sector with up to six layers of optimization starting from $|SC; 2, 1\rangle$ toward $|I; 2, 1\rangle$.
The numerical values of the results are given in Table~\ref{tab:L2Q0Z} and Table~\ref{tab:L2Q1Z}.
}
    \label{fig:L2nQ01Z}
\end{figure}
Figure~\ref{fig:L2nQ01Z} shows the convergence of the $\langle \hat Z_j \rangle$ as a function of the fermion site $j$, starting from $|SC; 2, 0\rangle$ in the $n_Q=0$ sector. 
The small icons added to the figures throughout this manuscript represent wether they have been obtained using a classical simulator (purple triangle) or with quantum hardware (blue diamond)~\cite{Klco:2019xro}.
The meson layers are seen to converge the (single qubit) $\langle \hat Z_j \rangle$ to be close to their exact values, but the infidelity of the wavefunction is above $\sim 2\%$.
Including the baryon-antibaryon operators reduces the infidelity to below $\sim 0.1\%$.

The situation is more complicated in the $n_Q=1$ sector because of the lack of translational invariance.
The structure around the heavy quark emerges from the variational optimization of the operator pool, with an optimized wavefunction of the form

\begin{equation}
\hat U^{n_Q=1}_{L=2} = 
e^{-i \theta_6 \hat O_{M0}^{(0)}}\ 
e^{-i \theta_5 \hat O_{B1}^{(0,1)}}\
e^{-i \theta_4 \hat O_{B0}^{(1)}}\ 
e^{-i \theta_3 \hat O_{M0}^{(1)}}\ 
e^{-i \theta_2 \hat O_{M1}^{(0,1)}}\ 
e^{-i \theta_1 \hat O_{M0}^{(0)}}
\ ,
\label{eq:L2UNITQ1}
\end{equation}
applied to the SC state $|SC; 2, 1\rangle$.
The convergence of the fidelity of this six-parameter variational wavefunction is given in Table~\ref{tab:L2VQEmanuelQ1}, and of the associated single-qubit $\langle \hat Z_j \rangle$ in Table~\ref{tab:L2Q1Z}.
Figure~\ref{fig:L2nQ01Z} shows the convergence of the $\langle \hat Z_j \rangle$ as a function of the fermion site $j$, starting from $|SC; 2, 1\rangle$.
While the SC ground state has the light quarks screening the heavy quark color, it is the meson correlations that dominate the interacting ground state, as seen from the operator ordering.  This is discussed more in App.~\ref{app:L2}.

\subsection{Implementing Domain Decomposition}
\label{sec:L3SP}
\noindent
The $L=3$ system, supported on 18 qubits, 
provides an opportunity to demonstrate a simple implementation of DDec by combining results and optimizations from the $L=1$ system, detailed in Sec.~\ref{sec:L1},
and the $L=2$ system, detailed in 
Sec.~\ref{sec:L2} and App.~\ref{app:L2}.
This is also the simplest system in which to implement heavy-quark motion.

Working with $g=1.0$ and $m_q=0.1$, the numerically exact wavefunctions for the $L=3$ system in the $n_Q=0,1$ systems are obtained using Lanczos algorithm, and are shown in App.~\ref{app:L3}.
Also given in that appendix are the energy densities associated with the components of the Hamiltonian and the associated $\langle \hat Z_j \rangle$.
The same methods used for $L=1$ and $L=2$ can be used to create the interacting ground state variationally with subpercent infidelity.  With $L=3$ there are additional operators that are added to the operator pool, 
which introduce meson and baryon-antibaryon correlations
between adjacent sites, 
\begin{equation}
\hat{O}_{M2} \equiv (\hat X \hat Z^7\hat Y - \hat Y \hat Z^7 \hat X )   \hat I+ \hat I   (\hat X \hat Z^7\hat Y - \hat Y \hat Z^7 \hat X )
\ ,\ 
    \hat O_{B2} \equiv  -4i(\hat\sigma^+\hat\sigma^+ \hat{I}^6 \hat\sigma^-\hat\sigma^-
    - {\rm h.c.})
\ \ ,
\label{eq:OM2B2def}
\end{equation}
in addition to the shorter-range correlations from the operators in Eq.~\eqref{eq:OPxyL2def} and Eq.~\eqref{eq:mesOpsL1}.
Thus, the operator pool is extended to
\begin{equation}
\hat O_{M0}^{(0)}, \ 
\hat O_{M0}^{(1)}, \ 
\hat O_{M0}^{(2)}, \ 
\hat O_{M1}^{(0,1)}, \ 
\hat O_{M1}^{(1,2)}, \ 
\hat O_{M2}^{(0,1)}, \ 
\hat O_{M2}^{(1,2)}, \ 
\hat O_{B0}^{(0)}, \ 
\hat O_{B0}^{(1)}, \ 
\hat O_{B0}^{(2)}, \ 
\hat O_{B1}^{(0,1)} \ 
\hat O_{B1}^{(1,2)} \ 
\hat O_{B2}^{(0,2)} \ 
\ .
\label{eq:OPxyL3}
\end{equation}
In $\hat{O}_{Md}$, $d$ indicates that there are $d^2 + d + 1$ number of $\hat Z$s in between the $\hat X$ and $\hat Y$; 
in $\hat{O}_{Bd}$, 
the number of $\hat I$s is $3d$ for even $d$ and $3d-1$ for odd $d$. 
As preparing the ground state in the $n_Q=0$ sector is not the focus of this work, 
we do not provide details or discussions (see App.~\ref{app:L3}), 
and focus on the $n_Q=1$ sector.

\begin{figure}[!ht]
    \centering \includegraphics[width=0.75\linewidth]{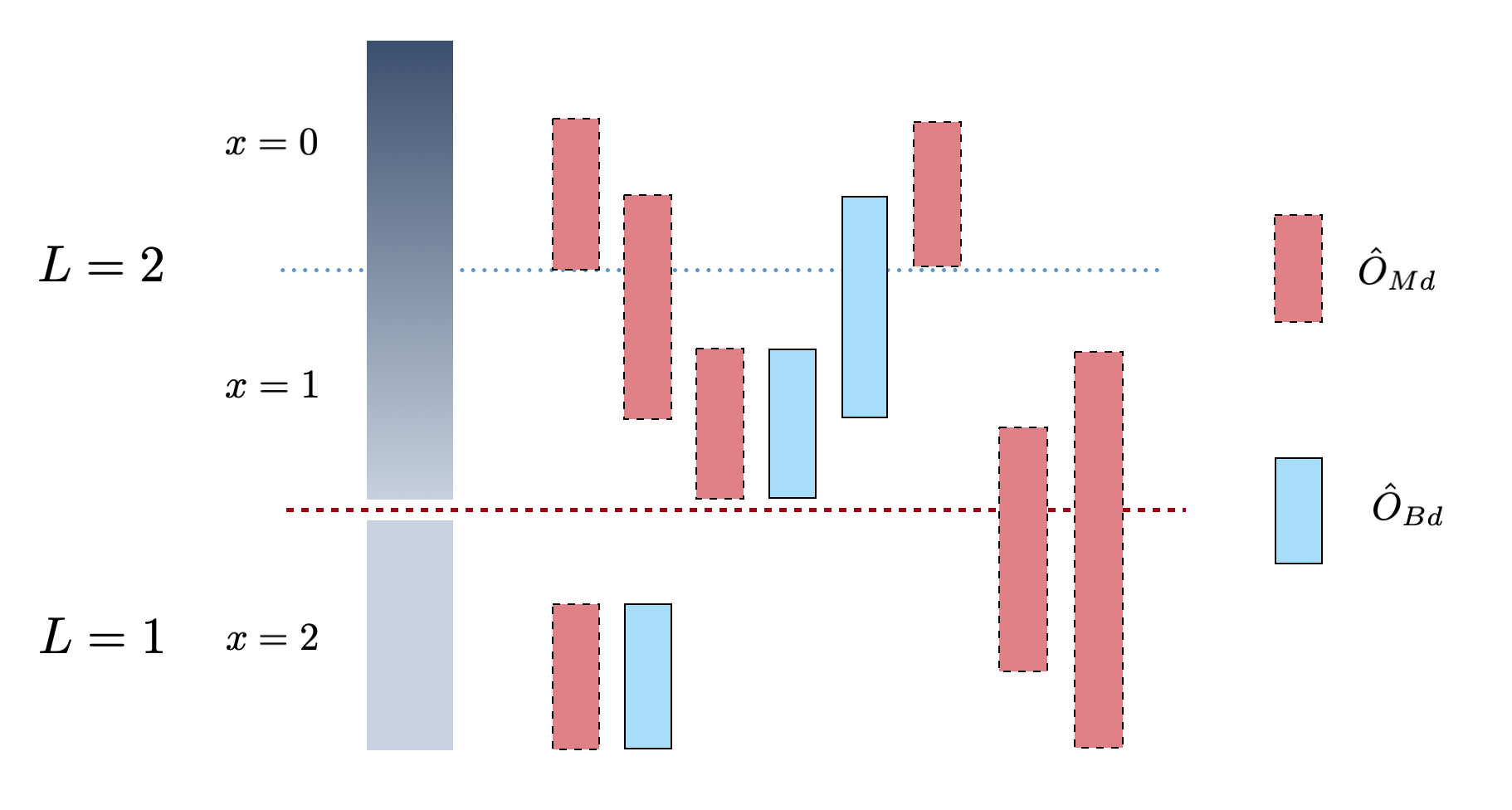}
    \caption{A block diagram for the DDec of the variational wavefunction in the $n_Q=1$ sector for $L=3$. 
    The $L=3$ wavefunction is initialized from the variationally optimized $n_Q=1, L=2$ (with the heavy quark located at $x=0$) and  $n_Q=0, L=1$ wavefunctions, as indicated by the operators acting either side of the dashed red line.  
    The re-optimized operators in each sector are implemented in parallel, followed by the application of boundary operators that entangle the two sectors, which are also optimized (with all of the operators globally optimized).
    }
    \label{fig:DDecL3}
\end{figure}

\begin{figure}[ht!]
    \centering
\includegraphics[width=0.8\linewidth]{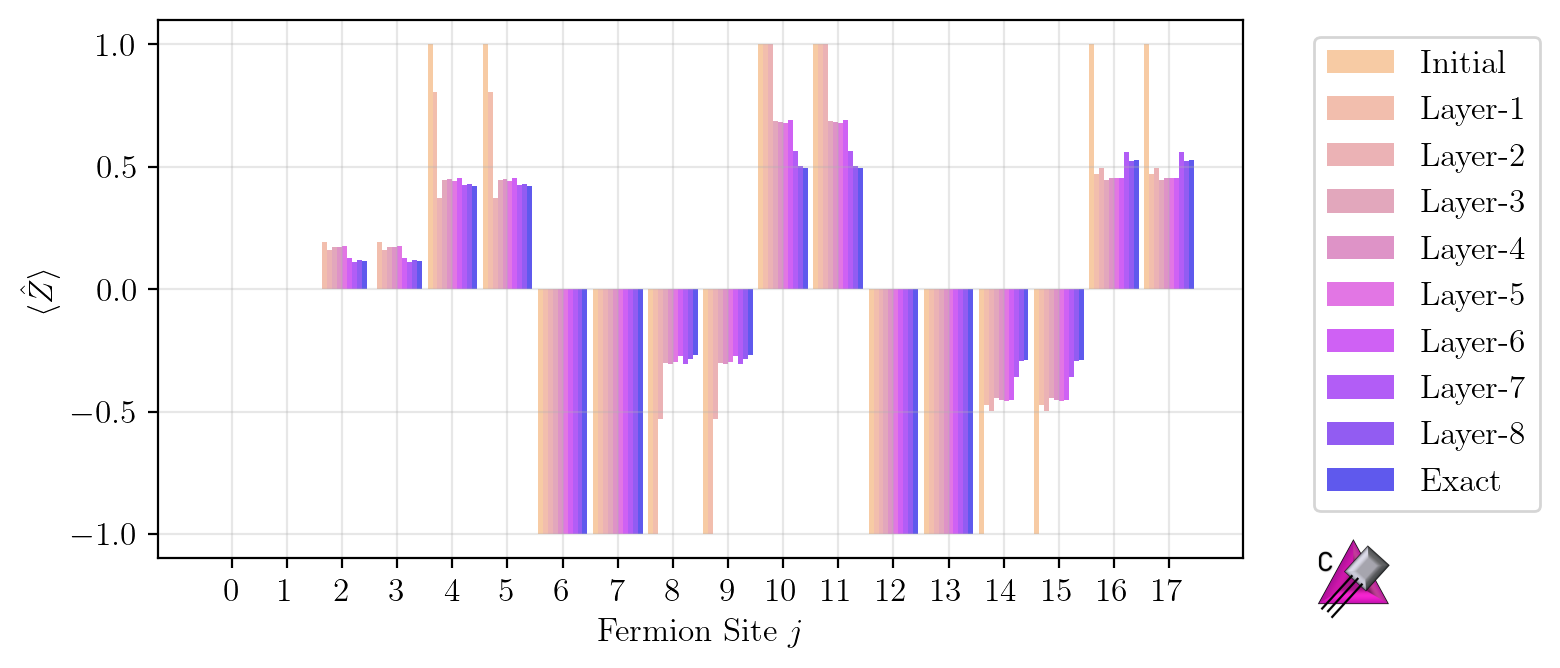}
\caption{The $\langle \hat Z_j \rangle$ as a function of fermion site $j$,  from $|SC; 3, 1, x=0\rangle$,  
through the 8 classically-optimized layers (corresponding to the application of the 10 operators in Eq.~\eqref{eq:L3U}, as shown in Fig.~\ref{fig:DDecL3}), 
through to the exact values from $|I; 3, 1, x=0\rangle$ for $m_q=0.1$ and $g=1.0$.  
The heavy quark is located at $x=0$ in sites $j=\{0,1\}$, and  $j=\{6,7,12,13\}$ correspond to unoccupied heavy-quark sites.
The numerical values of the results are given in Table~\ref{tab:L3Q1Z}.
}
    \label{fig:L3nQ1Z}
\end{figure}

To prepare the ground state of the $n_Q=1$ sector with a heavy quark located at $x=0$
using DDec,
the $L=3$ wavefunction is initialized with the operator sequences and angles
determined from the variationally-optimized 
$n_Q=1, L=2$, with the heavy quark located at $x=0$, 
and  $n_Q=0, L=1$ systems, as shown in the schematic diagram in Fig.~\ref{fig:DDecL3}.
As the $L=3$ target wavefunction differs from the tensor product of the 
$L=2$ and $L=1$ wavefunctions, 
with modifications by the boundary that
penetrate into both sectors determined by the confinement scale,
the fit angles in both sectors are re-optimized layer-by-layer.\footnote{Such re-optimizations in the bulk of the sectors
will become decreasingly significant with increasing lattice subvolumes, as the fit parameters asymptotically converge to fixed-point values in a confining theory~\cite{Klco:2019yrb,Klco:2020aud,Farrell:2023fgd,Farrell:2024fit}.}
This is followed by optimizing boundary operators that entangle the two sectors.
Through optimization, we have identified a variational wavefunction that is prepared by a unitary operator defined in terms of 10 angles,
\begin{align}
\hat U^{n_Q=1}_{L=3} = \ & 
e^{-i \theta_{10} \hat O_{M2}^{(1,2)} }\ 
e^{-i \theta_{9} \hat O_{M1}^{(1,2)} }\ 
e^{-i \theta_{8} \hat O_{M0}^{(0)} }\ 
e^{-i \theta_{7} \hat O_{B1}^{(0,1)} }\ 
e^{-i \theta_{6} \hat O_{B0}^{(1)} }\ 
\nonumber\\
&
\times e^{-i \theta_{5} \hat O_{M0}^{(1)} }\ 
e^{-i \theta_{4} \hat O_{B0}^{(2)} }\ 
e^{-i \theta_{3} \hat O_{M1}^{(0,1)} }\ 
e^{-i \theta_{2} \hat O_{M0}^{(2)} }\ 
e^{-i \theta_{1} \hat O_{M0}^{(0)} }
\ ,
\label{eq:L3U}
\end{align}
acting on the $n_Q=1$ SC ground state, $|SC; 3, 1, x=0\rangle$,
given in Eq.~\eqref{eq:SCL3n1x0}, with a third index to denote the location of the heavy quark.
As shown in Fig.~\ref{fig:DDecL3}, the first two layers interleave operations on the $L=1$ and $L=2$ sections, followed by four layers acting on the $L=2$ section, and followed by two entangling layers at the boundary.
As each layer is applied, the fit angles from the individual sections provide starting points for further global optimizations.
The results of minimizing the infidelity of the  wavefunction prepared with the unitary transformation defined in Eq.~\eqref{eq:L3U}, with respect to the exact wavefunction, are shown in Table~\ref{tab:L3DoDeQ1}.
With the two layers of the $L=1$ applied in parallel with the four layers of the $L=2$ region, followed by two layers to heal the boundary, an infidelity density below $0.5\%$ is achieved.
The $\langle \hat Z_j\rangle$ obtained with each layer are  displayed in Fig.~\ref{fig:L3nQ1Z}, and given in Table~\ref{tab:L3Q1Z}.
The impact of optimization within each section is clear from the behavior of the $\langle \hat Z_j \rangle$, which converge rapidly within each sector, but slower in the boundary region $j=\{10,11\}$.

The main reason that DDec is effective in this small system when combining the ground states of the $L=2$ and $L=1$ sections as a first approximation to the $L=3$ wavefunction is that the coupling, 
$g=1$, is relatively strong, giving rise to short correlation lengths.
This implies that the boundary corrections are modest and easily captured. 
Approaching the continuum limit, where $g\rightarrow 0$, will require increasing spatial volumes to accommodate the increasing correlation length(s).

\subsubsection{Color Complexity}
\label{sec:L3taus}
\noindent
Because of the key role of color entanglement in non-Abelian systems, we further explore its structure among the heavy quark and the light quarks, along with the quantum complexity measured by stabilizer R\'{e}nyi entropies (SRE)~\cite{Leone:2021rzd,Haug:2022vpg,Haug:2023hcs,Leone:2024lfr,Bittel:2025yhq}.
Of particular interest is the behavior of entanglement as a function of distance from the heavy quark to identify long-range quantum correlations, for which we chose to compute the mutual information $I^{(0,x)}_f$, which contains both classical and quantum correlations, and the 4-tangle~\cite{Wong:2000cmz}, $\tau_{4f}^{(0,x)}$, for $f=q,\overline{q}$ (light quarks and anti-quarks) separately.

The mutual information, $I^{(0,x)}_f$, between the heavy quark and light quarks or anti-quarks at spatial site $x$ is defined to be 
\begin{equation}
I^{(0,x)}_f = S(\hat\rho_0^{Q}) + S(\hat\rho_x^{f}) - S(\hat\rho_{0,x}^{Qf})
\ ,
\label{eq:I0x}
\end{equation}
where $\hat\rho_x^{f}$ is the ($4\times 4$) reduced density matrix of fermion-type $f$ at spatial site $x$, constructed by $\hat\rho_x^{f}={\rm Tr}_{\rm rest}[\hat\rho]$. 
$\hat\rho_{0,x}^{Qf}$ is the ($16\times 16$) reduced-density matrix from the pair of fermions, and $S(\hat\rho)$ is the von Neumann entropy, $S(\hat\rho)=-{\rm Tr}[\hat\rho\log_2\hat\rho]$.
The mutual information between the heavy quark and light anti-quarks is similarly defined.
The 4-tangle, $\tau_{4f}^{(0,x)}$, between the heavy quark at spatial site 0 and the light quarks (or anti-quarks) at spatial site $x$ is defined by matrix elements of the product $\hat Y_0^{Q_r} \hat Y_0^{Q_g} \hat Y_x^{f_r} \hat Y_x^{f_g}$,
\begin{equation}
\tau_{4f}^{(0,x)} = 
|\langle\psi| 
\hat Y_0^{Q_r} \hat Y_0^{Q_g} \hat Y_x^{f_r} \hat Y_x^{f_g}
| \psi^*\rangle |^2
\ .
\label{eq:tau40x}
\end{equation}
Typically, sub-tangles among the contributing sites, such as $\tau_{2q}^{(0,x)}$ are also evaluated, but these all vanish in this case.

\begin{figure}[ht!]
    \centering
\includegraphics[width=0.95\linewidth]{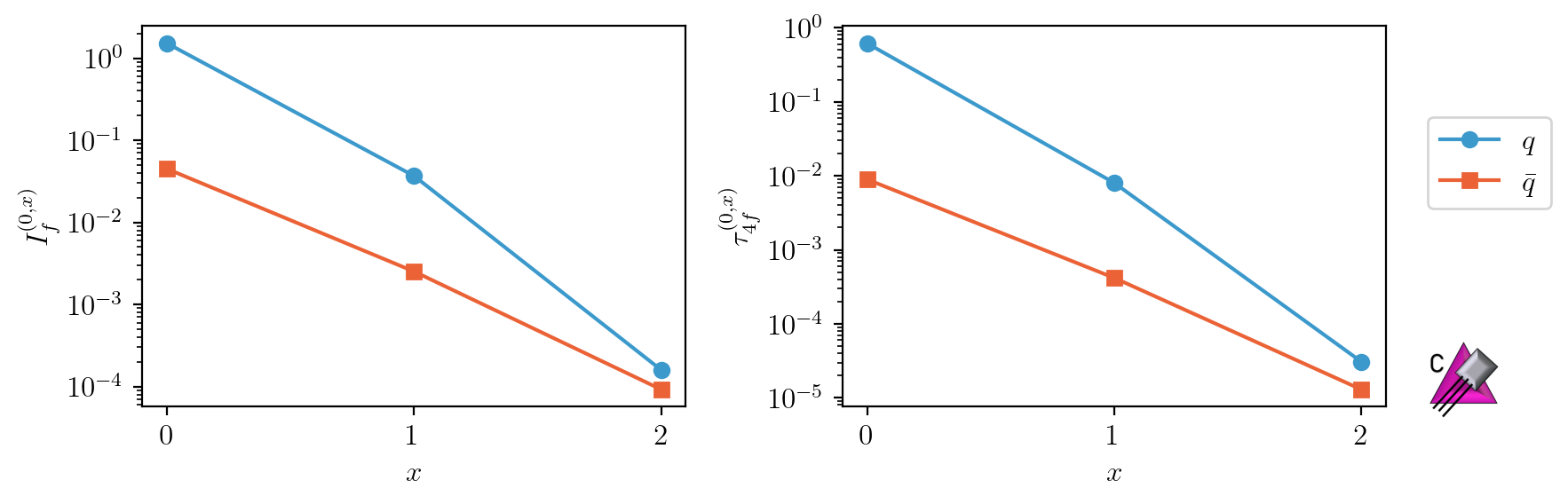}
\caption{The mutual information (left panel, defined in Eq.~\eqref{eq:I0x}) and 4-tangle (right panel, defined in Eq.~\eqref{eq:tau40x}) between the heavy quark at $x=0$ and the quarks and anti-quarks at $x=\{0,1,2\}$ in the $L=3$ system with $m_q=0.1$ and $g=1.0$.
Numerical values of the displayed points given in Table~\ref{tab:L3Q1Iijtau4}.
}
    \label{fig:L3nQ1Y4}
\end{figure}

Figure~\ref{fig:L3nQ1Y4} displays the mutual information and 4-tangles among the heavy and light quarks or anti-quarks as a function of spatial separation from the heavy quark, with numerical values given in Table~\ref{tab:L3Q1Iijtau4}.
As baryons preferentially screen the heavy-quark charges, the mutual information and 4-tangle with quarks are significantly larger than with anti-quarks.
The decrease of entanglement
with distance is consistent with exponential, as later confirmed in large systems in Sec.~\ref{sec:largeL}.
It is interesting to note that, in the variational wavefunction, the non-zero $x=2$ values, $\tau_{4q}^{(0,2)}$ and $\tau_{4\overline{q}}^{(0,2)}$, are only possible due to the boundary operators, and would vanish in their absence.

\begin{figure}[ht!]
    \centering
\includegraphics[width=0.4\linewidth]{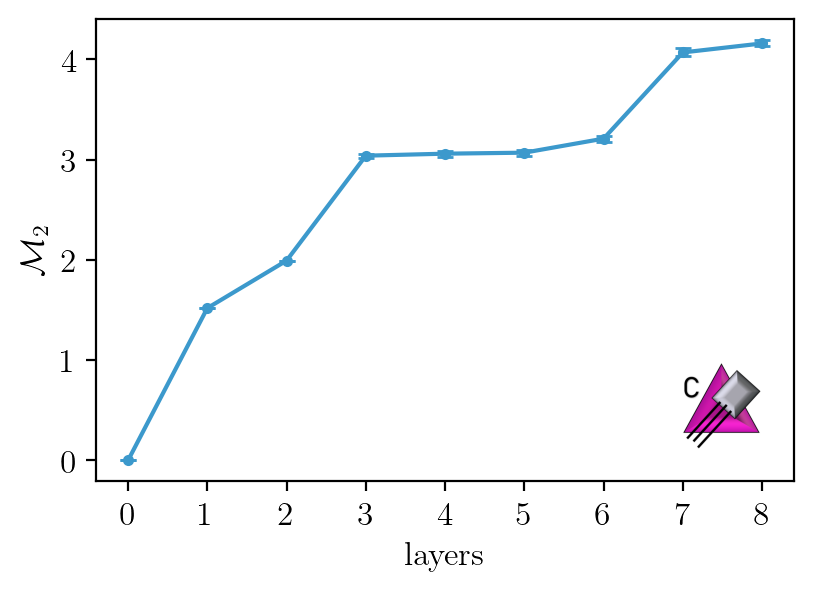}\ \ 
\caption{
The ${\cal M}_2$  non-stabilizerness in the $L=3$ wavefunction with $m_q=0.1$ and $g=1.0$ 
as a function of the number of variational layers for a  heavy quark at $x=0$.
Numerical values of the displayed points are given in Table~\ref{tab:L3DoDeQ1}.
}
    \label{fig:L3nQ1M2}
\end{figure}

A third quantity that we have examined is the quantum complexity, or quantum magic, measured by the SRE, 
${\cal M}_2$~\cite{Leone:2021rzd,Haug:2022vpg,Haug:2023hcs,Leone:2024lfr,Bittel:2025yhq}.
This provides a measure of the non-stabilizerness of a quantum state~\cite{gottesman1998a,Aaronson_2004,Bravyi_2016}, 
and how difficult it is for a classical computer to prepare it.  
A non-zero value of an SRE indicates that a state
lies outside of those
preparable 
by the Clifford gate set, 
and requires the inclusion of T-gates in quantum circuits.
It has recently been shown that the evolution of magic during 
Quantum Approximation Optimization Algorithm (QAOA) 
presents a complexity barrier, 
as the magic increases then decreases as the target state is approached with increasing numbers of unitary layers~\cite{Capecci:2025ull}.
To search for a complexity barrier in creating the ground states of the systems considered in this work, 
we compute the magic at each layer of operators creating the $L=3$, $n_Q=1$ ground state using DDec. 
For a system of $N$ qubits,
the SRE, ${\cal M}_\xi$, is defined as~\cite{Leone:2021rzd}
\begin{equation}
    {\cal M}_\xi = \frac{1}{1-\xi} 
    \log_2 \left( \sum_{\hat P\in \hat P_N} 
    \frac{\langle\psi| \hat P |\psi\rangle^{2\xi}}{2^N} 
    \right)
    \ \ ,
    \label{eq:SRE}
\end{equation}
where $\hat P$ is a string of Pauli operators, and the sum extends over all such Pauli strings for $N$ qubits, 
denoted by $\hat P_N$.\footnote{
For $\xi=2$, and writing the wavefunction in terms of eigenstates of the $\hat Z$ operator with eigenvalues 
$\pm 1$, $|\psi\rangle = \sum_{\bm\sigma} c_{\bm\sigma} |{\bm\sigma}\rangle$,
the expression in Eq.~\eqref{eq:SRE} 
reduces to a 4-replica summation~\cite{Tarabunga:2023xmv},
\begin{equation}
e^{-{\cal M}_2} = 
\sum_{ {\bm\sigma}^{(1)} , {\bm\sigma}^{(2)} , {\bm\sigma}^{(3)} , {\bm\sigma}^{(4)}}
c_{{\bm\sigma}^{(1)}}\ 
c_{{\bm\sigma}^{(2)}}\ 
c_{{\bm\sigma}^{(3)}}\ 
c_{   {\bm\sigma}^{(1)}   {\bm\sigma}^{(2)}   {\bm\sigma}^{(3)}   }\ 
c^*_{   {\bm\sigma}^{(1)}   {\bm\sigma}^{(2)}   {\bm\sigma}^{(4)}   }\ 
c^*_{   {\bm\sigma}^{(1)}   {\bm\sigma}^{(3)}   {\bm\sigma}^{(4)}   }\ 
c^*_{   {\bm\sigma}^{(2)}   {\bm\sigma}^{(3)}   {\bm\sigma}^{(4)}   }\ 
c^*_{{\bm\sigma}^{(4)}}
\ ,
\label{eq:Tara}
\end{equation}
where products of ${\bm\sigma}^{(s)} $ denote direct element by element multiplies.
This expression provides a parametric acceleration of evaluations in sparse wavefunctions, 
and is what we implemented in this work 
(via statistical sampling).
The summation in Eq.~\eqref{eq:SRE} scales as $ \sim 4^{n_Q}$ for $n_Q$ qubits, 
independent of the number of non-zero amplitudes contributing to $|\psi\rangle$.
The properties of the Pauli strings can be used to greatly reduce the number of evaluations.
In contrast, the summation in Eq.~(\ref{eq:Tara}) scales as $\sim {\cal N}_{\cal A}^4$, where 
${\cal N}_{\cal A}$ is the number of non-zero amplitudes.
}
The values of ${\cal M}_2$ for the $L=3$ system as a function of the variational layer are shown in Fig.~\ref{fig:L3nQ1M2} and given in Table~\ref{tab:L3DoDeQ1}.
One observes a monotonic increase (within sampling uncertainties) in magic with increasing numbers of layers.
The increase in ${\cal M}_2$ plateaus toward layer-6, but jumps when layer-7 is included, corresponding to the first boundary operator that joins the $L=2$ and $L=1$ regions.
In contrast to the QAOA observation, there is no evidence for a magic barrier in preparing the $L=3$ wavefunction.

\subsection{Moving Heavy Quarks}
\label{sec:L3Qmotion}
\noindent
With the preparation of the ground state of the $L=3$ system with a heavy quark complete, we begin to explore the time evolution of the system after the heavy quark is moved from $x=0$ to $x=1$.  

\subsubsection{Through the Lattice Vacuum}
\label{sec:L3QmotionVAC}
\begin{figure}[ht!]
    \centering
\includegraphics[width=0.99\linewidth]{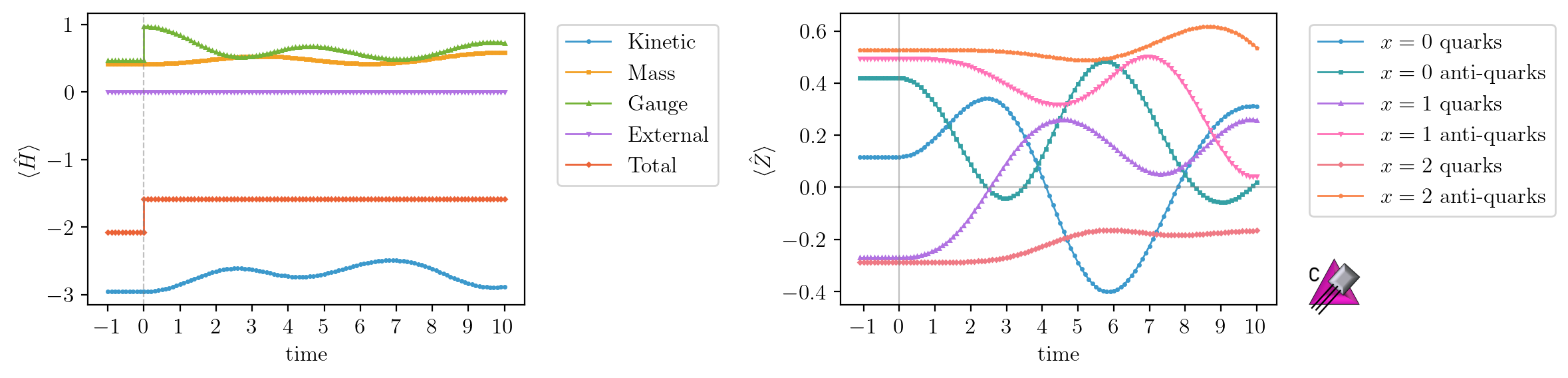}
\caption{The left panel shows the expectation values of the components of the Hamiltonian, and their total, as a function of time in the $L=3$ system.  
At $t=0$, the heavy quark is instantaneously moved from $x=0$ to $x=1$.
The right panels shows the associated time evolution of the $\langle \hat Z_j \rangle$. 
The right-justified dashed lines correspond to the values in the ground state with the heavy quark at $x=1$.
}
    \label{fig:L3Hi}
\end{figure}

\noindent
The exact (classically-determined) ground state of the $L=3$ system is prepared with a heavy-quark positioned at $x=0$.  
At $t=0$, the heavy-quark is instantaneously moved from $x=0$ to $x=1$ via the FSWAP operations defined in Eq.~(\ref{eq:FSWAP6sites}).
The system is then evolved under the exact Hamiltonian in discrete time steps, but without Trotterization with respect to the individual terms contributing to the Hamiltonian.
Using the same Hamiltonian parameters, $g=1.0$ and $m_q=0.1$, and a time step of $\delta t = 0.2$, the different contributions to $\hat H$, as well as their sum, are computed as a function of time and are displayed in the left panel of Fig.~\ref{fig:L3Hi}. 
The change of heavy-quark position at $t=0$ by one spatial site
injects energy into the system, by increasing the energy in the gauge field (at that instant).\footnote{In the infinite-volume and continuum limits, 
the magnitude of the discontinuity vanishes 
as $g\rightarrow 0$ ($a\rightarrow 0$, and the correlation length become 
infinitely large (in lattice units)), 
and with the boundaries infinitely far away from the heavy-quark position(s), 
the net energy injected vanishes by Lorentz invariance.
In contrast, as considered in our previous work~\cite{Farrell:2024mgu}, 
in the presence of matter, 
the net injected energy will not vanish, distributed into internal excitations and fragmented hadronic particles.
The extent to which the 
injected energy remains near the heavy quark in the form of stable hadronic excitation, or is radiated into propagating color-singlet states requires simulations of larger systems, 
and both effects are expected to vanish in the continuum and infinite-volume limits 
in the absence of a background medium. }
The average occupations of the quarks and anti-quarks,
$\langle \hat Z_j \rangle$,
are shown in the right panel of Fig.~\ref{fig:L3Hi}.
As expected, the responses of the quarks and anti-quarks closest to the heavy-quark's position are the most immediate and largest, while the responses at $x=2$ are delayed by the speed of light and suppressed
by distance.

The time evolution of the entanglement between the heavy quark and the light quarks or anti-quarks can also be explored.  
The 4-tangles of the initial state evolve with time, in such a way that in the continuum and infinite volume limits, the measures of entanglement will become Lorentz invariant.
That is to say, that the entanglement structure in the light quarks is expected to move with the heavy quark (at constant velocity).
\begin{figure}[ht!]
    \centering
 \includegraphics[width=0.9\linewidth]{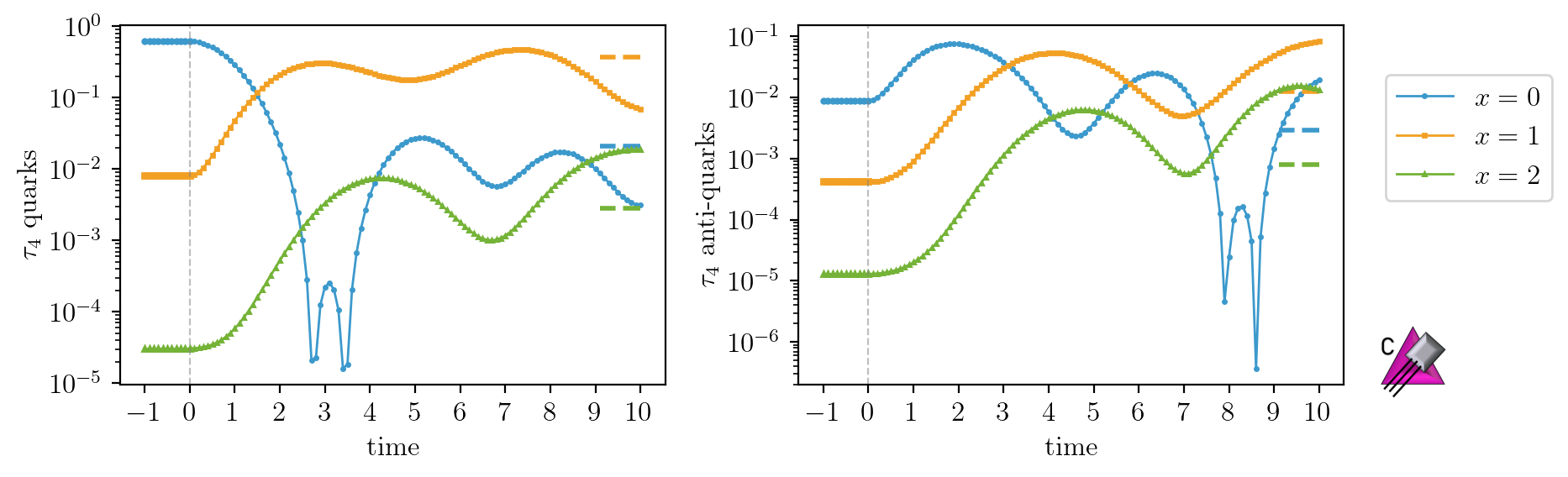}
\caption{The 4-tangles of the light quarks and anti-quarks with the heavy quark.
At $t=0$, the heavy quark is instantaneously moved from $x=0$ to $x=1$, and the definition of $\tau_4$ changes from
$\tau_{4f}^{(0,x)}$ to $\tau_{4f}^{(1,x)} $
The left panels shows the 4-tangle for the quarks, and the right panel for the anti-quarks.
The right-justified dashed lines correspond to the values in the ground state with the heavy quark at $x=1$, with numerical values given in Table~\ref{tab:L3Q1tau4x1}.
}
    \label{fig:tau4L3time}
\end{figure}
The 4-tangles, $\tau_{4f}^{(0,x)}(t)$,  defined in Eq.~\eqref{eq:tau40x} at $t=0$ as discussed above, becomes $\tau_{4f}^{(1,x)}(t)$ after the heavy quark is moved from $x=0$ to $x=1$.

Figure~\ref{fig:tau4L3time} shows the time evolution of the quark and anti-quark $\tau_4(t)$ for $L=3$ computed from the exact wavefunction(s).
Given the initial configuration, we anticipate that in the ideal, infinite-volume system, that the values for $x=0$ and $x=2$ should become approximately equal, up to internal excitations and reflections from the boundaries, due to the symmetry of the final state system.

\subsubsection{In-Medium: Energy Loss and $\frac{dE}{dx}$}
\label{sec:L3QmotionIM}
\begin{figure}[ht!]
    \centering
\includegraphics[width=0.9\linewidth]{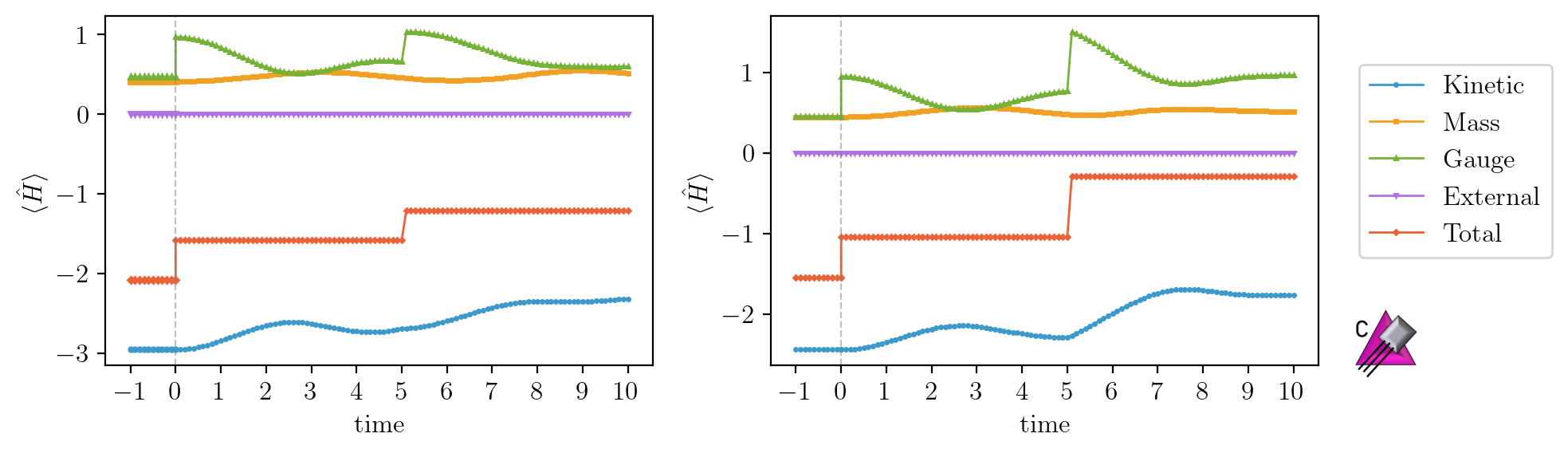}
\caption{
The contributions to the energy, and the total energy,
in the $L=3$ system, as the heavy quark initially at $x=0$ is moved to $x=1$ at $t=0$ and to $x=2$ at $t=5$.
The left panel shows the total energy in the motion across the vacuum, 
while the right panel shows total energy in the presence of a second heavy quark at $x=2$.
}
    \label{fig:Hix012}
\end{figure}

\noindent
The $L=3$ system, despite its limitations, permits a first exploration of the energy loss in the presence of matter.
To address this, and to estimate the instantaneous energy loss, $dE/dx$, in these systems with matter present, we perform two classical simulations, following the method presented in Ref.~\cite{Farrell:2024mgu}.  
Two systems are prepared: the ground state of the system with one heavy quark at $x=0$, and that with two heavy quarks, one at $x=0$ and the other at $x=2$.
In each, the heavy quark is moved from $x=0$ to $x=1$ at $t=0$ and then to $x=2$ and $t=5$, the latter time being arbitrarily selected.
The total energy of each system is determined in the three time intervals, ($t<0$, $0\le t < 5$ and $t\ge 5$) from which the difference between them can be translated into an estimator for $dE/dx$.
The contributions to the energy from components of the Hamiltonian, and the total energy as a function of time are shown in Fig.~\ref{fig:Hix012}.
The total energies in the three intervals are
\begin{equation}
    E_{\rm vac} = \{\ 0,\ 0.5030,\ 0.8721 \}   \ -\ 2.0806
    \ \ ,\ \ 
     E_{\rm med} = \{\ 0,\ 0.5002,\ 1.2595 \}\ -\ 1.5485
     \ .
\end{equation}
Estimates of $dE/dx$ are determined by the differences in total energy between adjacent plateaus, followed by a subtraction of the differences from moving in the lattice vacuum, giving,
\begin{align}
\frac{dE}{dx}^{({\rm vac})}\Big|_{x=(1,0)}^{(\Delta t=5)} & = 0.5002\ ,\ 
\frac{dE}{dx}^{({\rm vac})}\Big|_{x=(2,1)}^{(\Delta t=5)} \ =\ 0.3719\ \nonumber\\
\frac{dE}{dx}^{({\rm med})}\Big|_{x=(1,0)}^{(\Delta t=5)} & = 0.5030\ ,\ 
\frac{dE}{dx}^{({\rm med})}\Big|_{x=(2,1)}^{(\Delta t=5)} \ =\ 0.7564\ 
     \ .
\end{align}
Combining these values gives, 
\begin{align}
\langle \frac{dE}{dx}\rangle \Big|_{x=(1,0)}^{(\Delta t=5)} & = 
\frac{dE}{dx}^{({\rm med})}\Big|_{x=(1,0)}^{(\Delta t=5)}  -  \frac{dE}{dx}^{({\rm vac})}\Big|_{x=(1,0)}^{(\Delta t=5)} 
\ =\ 0.0028
\nonumber\\
\langle \frac{dE}{dx}\rangle \Big|_{x=(2,1)}^{(\Delta t=5)} & = 
\frac{dE}{dx}^{({\rm med})}\Big|_{x=(2,1)}^{(\Delta t=5)}  -  \frac{dE}{dx}^{({\rm vac})}\Big|_{x=(2,1)}^{(\Delta t=5)}
\ =\ 0.3846
     \ .
\end{align}
As the energy loss in the vacuum is entirely a lattice artifact, its subtraction from the in-medium energy loss removes this leading unphysical contribution.
These values show the impact of the matter (heavy quark) located at $x=2$ on the energy loss experienced by the moving heavy quark.
The truncations and approximations employed in this simple calculation are significant, and simulations are not expected to provide robust estimates of the true energy loss in the continuum theory.  
However, these results demonstrate the potential utility of this method.

\subsection{Large-$L$ Tensor-Network Calculations}
\label{sec:largeL}
\noindent
In order to go beyond lattice sizes limited by state vector simulations, we use tensor network techniques to study the static and dynamical properties of systems with heavy quarks. 
We represent the states as matrix product states (MPS), and use the density matrix renormalization group (DMRG) algorithm to prepare them in the correct sector, and the time dependent variational principle (TDVP) algorithm to perform time evolution.
This is implemented via the {\tt ITensor}~\cite{fishman2022itensor} and {\tt ITensorMPOContruction}~\cite{Corbett:2025flm} packages. While for ground-state preparation and static property studies a maximum bond dimension of 200 was sufficient (the variance $\langle \psi | \hat{H}^2 | \psi \rangle - \langle \psi | \hat{H} | \psi \rangle^2$ was below $10^{-5}$), for time evolution it had to be increased to 600. 
The real-time dynamics was found to be considerably harder to simulate than the static properties.

\begin{figure}[ht!]
    \centering
\includegraphics[width=0.9\linewidth]{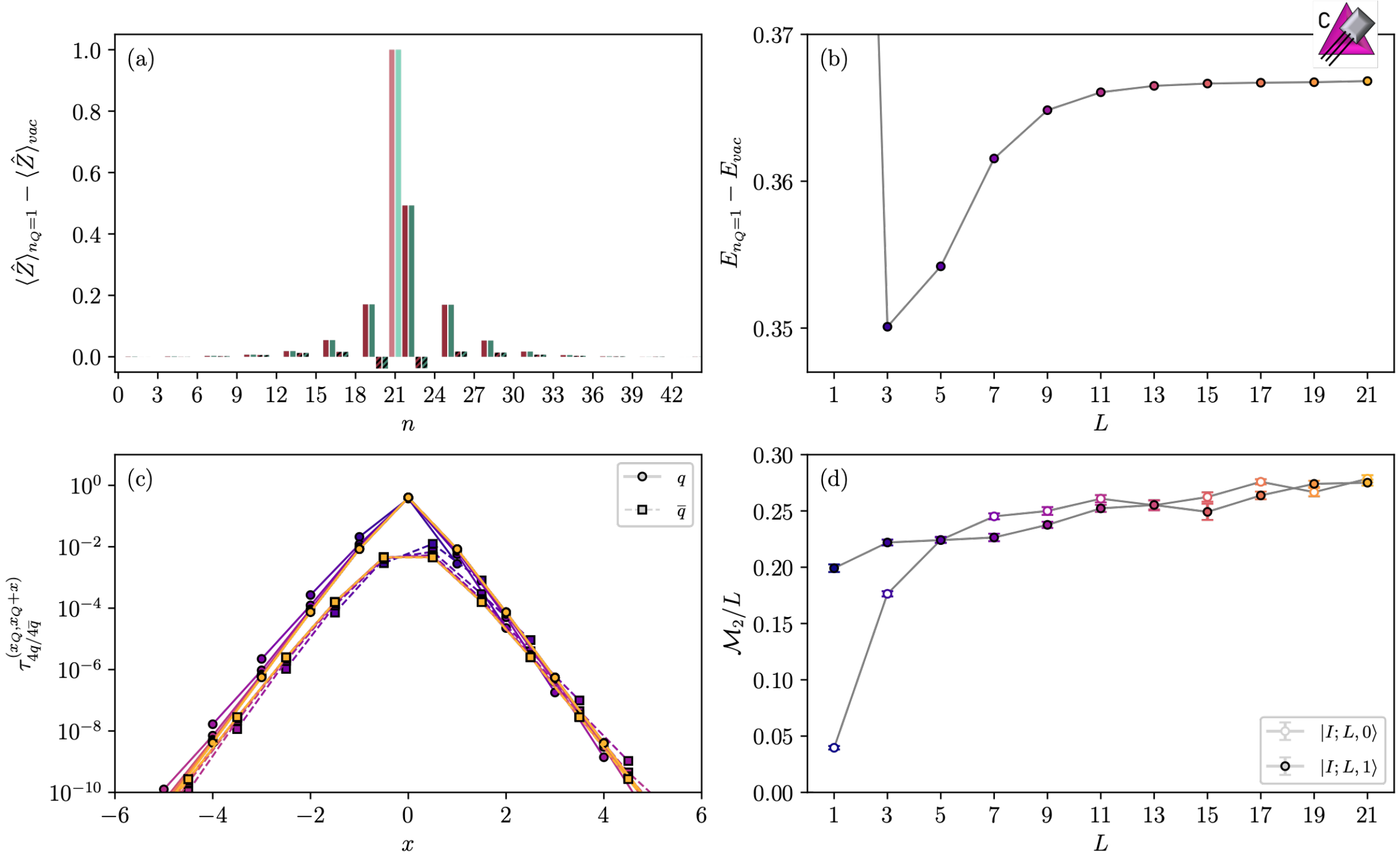}
\caption{Static properties of the system with and without a heavy quark at the center of the lattice, with $m_q=0.1$ and $g=1.0$ for large systems.
(a) The difference in the distribution of charges between the state with a heavy quark in the middle and the vacuum state, with $L=15$.  
(b) The energy difference with the vacuum as a function of the extent of the lattice $L$, reaching a plateau value of the infinite-volume gap. 
This is consistent with the expected exponential convergence of the mass of a single heavy meson with increasing lattice volume.
(c) 
The 4-tangles for quarks and anti-quarks, 
$\tau^{(x_Q,x_Q+x)}_{4q}$ and 
$\tau^{(x_Q,x_Q+x)}_{4\overline{q}}$,
for a series of $L$, where the position of the heavy quark $x_Q$ is at the middle of the lattice, displaying the exponential decay of correlations. The points for the anti-quarks have been shifted to the right for clarity. 
(d) 
The ${\cal M}_2/L$ SRE densities for both vacuum and heavy-quark states as a function of $L$.
The convergence is due to the heavy-quark wavefunction having support over a limited number of lattice sites.
}
    \label{fig:static_L}
\end{figure}
Figure~\ref{fig:static_L} shows the results obtained in a series of calculations performed on both the vacuum state, $|I,L,0\rangle$, and the state with a heavy-quark positioned at the mid-point of the lattice, $|I,L,1\rangle$, with $L\in\{1,3,\ldots,21\}$.
To observe how the light quarks screen and confine the heavy quark, in Fig.~\ref{fig:static_L}(a), 
the difference of the charges of each quark and color, measured by $\langle \hat Z \rangle$, 
between the state with and without a heavy quark, is shown.
Since the heavy quark is far away from the boundaries, a clean exponential decay in $\langle \hat Z \rangle$ is observed.

By positioning the heavy quark at the center of the lattice, the energy gap to the vacuum state can be extracted with exponential convergence in lattice size, as shown in Fig.~\ref{fig:static_L}(b).
Despite working with OBCs, where $1/L$ volume effects are present for typical quantities, this leading contribution cancels in the difference, as the heavy quark (and the screening light quarks) are sufficiently far from the boundaries.
This exponential decay of correlations is also seen in the mutual information and 4-tangles as a function of distance from the heavy quark. 
The 4-tangles $\tau^{(x_Q,x_Q+x)}_{4f}$ ($f = q, \bar q $), given in Eq.~\eqref{eq:tau40x}, are shown in Fig.~\ref{fig:static_L}(c) for a selection of (large) lattice volumes as a function of (spatial) distance to the heavy quark.
As lattice volumes increase, the correlation lengths converge.
Figure~\ref{fig:static_L}(d) shows the ${\cal M}_2/L$ SRE densities for both vacuum and heavy-quark states, computed via the perfect Pauli sampling method~\cite{Lami:2023naw,Haug:2023hcs} using 2000 samples. 
These densities are seen to converge with increasing volume, consistent with the heavy hadron having only limited spatial support on the lattice.
For the small $L$ values, as explained in Ref.~\cite{Hoshino:2025ine}, systems with OBCs should display a logarithmic dependence of $\mathcal{M}_2$ with $L$, and that is case, as seen in Fig.~\ref{fig:static_L}(d).

\begin{figure}[ht!]
    \centering
\includegraphics[width=\linewidth]{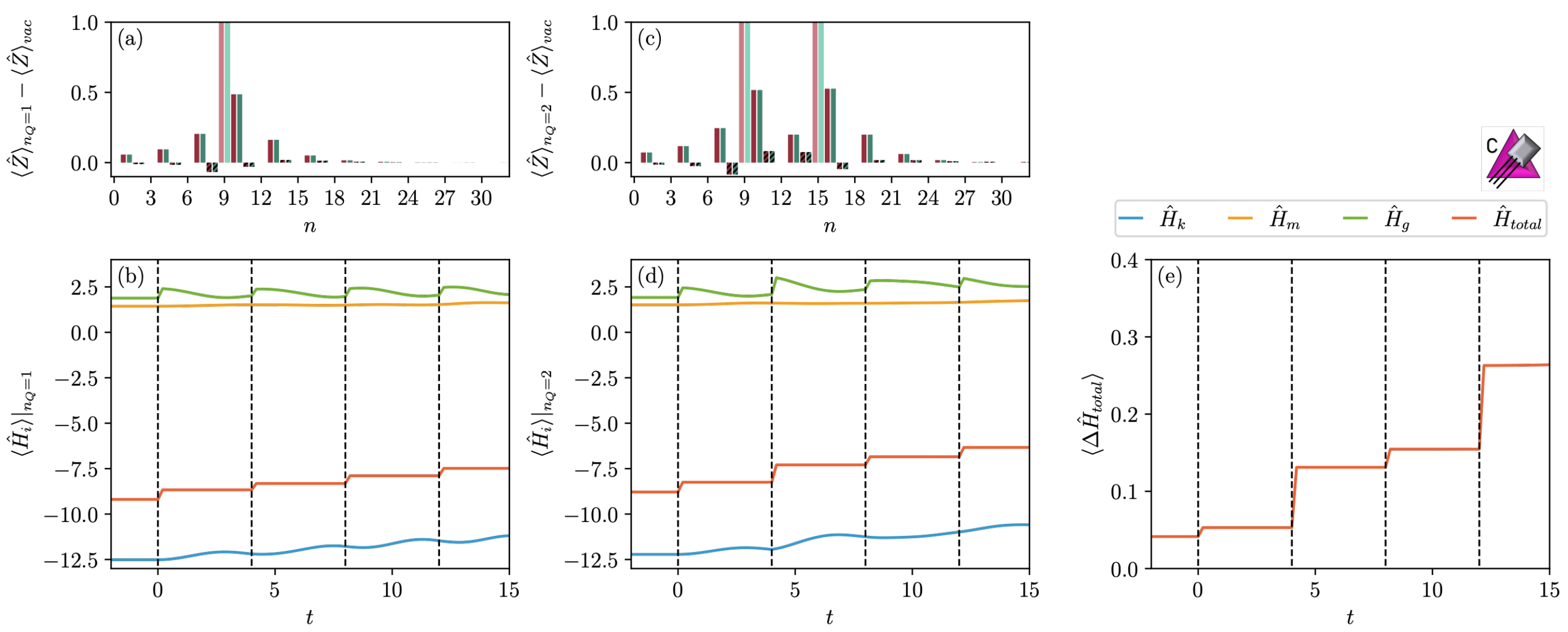}
\caption{Time evolution of the system with one and two heavy quarks, with $m_q=0.1$, $g=1.0$ and $L=11$. 
Panel (a) shows the initial charge distribution for a single heavy quark at $x=3$, and  
panel (b) shows the time evolution of the total energy and individual terms of the 
Hamiltonian as the charge is moved one site every $\Delta t = 4$ (dashed vertical lines). 
Panel (c) shows the initial charge distribution with two heavy quarks, at $x=3$ and $x=5$. 
(d) Same as (b), but moving the heavy quark initially at $x=3$, with another heavy quark static at $x = 5$.
(e) The difference in the energy contributions to the 
systems with and without a second heavy quark, defined in Eq.~\eqref{eq:diffH}.
}
    \label{fig:dynamic_L}
\end{figure}
Finally, we compute the associated energy differences when one heavy quark is moved across the lattice in the presence of a second heavy quark.
Figure~\ref{fig:dynamic_L} shows the initial charge distributions and time-dependent energies as one heavy quark is moved to adjacent spatial sites at regular time intervals, $\Delta t=4$, for an $L=11$ lattice.

Panels (a) and (b) show the results obtained without the second heavy quark, where any shift in energy is due to lattice artifacts (broken Lorentz symmetry).
Similarly, panels (c) and (d) show these quantities in the presence of a second heavy quark, where a difference in energy loss due to the collision is observed
With vacuum and single-heavy quark contributions removed, panel (e) shows $\langle \Delta \hat{H}_{total} \rangle$ for the two heavy-quark system as a function of time, defined as
\begin{equation}
    \langle \Delta \hat{H}_{total} \rangle\  =\  
    \langle \hat{H}_{total} \rangle|_{n_Q=2} 
    \ -\  \langle \hat{H}_{total} \rangle|_{n_Q=1,\ x(0)=3}
    \ -\  \langle \hat{H}_{total} \rangle|_{n_Q=1,\ x=5}
    \ +\  \langle \hat{H}_{total} \rangle|_{n_Q=0} 
    \ ,
\label{eq:diffH}
\end{equation}
where $\langle \hat{H}_{total} \rangle|_{n_Q=2} $ refers to the total energy of the system with both quarks, and $\langle \hat{H}_{total} \rangle|_{n_Q=1,\ x(0)=3}$ refers to the energy with one moving quark starting at $x=3$.
Both $\langle \hat{H}_{total} \rangle|_{n_Q=2} $ and $\langle \hat{H}_{total} \rangle|_{n_Q=1,\ x(0)=3}$ are time dependent, depending upon the location and speed of the moving heavy quark. 
$\langle \hat{H}_{total} \rangle|_{n_Q=1,\ x=5}$ refers to the energy with a static heavy quark at $x=5$, and $\langle \hat{H}_{total} \rangle|_{n_Q=0}$ refer to the energy of the vacuum, both independent of time.
This difference between calculated energies removes the leading-order lattice artifacts, and would vanish if there were no interactions between the heavy mesons.

\section{Quantum Simulation of SU(2) in 1+1D}
\label{sec:QSim}
\noindent
This section presents the methods, circuits and error-mitigation techniques that we have used to perform quantum simulations of dynamics in the system described above on three physical sites $L=3$ (which maps to 18 qubits), and the results we have obtained using IBM's {\tt heron} processor quantum computer {\tt ibm\_pittsburgh}.
The ground state of the system with a heavy quark present is first prepared via ADAPT-VQE and DDec, then the heavy quark is translated to an adjacent physical site by FSWAP operations. After translation, the state is time-evolved with the system Hamiltonian to simulate the dynamic response of the light quarks. 
The quantum circuits are designed with the objective of minimizing CNOT depth and count in the context of nearest-neighbor connectivity, as applicable to superconducting-qubit architectures. 
A resource analysis,  including the scaling of circuit resources with regard to the number of physical sites and time evolution steps,  is discussed. 
We measure the associated charge distributions and energy losses using IBM's {\tt ibm\_pittsburgh}, and compare with classically calculated results.

\subsection{Algorithms, Techniques and Quantum Circuits}
\label{sec:ApTpC}
\noindent
There are two types of operators involved in the state preparation and time evolution: those with two $\hat \sigma^\pm$s, with $-\hat{Z}$s inserted in between to accommodate the fermion-exchange anti-symmetry, and those with four $\hat \sigma^\pm$s.
In the two $\hat \sigma^\pm$s case, the form $(\hat{\sigma}^{-}\hat Z^n\hat{\sigma}^{+} - {\rm h.c.})$ is used in the meson operators in state preparation (to furnish real wavefunctions), while $(\hat{\sigma}^{-}\hat Z^n\hat{\sigma}^{+}+ {\rm h.c.})$ is used in time evolution from the kinetic term. 
Operators with four $\hat \sigma^\pm$s can be implemented using GHZ transformations, as presented in Refs.~\cite{Stetina:2020abi,Farrell:2022wyt}. 
Baryon operators of the form $\hat\sigma^+\hat\sigma^+\hat\sigma^-\hat\sigma^- - {\rm h.c.}$ are used in state preparation, and those of the form $\hat\sigma^+\hat\sigma^-\hat\sigma^-\hat\sigma^+ + {\rm h.c.}$ are used in time evolution from the gauge Hamiltonian, implemented by GHZ transformation circuits.
The GHZ transformation used to implement the gauge Hamiltonian evolution is also used in measuring the energies.

\subsubsection{R-box Constructions}
\label{sec:Rboxes}
\begin{figure}[htpb]
    \centering
    \includegraphics[width=0.75\textwidth]{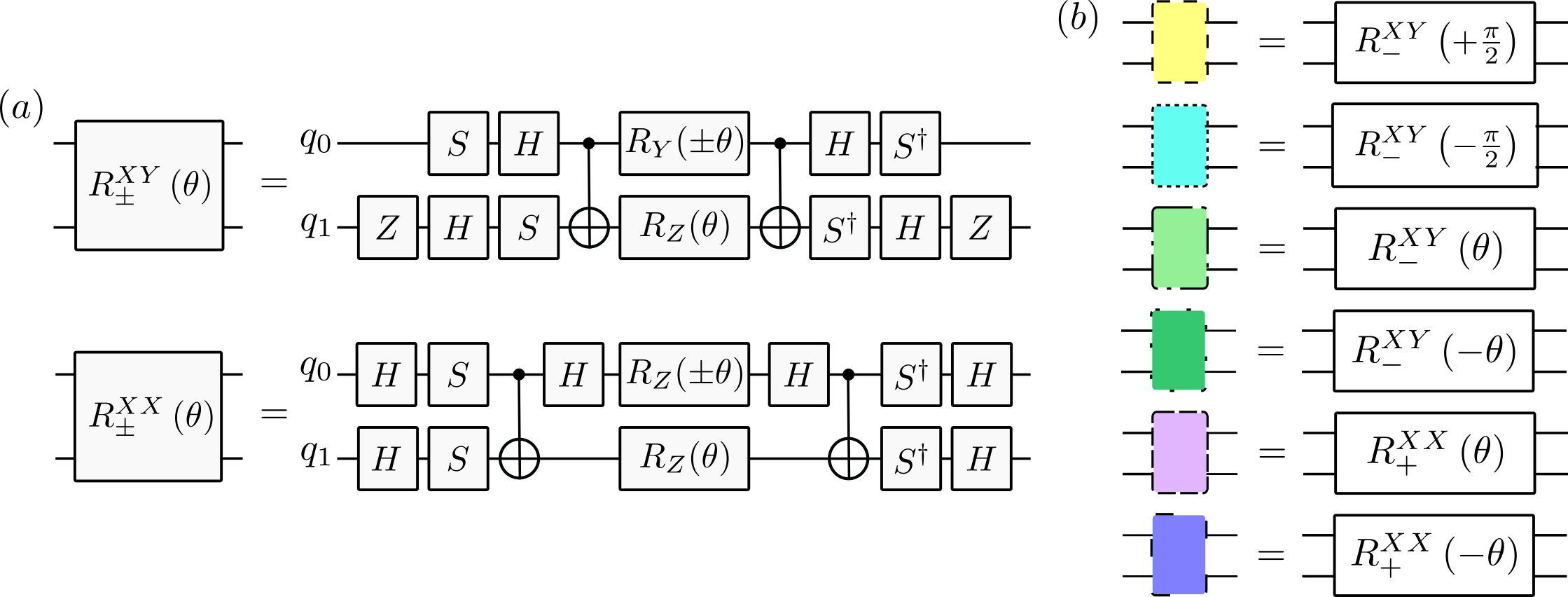}
    \caption{Panel (a) shows the circuits implementing $R_{\pm}^{XY}$ and $R_{\pm}^{XX}(\theta)$, and
    panel (b) shows the shorthand notation that will be used in subsequent (composite) circuits.}
    \label{fig:rbox}
\end{figure}

\noindent
Building upon the work in Refs.~\cite{Algaba:2023enr,Farrell:2023fgd,Farrell:2024fit}, the circuits for operators with two $\hat \sigma^\pm$s can be implemented for nearest-neighbor connectivity with two-qubit blocks, referred to as R-boxes, defined as
\begin{equation}
    R_{\pm}^{XY}(\theta) \equiv   e^{-i\frac{\theta}{2}(\hat{Y}\hat{X} \pm \hat{X} \hat{Y})} \ , \
    R_{\pm}^{XX}(\theta)  \equiv  e^{-i\frac{\theta}{2}(\hat{X}\hat{X} \pm \hat{Y} \hat{Y})} \ .
    \label{eq:RboxDef}
\end{equation}
Circuits for implementing these building blocks are shown in Fig.~\ref{fig:rbox}(a), and the shorthand notation we will use to represent them in subsequent (larger) circuits are given in Fig.~\ref{fig:rbox}(b).
These R-boxes can be sequentially layered to implement long-ranged operators in the form of $e^{-i\theta (\hat{\sigma}^{\pm}\hat Z^n\hat{\sigma}^{\mp} \pm {\rm h.c.})}$, examples of which are given in App.~\ref{appen:cir}.

\subsubsection{State Preparation}
\label{sec:StatePrepCir}
\noindent 
A few steps are taken to prepare the ground state of the interacting theory in the presence of a heavy quark.
First, the  SC ground state is prepared, as given in Eq.~\eqref{eq:SU2SCQ1} for one spatial site, and Eqs.~\eqref{eq:nq2L2} and \eqref{eq:SCL3n1x0} for $L=2$ and $L=3$ sites, respectively.
The corresponding circuit to prepare the SC ground state for the spatial site with the heavy quark is given by
\begin{equation}
 \begin{tikzpicture}[baseline={([yshift=1.0 ex]current bounding box.center)}]
  \node[scale=0.8] {
\begin{quantikz}[row sep={0.9cm,between origins}]
    \lstick{$\ket{Q_r}$}      & \gate{X}\slice[style=colorslice]{} &          &           & \targ{}   &   \\
    \lstick{$\ket{Q_g}$}      & \gate{X}          & \gate{X} & \targ{}   & \ctrl{-1} &   \\   
    \lstick{$\ket{q_r}$}      & \gate{X}          & \gate{H} & \ctrl{-1} & \ctrl{1}  &   \\
    \lstick{$\ket{q_g}$}      & \gate{X}          &          &           & \targ{}   &   \\
    \lstick{$\ket{\bar q_r}$} &                   &          &           &           &   \\
    \lstick{$\ket{\bar q_g}$} &                   &          &           &           & 
\end{quantikz}
};
\end{tikzpicture}
\label{eq:stateprep1}
\end{equation}
with the remaining $L-1$ spatial sites are prepared with 
$\hat X\hat X\hat X\hat X\hat I\hat I$
acting on $|0\rangle^{\otimes 6 (L-1)}$.

DDec is used to prepare the interacting  ground state from the SC ground state,
as discussed in Sec.~\ref{sec:L3SP}.
For each lattice segment of length $l_i$, classical ADAPT-VQE is used to establish the sequence of unitary operators and their corresponding coefficients that prepare the ground state with fidelity greater than a prescribed value.
The lengths of these lattice segments should ideally be chosen to be larger than the longest correlation length in the theory.
After each lattice segment is optimized by minimizing the energy, the ground state of the full system is established by first applying the operators responsible for  preparing each segment, then addressing the interactions and boundary effects between segments by applying operators across each boundary. 
The coefficients of each of the operators are re-optimized at each stage of ``the build'' for the full ground state.
Because of the translation symmetry of regions of the state when far from the heavy quark, scalable operators can be implemented in these sections of the lattice.

Preparing the interacting heavy-quark ground state from the SC heavy-quark ground state differs from preparing it from the interacting vacuum.
That is because the segment containing the heavy quark is distinct from the vacuum, requiring a different operator sequence and set of angles to accommodate modifications to the light-quark wavefunction within correlation lengths of the heavy quark.
The same operator pool with the meson and baryon operators can be used to prepare the interacting vacuum and heavy-quark states from their corresponding SC states, but the operator sequences selected by ADAPT-VQE are different.

Collecting the results from Sec.~\ref{sec:CSim}, a general operator in the state preparation operator pool is a color singlet with significant overlap with states in the hadronic spectrum, i.e., a set of interpolating operators.   
This translates into a pool of relatively local meson and baryon operators (with size set by the confinement scale).  
These operators have their fermion operators separated by a modest number of sites, with the impact of the latter  suppressed by the confinement scale.
A set of interpolating meson operators that is found to be effective as circuit elements is (compiled from Eqs.~\eqref{eq:mesOpsL1}, \eqref{eq:OPxyL2def} and \eqref{eq:OM2B2def}):
\begin{align}
    \hat{O}_{M0} & \equiv (\hat X \hat Z\hat Y - \hat Y \hat Z\hat X )   \hat I + \hat I  (\hat X \hat Z\hat Y - \hat Y \hat Z\hat X )
    \ ,\nonumber\\
    \hat{O}_{M1} & \equiv (\hat Y \hat Z^3\hat X - \hat X \hat Z^3 \hat Y )  \hat I+ \hat I   (\hat Y \hat Z^3\hat X - \hat X \hat Z^3 \hat Y )\ ,\nonumber\\
    \hat{O}_{M2} & \equiv (\hat X \hat Z^7\hat Y - \hat Y \hat Z^7 \hat X )   \hat I+ \hat I   (\hat X \hat Z^7\hat Y - \hat Y \hat Z^7 \hat X )
     \ , 
     \label{eq:mesOps}
\end{align}
where it is understood that the $\hat X, \hat Y$ act on the light quarks and anti-quarks only.
$\hat{O}_{M0}$ acts on $r\bar r$ and $g\bar g$ on the same physical site, $\hat{O}_{M1}$ acts on $\bar r r$ and $\bar g g$ on two neighboring sites, and $\hat{O}_{M2}$ acts on $r\bar r$ and $g\bar g$ also on two neighboring sites. 
Note that $\hat O_{M1}$ has the opposite relative sign from $\hat O_{M0}$ and $\hat O_{M2}$ because of the number of $\hat Z$s it encounters when constructed from an operator commuting with the mass term.
The implementation of the $\hat U_{Md} = e^{-i\theta\hat{O}_{Md}}$ circuits can be found in App.~\ref{appen:cir}.
In terms of entangling gates, $\hat U_{M0}$ has depth 6 with a total number of 8 CNOT gates, $\hat U_{M1}$ has depth $14$ with a total number of 28 CNOT gates, and $\hat U_{M2}$ has depth 26 with a total number of 60 CNOT gates.

\begin{figure}[htpb]
    \centering \includegraphics[width=0.99\textwidth]{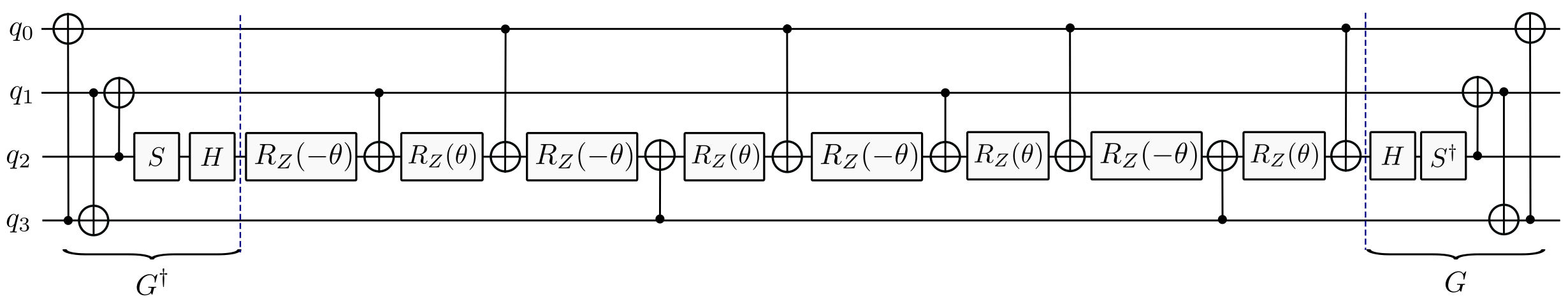}
    \caption{A circuit implementing $e^{-i\theta\hat O_{B0}}$, given
    in Eq.~\eqref{eq:B01}, where
    $G$ and $G^{\dagger}$ 
    implement a GHZ transformation. Additional details about constructing this circuit
    can be found in App.~\ref{appen:cir}.
    }
    \label{fig:baryon}
\end{figure}

The fundamental local baryon operators that create baryon-anti-baryon pairs on the same spatial site and  separated by one spatial site are,
\begin{equation}
    \hat O_{B0} = 4i(\hat\sigma^+\hat\sigma^+\hat\sigma^-\hat\sigma^- - {\rm h.c.}) 
    \ ,\ 
    \hat O_{B1} = -4i(\hat\sigma^+\hat\sigma^+ \hat{I}^2 \hat\sigma^-\hat\sigma^-
    - {\rm h.c.})
    \ \ .
    \label{eq:B01}
\end{equation}
The circuit for ${\exp}(-i\theta\hat O_{B0})$ is shown in Fig.~\ref{fig:baryon}, and ${\exp}(-i\theta\hat O_{B1})$ can be implemented with two additional qubits in the middle. Additional discussions can be found in App.~\ref{appen:cir}.
The CNOT depth of $\hat U_{Bd}$ on an all-to-all connected quantum device (all-to-all CNOT depth) is 14.

With the fundamental building blocks of the circuits established, the overall sequence of state preparation operators can be developed.
As described before, for notational clarity and specificity, an argument $(x)$ is added to $\hat O$ (for example, $O_{M0}^{(x)}$) to indicate the physical site the operator acts on (and $(x,y)$ if the operator acts on two sites).
The classical optimization of the sub-lattice operators and the implementation of DDec is detailed above Sec.~\ref{sec:L3SP}.
We repeat elements of that discussion for completeness.
As sketched in Fig.~\ref{fig:DDecL3}, preparing the $Q = 1$ ground state on the $x=0,1$ spatial sites of the $L=3$ lattice with sub-percent precision is accomplished by,
\begin{equation}
|\psi_1\rangle = 
e^{-i (-0.2000) O_{M0}^{(0)}}
e^{-i (0.0407) O_{B1}^{(0,1)}}
e^{-i (0.0196) O_{B0}^{(1)}}
e^{-i (0.1820) O_{M0}^{(1)}}
e^{-i (0.2642) O_{M1}^{(0,1)}}
e^{-i (0.3802) O_{M0}^{(0)}}\ 
|SC,3,1\rangle
\ ,
\end{equation}
with the fit angles given in Table~\ref{tab:L2VQEmanuelQ1}.
The $Q = 0$ ground state on the $x=2$ site is accomplished by
\begin{equation}
    |\psi_2\rangle = e^{-i (0.0270) O_{B0}^{(2)}} e^{-i (0.2200) O_{M0}^{(2)}} \ 
    |\psi_1\rangle
    \ ,
\end{equation}
and sewing the two segments together to furnish the $L=3$ ground state with $n_Q=1$ is accomplished using
\begin{equation}
    \ket{\psi_{f}} = e^{-i (-0.0995) O_{M2}^{(1,2)}}e^{-i (0.2314) O_{M1}^{(1,2)}}\ket{\psi_{2}}
    \ .
\end{equation}

The all-to-all CNOT depth of the state-preparation circuit is depth 2 for the SC vacuum (with 3 CNOT gates) for a lattice of any size, is depth 50 for the DDec of the $L=2$ sublattice with cancellations and depth 20 for the $L=1$ sublattice (which can be performed in parallel).
The boundary operators require an additional depth of 38, which can be partially performed in parallel.
The total all-to-all CNOT depth for state preparation of the $L=3$ system with $n_Q=1$ is depth 72.

\subsubsection{Heavy-Quark Translation and Time Evolution}
\noindent
To provide examples of the circuits that can implement the motion of heavy quarks, we move a heavy quark from the first physical site to the second using FSWAP operators with $-\hat Z$s inserted in the middle for fermion anti-symmetry.
Expression of the FSWAP operator for a single color is given in Eq.~\eqref{eq:FSWAP6sites}. 
The R-box constructions analogous to those used for state preparation can be used to implement this motion. 
For nearest-neighbor connectivity, the FSWAP circuit CNOT-depth is 22.
\begin{figure}[htpb]
    \centering
    \includegraphics{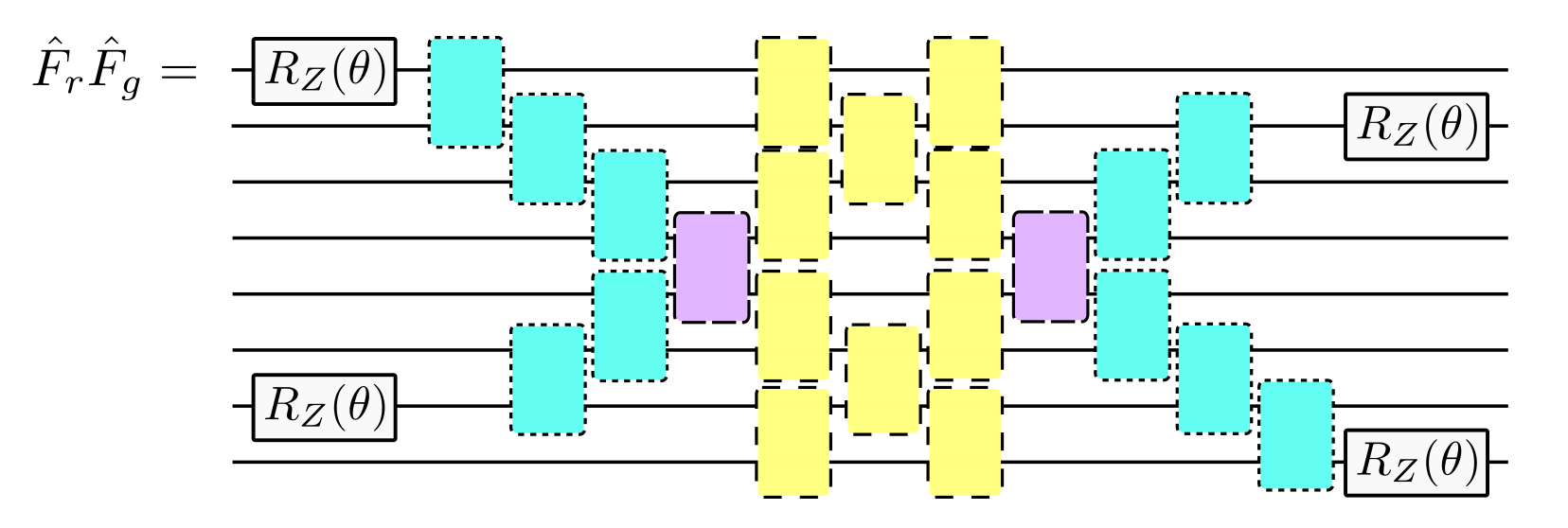}
    \caption{Circuits to implement the spatial translation of a heavy quark between adjacent spatial lattice sites using FSWAP circuits. 
    The $\theta$s appearing in circuit elements take the value of $-\pi/2$.
    }
    \label{fig:fswap}
\end{figure}
Figure~\ref{fig:fswap} shows the composite quantum circuit that implements motion of a heavy quark, where SU(2) symmetry is used to compress the depth of the circuit. Construction details can be found in App.~\ref{appen:cir}.

After translating the heavy quark by one spatial site, the state is time evolved with a Trotterized Hamiltonian. 
Implementation of the mass term is straight forward, as it is a product of single-qubit $Z$-rotations:
\begin{align}
    \hat U_{m,q}(t) & = e^{-i \frac{t m_q}{2} \hat Z_j} = R_Z(t m_q )\ \ :\ \ {\rm quarks} \ ,
    \nonumber\\
    \hat U_{m,\overline{q}} (t) & = e^{+i \frac{t m_q}{2} \hat Z_j} = R_Z(-t m_q )\ \ :\ \ {\rm antiquarks}    
    \ .
    \label{eq:Um}
\end{align}
For a given spatial site, these evolution operators are implemented by the circuit shown in Fig.~\ref{fig:h_full}.
As these are single-qubit gates, they do not contribute to the depth of the circuit.
We have not included the time evolution of mass Hamiltonian associated with the heavy quark sites. 
As their kinetic term is not included, their mass terms add only an overall phase to the wavefunction that is position independent.

The kinetic terms correspond to two types of operators, those with one $\hat Z$ in the middle (denoted $\hat{H}_{k0}$) and those with three (denoted $\hat{H}_{k1}$), as given in Eq.~\eqref{eq:UHK},
\begin{align}
        \hat{H}_{k0} & \equiv (\hat X \hat Z\hat X + \hat Y \hat Z\hat Y ) \otimes \hat I + \hat I \otimes (\hat X \hat Z\hat X + \hat Y \hat Z\hat Y )\ , \nonumber \\
        \hat{H}_{k1} & \equiv  (\hat X \hat Z ^3 \hat X + \hat Y \hat Z ^ 3 \hat Y ) \otimes \hat I + \hat I \otimes (\hat X \hat Z ^3 \hat X + \hat Y \hat Z ^ 3 \hat Y )
        \ .
        \label{eq:UHK}
\end{align}
$\hat{H}_{k0}$ acts on $r\bar r$ and $g\bar g$ on the same physical site, while $\hat{H}_{k1}$ acts on $\bar r r$ and $\bar g g$ on two neighboring sites.
When acting on the same spatial sites, $\hat{H}_{k0}$ and $\hat{H}_{k1}$ do not commute. 
An additional site label $(x)$ is included to indicate which site the operator acts on. 
The complete kinetic circuit is implemented using first order Trotterization, where a layer of $\hat{H}_{k0}$ evolution is first applied, followed by $\hat{H}_{k1}$ evolution, as given in Eq.~\eqref{eq:UHKevol},
\begin{equation}
        e^{-it\hat H_k} \rightarrow  \exp\left(i \frac{t}{4}\sum_{x = 0}^{L-1}\hat{H}_{k0}^{(x)}\right) \exp\left(i \frac{t}{4}\sum_{x = 0}^{L-2}\hat{H}_{k1}^{(x,x+1)}\right)\ .
        \label{eq:UHKevol}
\end{equation}
Time evolution by $\hat{H}_{k0}$ is implemented by a circuit with a two-qubit depth of 6, while evolution by $\hat{H}_{k1}$ is implemented by a circuit of depth 14.
The leading-order Trotterized kinetic-energy evolution operator acting on three physical sites is shown in Fig.~\ref{fig:h_full}.
This can be straightforwardly generalized to larger systems, as well as higher order Trotterizations.

\begin{figure}[htpb]
    \centering
    \includegraphics[width=\textwidth]{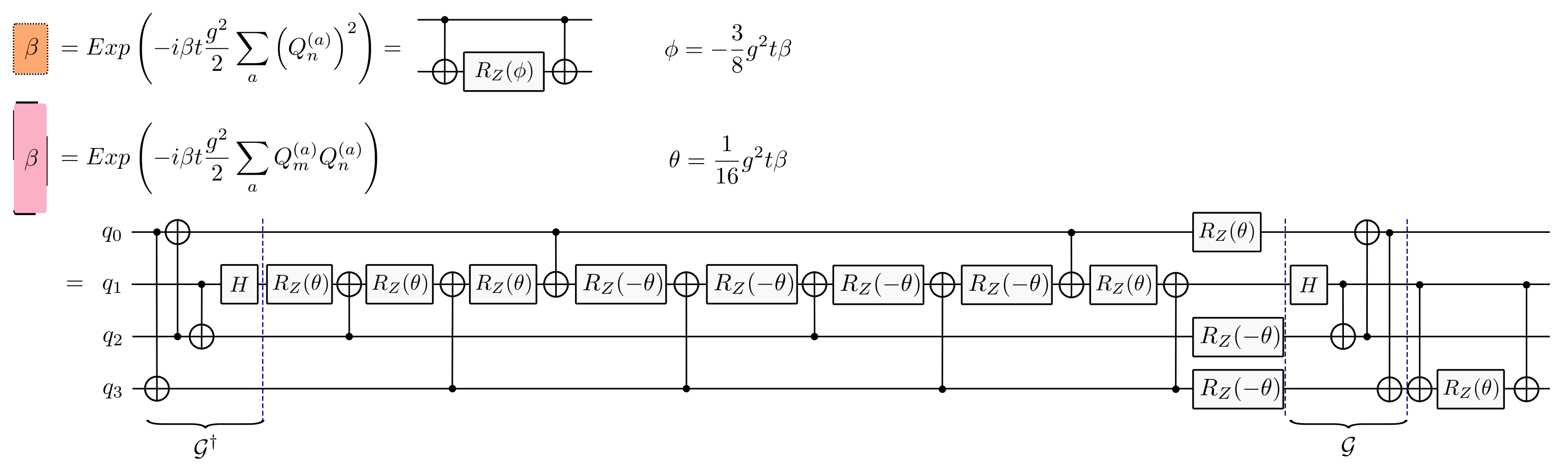}
    \caption{Circuit elements implementing evolution by the chromo-electric Hamiltonian,
    given in Eqs.~\eqref{eq:HgL3} and \eqref{eq:Qops}.
    The orange elements implement single-site contributions, while the  pink implements contributions from different spatial sites. 
}
\label{fig:hgauge_seg}
\end{figure}

\begin{figure}[htpb]
    \centering
    \includegraphics{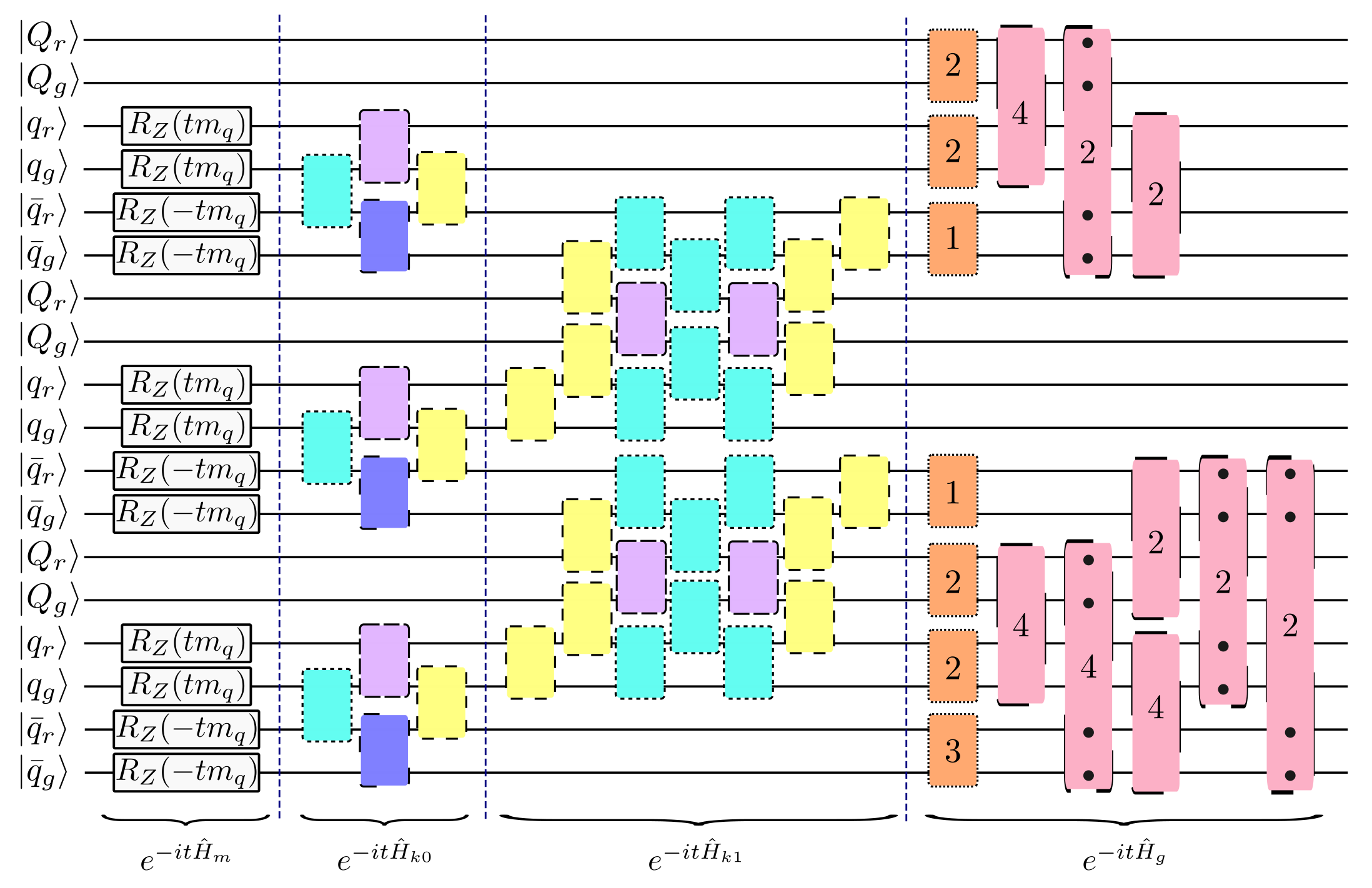}
    \caption{A quantum circuit implementing the terms involved in a Trotterized time evolution. The left-most block implements the mass terms,
    given in Eq.~\eqref{eq:Um}. 
    This is followed by
    the blocks related to the kinetic-energy operators, given in Eq.~\eqref{eq:UHKevol}.
    The angle defining the circuit elements, given in Fig.~\ref{fig:rbox}, is given by $\theta = t/2$. 
    Finally, the right-most block implements
    the complete chromo-electric contribution to the Hamiltonian (without optimization), given in Eqs.~\eqref{eq:HgL3} and \eqref{eq:Qops}.
    The orange and pink circuit elements are defined in Fig.~\ref{fig:hgauge_seg}, and the black dots on the pink elements indicate the engaged qubits.
    }
    \label{fig:h_full}
\end{figure}

The contribution from the energy in the chromo-electric field can be usefully expressed in its symmetric form. 
Making use of the lattice having overall vanishing color charge, the chromo-electric field at each of the sites can be obtained symmetrically from both ends of the lattice (similar to the construction in App.~\ref{app:L2} for $L=2$, and the general case in App.~\ref{app:explicitham}).
Using
\begin{align}
    E^{(a)}_0 & = \hat{\mathcal{Q}}^{(a)}_{0}+ \hat{\mathcal{Q}}^{(a)}_{1}\ ,\ E^{(a)}_1 =  E^{(a)}_0 + \hat{\mathcal{Q}}^{(a)}_{2}\ , & &{\rm starting\ from\ the}\ n=0\ {\rm site} \ ,\ \nonumber\\
    E^{(a)}_5 & = \hat{\mathcal{Q}}^{(a)}_{8}\ ,\ E^{(a)}_4 =  E^{(a)}_5 +\hat{\mathcal{Q}}^{(a)}_{6} + \hat{\mathcal{Q}}^{(a)}_{7}\ ,\ E^{(a)}_3 = E^{(a)}_4 + \hat{\mathcal{Q}}^{(a)}_{5}\ , & &{\rm starting\ from\ the}\ n=8\ {\rm site}
\ , 
\end{align}
provides for the chromo-electric component of the Hamiltonian to be expressed as 
\begin{align}
H_g  =  \frac{g^2}{2} \sum_{a} & 
\left[ 2 \hat{\mathcal{Q}}_{0}^2 
+ 2 \hat{\mathcal{Q}}_{1}^2 
+ \hat{\mathcal{Q}}_{2}^2 
+ 4 \hat{\mathcal{Q}}_{0} \hat{\mathcal{Q}}_{1}
+ 2 \hat{\mathcal{Q}}_{0} \hat{\mathcal{Q}}_{2}
+ 2 \hat{\mathcal{Q}}_{1} \hat{\mathcal{Q}}_{2}
\right.\nonumber\\
& \ + 2 \hat{\mathcal{Q}}_{6}^2 + 2 \hat{\mathcal{Q}}_{7}^2 + 3 \hat{\mathcal{Q}}_{8}^2
+ 4 \hat{\mathcal{Q}}_{6} \hat{\mathcal{Q}}_{7} + 4 \hat{\mathcal{Q}}_{6} \hat{\mathcal{Q}}_{8}
+ 4\hat{\mathcal{Q}}_{7} \hat{\mathcal{Q}}_{8} \nonumber\\
& \ \left. + \hat{\mathcal{Q}}_{5}^2 + 2 \hat{\mathcal{Q}}_{5} \left( \hat{\mathcal{Q}}_{6} + \hat{\mathcal{Q}}_{7} + \hat{\mathcal{Q}}_{8} \right)
\right]
\ .
\label{eq:HgL3}
\end{align}
where the generator indices $a$ have been suppressed in Eq.~\eqref{eq:HgL3} for simplicity.
The distinct structures given in Eq.~\eqref{eq:HgL3}, acting on $r$-$g$ pairs, can be implemented by the two general operators 
\begin{align}
\sum_a \left(\hat{\mathcal{Q}}_n^{(a)}\right)^2 & = \frac{3}{8} \left( \hat I - \hat Z_{2n} \hat Z_{2n+1} \right) \ , \nonumber\\
\sum_a  \hat{\mathcal{Q}}_n^{(a)} \hat{\mathcal{Q}}_m^{(a)}  & = \frac{1}{16}\left(\hat Z_{2n} -\hat Z_{2n+1}\right)\left(\hat Z_{2m} -\hat Z_{2m+1}\right) + \frac{1}{2} \left(\hat\sigma_{2n}^+\hat\sigma_{2n+1}^- \hat\sigma_{2m}^-\hat\sigma_{2m+1}^+ \ +\ {\rm h.c.} \right) \ ,
\label{eq:Qops}
\end{align}
where $(\hat{\mathcal{Q}}_n^{(a)})^2$ corresponds to single-staggered-site contributions, 
while $\hat{\mathcal{Q}}_n^{(a)} \hat{\mathcal{Q}}_m^{(a)}$ corresponds to contributions from different staggered sites.
Same-site interactions contain only $\hat Z \hat Z$ operators and can be implemented straightforwardly. Interactions between different staggered sites can be implemented efficiently using a GHZ transformation~\cite{Farrell:2022vyh}.\footnote{The GHZ transformation $\mathcal{G}$ used here is to be distinguished from the GHZ transformation $G$ use earlier in the state preparation circuits. They facilitate the same goal of implementing a target operator by transforming the basis of efficient circuits in the computational basis. Since the target operators in state preparation and time evolution have different forms, the GHZ transformations are slightly different} The operator order from left to right correspond to circuit wires from top to bottom. 
\begin{align}
    \hat\sigma^+\hat\sigma^-\hat\sigma^-\hat\sigma^+ + {\rm h.c.} & = 
    \frac{1}{8}\left[
    \hat X\hat X\hat X\hat X
    +
    \hat X\hat X\hat Y\hat Y
    +
    \hat X\hat Y\hat X\hat Y
    -
    \hat X\hat Y\hat Y\hat X
    -
    \hat Y\hat X\hat X\hat Y
    +
    \hat Y\hat X\hat Y\hat X
    +
    \hat Y\hat Y\hat X\hat X
    +
    \hat Y\hat Y\hat Y\hat Y
     \right] ,
    \nonumber\\
    \hat {\cal G}^\dagger \left(\hat\sigma^+\hat\sigma^-\hat\sigma^-\hat\sigma^+ + {\rm h.c.} 
    \right) \hat {\cal G} & = 
        \frac{1}{8}\left[
        -\hat Z\hat Z\hat Z\hat I - 
        \hat Z\hat Z\hat I\hat Z + 
        \hat I\hat Z\hat Z\hat Z + 
        \hat I\hat Z\hat I\hat I -
        \hat Z\hat Z\hat Z\hat Z - 
        \hat Z\hat Z\hat I\hat I + 
        \hat I\hat Z\hat Z\hat I + 
        \hat I\hat Z\hat I\hat Z  
     \right]\ .
    \label{eq:XYTOIZaa}
\end{align}
Additionally, the $\hat Z \hat Z$ interactions between staggered sites can also be transformed into a GHZ basis. 
The $\hat Z \hat Z$ operators in a GHZ basis on four qubits are\footnote{Note that one four-qubit GHZ transformation can only transform three $\hat Z \hat Z$ operators into a single $\hat Z$, so one of the four 
remains untransformed.
In the SU(3) case, all the $\hat Z \hat Z$ operators can be absorbed into the GHZ basis because there are $\binom{3}{2} = 3$ sets of 4-qubit GHZ transformations and $3 \times 3 = 9$ $\hat Z \hat Z$ operators.
The ratio between the number of $\hat Z \hat Z$ operators and GHZ transformations is $3$, the maximum number that can be absorbed. 
In the SU(2) case, there are $\binom{2}{2} = 1$ set of 4-qubit GHZ transformations and $2 \times 2 = 4$ $\hat Z \hat Z$ operators. 
The ratio between the number of $\hat Z \hat Z$ operators and GHZ transformations is $4$.}
\begin{equation}
\left(\hat Z_0 -\hat Z_{1}\right)\left(\hat Z_2 -\hat Z_{3}\right) = \hat Z_0\hat Z_2 - \hat Z_0\hat Z_3 -\hat Z_1\hat Z_2+\hat Z_1\hat Z_3 = \mathcal{G}(\hat Z_0 -\hat Z_3 - \hat Z_2)\mathcal{G}^{\dagger} + \hat Z_1\hat Z_3 \ .
\end{equation}
The associated quantum circuits for $\mathcal{G}$, $\mathcal{G}^{\dagger}$, and for the two types of interactions, are given in Fig.~\ref{fig:hgauge_seg}.
The two-qubit orange circuit element has depth 2, and each four-qubit pink circuit element has depth 16.
The total depth of the Trotterized time-evolution operator implementing the chromo-electric Hamiltonian (without optimization) is 78, as shown in Fig.~\ref{fig:h_full}.
This does not include truncations of the chromo-electric contributions made possible by confinement~\cite{Farrell:2024fit}, as the chromo-electric Hamiltonian is symmetrically implemented, and  interactions act on a maximum of two spatial sites.
We perform a second order Trotterization, where each step of the time evolution is
\begin{equation}
    U_2(t) = e^{-i\frac{t}{2}\hat{H}_{k1}}e^{-i\frac{t}{2}\hat{H}_{k0}}e^{-it\hat{H}_{m}}e^{-it\hat{H}_{g}}e^{-i\frac{t}{2}\hat{H}_{k0}} e^{-i\frac{t}{2}\hat{H}_{k1}}
    \ .
\end{equation}
Combining all of the contributions and without optimization, the total two-qubit gate depth of the circuit implementing one application of the complete second Trotterized time-evolution operator is 
\begin{equation}
{\rm depth} = 20 \ ({\rm Kinetic}) \times 2 + 0 \ ({\rm Mass}) + 78\  ({\rm Gauge})
\ =
 118
 \ .
\end{equation}
Since some parts of the circuits can be performed in parallel, the actual depth is 112. When performing multiple second-order Trotter steps, the $e^{-i\frac{t}{2}\hat{H}_{k1}}$ terms between consecutive steps can be combined. 
Therefore, the steps after the first one are shallower by the depth of $e^{-it\hat{H}_{k1}}$ (102 compared to 112). 
An analysis of the systematic errors in first- and second-order Trotterization can be found in App.~\ref{appen:resources}.

\subsection{Resource Projection and Mapping to Device}
\label{sec:resproj}
\noindent
The scaling of quantum resources is considered with respect to two simulation parameters, increasing the lattice size and increasing the number of Trotter steps in the time evolution (which permits a simulation to reach longer times with limited systematic errors).
On the axis of increasing the number of Trotter steps, state preparation is not impacted, and as each step of the time-evolution circuit has the same depth, the total depth of the time evolution grows as $\mathcal{O}(n)$ where $n$ is the number of steps.
For increasing the system size, the circuit depth for initial state preparation depends on the imposed constraints. 
It remains constant when the heavy quark is initialized 
near a boundary with a short  correlation length, but increases when higher precision or longer-range correlations are required.
The depth of each step of the Trotterized time evolution is affected differently by each term in the Hamiltonian. 
For the mass and kinetic terms, increasing the number of spatial lattice sites does not increase the depth of circuit elements required to implement one Trotter step.
However, the contribution from the chromo-electric term requires more attention because of its naive all-to-all connectivity.
To permit scalable simulation, the chromo-electric term needs to be truncated at a distance determined by the confinement scale, introducing an additional, but exponentially small, truncation error.

\begin{figure}[htpb]
    \centering
    \includegraphics[width=0.3\linewidth]{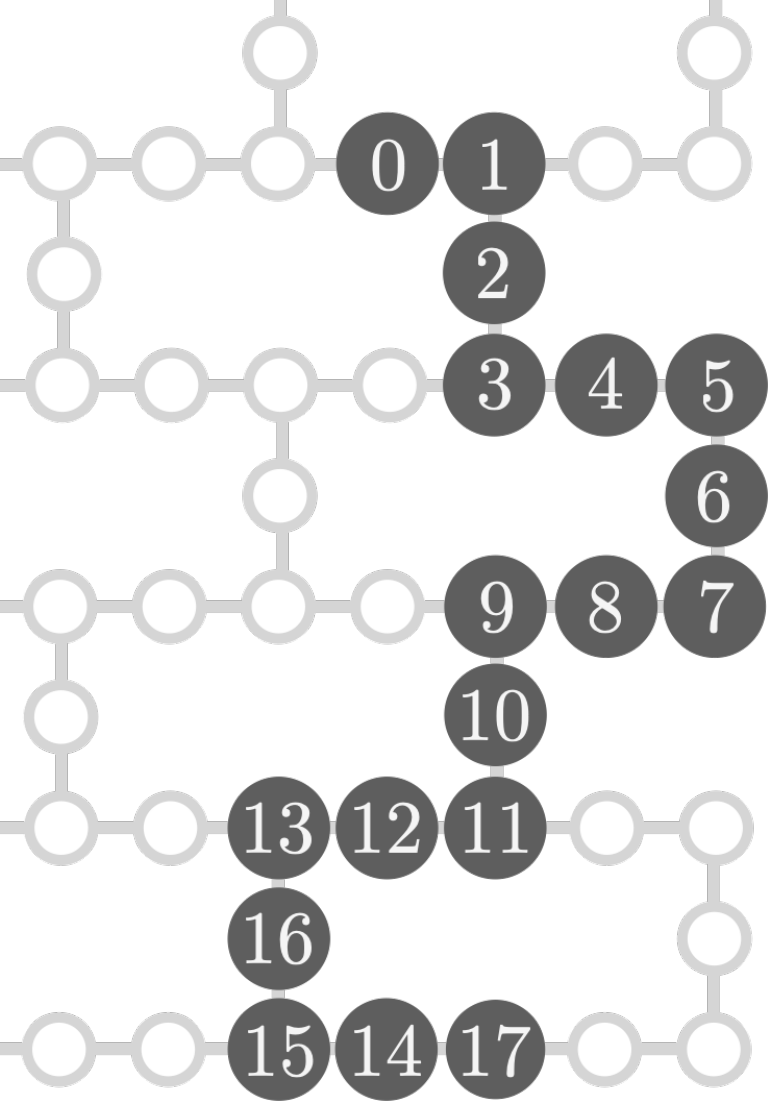}
    \caption{
    Qubit layout of the $L=3$ system on {\tt ibm\_pittsburgh}.
}
    \label{fig:layout}
\end{figure}

For transpilation to a target device, the circuits are adapted to the native gate set and the connectivity of the device.
Our target device is IBM's  {\tt ibm\_pittsburgh} with a heavy-hex connectivity.
Each CNOT gate is transpiled to a CZ gate plus single qubit rotations, so they are equivalent in the consideration of two-qubit-gate count. 
To optimize the depth of the transpiled time-evolution circuit, in particular the gauge term, we use the layout displayed in Fig.~\ref{fig:layout}, where the last qubits have been rearranged.
This is not necessarily the layout that gives the shallowest possible depth, since the goal is not to over-optimize this 18 qubit system. 
The qubits in Fig.~\ref{fig:layout} were selected by imposing a maximum error in the CZ gate of 0.005 on {\tt ibm\_pittsburgh} with a median CZ error of 0.0017. 
The resource requirements for each component of the simulation (state preparation and time evolution) are shown in Table~\ref{tab:component_resource}.
Further details, such as CZ depth distribution across the lattice, can be found in App.~\ref{appen:resources}.
\begin{table}[htpb]
\renewcommand{\arraystretch}{1.1}
    \centering
    \begin{tabular}{|r|c|c|}
        \hline
        Term & CZ All-to-All & CZ transpiled \\
        \hline
        \hline
        Strong Coupling & 2 & 2\\
        VQE Layers & 70 & 126\\
        Fermionic SWAP & 22 & 22 \\
        Time Evolution & 112 & 254\\\hline
        Total & 182 & 398 \\\hline
        Additional time step & 102 & 248 \\\hline\hline
    \end{tabular}
    \caption{
    Quantum resource requirements 
    (entangling gates)
    for preparing the initial states and for time evolution.
    The transpiled requirements are for the fixed qubit layout shown in Fig.~\ref{fig:layout}.
    The total resource requirements are less than the sum of the listed components 
    because some of the operations can be performed in parallel.
    }
    \label{tab:component_resource}
\end{table}
%

\subsection{Results from Quantum Simulations}
\label{sec:QSresults}
\noindent
A suite of state preparation and time evolution circuits, along with the associated error-mitigation circuits, that have been detailed in previous sections, were implemented with IBM's {\tt qiskit}~\cite{qiskit} and executed on IBM's quantum computer {\tt ibm\_pittsburgh}.
The results of these simulations were analyzed classically with standard statistical techniques using {\tt python}, {\tt julia} and {\tt Mathematica}.
We focus on two observables related to heavy-quark motion: charge distribution and energy loss.

\subsubsection{Charge Distribution}
\label{sec:chargeD}
\noindent
An initial state of the $L=3$ ($n_{\rm qubits} = 18$) system using Hamiltonian parameters $m_q=0.1$ and $g=1.0$ was prepared with a heavy quark residing on the  $x=0$ spatial site. 
The heavy quark is then moved to the $x=1$ spatial site and the system is time evolved forward to $t=1.0$ using a single second-order Trotter step.
Several error mitigation methods were applied to compensate for the noisy hardware. 
Pauli twirling~\cite{Wallman:2015uzh} was performed around the entangling gates, with 64 distinct twirled circuits for the production runs.
Dynamical decoupling~\cite{Viola:1998jx,Ezzell:2022uat} and measurement mitigation~\cite{Berg:2020ibi} were applied through the {\tt qiskit}~\cite{qiskit} Estimator primitive. 
To remove the remaining bias due to incoherence of the device, we apply operator-decoherence renormalization (ODR)~\cite{Urbanek:2021oej,ARahman:2022tkr,Farrell:2023fgd}. 
For that, we ran two types of circuits: the ``physics'' circuit that implements the process of interest, and two ``mitigation'' circuits.
The mitigation circuits are set up such that they share the same structure as the physics circuits, but are easy to simulate classically in that they only contain Clifford circuits. In the state preparation, the non-Clifford angles in the rotation gates\footnote{The gates $R_{\pm}^{XY}(\theta)$ and $R_{\pm}^{XX}(\theta)$ with $\theta = \pm \frac{\pi}{2}$ in Fig.~\ref{fig:rbox} do not get modified for the mitigation circuits.} are set to $\theta_{\rm mit}=0, \pi/2$.
In the time evolution, the mitigation circuits evolve the system forward then backward in time such that there is no change in an ideal simulation.\footnote{For one step of second-order Trotterized time evolution, the mitigation circuit becomes
\begin{equation}
    e^{i\frac{t}{2}\hat{H}_{k1}}e^{i\frac{t}{2}\hat{H}_{k0}}e^{-i0\hat{H}_{m}}e^{-i0\hat{H}_{g}}e^{-i\frac{t}{2}\hat{H}_{k0}} e^{-i\frac{t}{2}\hat{H}_{k1}} \ .
\end{equation}}
With these quantities, the rescaling factor to compensate for the decoherence can be computed:
\begin{equation}
    \eta_{\hat O} = \frac{\langle \hat O \rangle^{\rm mit}_{\rm meas}}{\langle \hat O \rangle^{\rm mit}_{\rm pred}} 
    \ \ \ \ \rightarrow \ \ \ \ 
    \langle \hat O \rangle^{\rm physics}_{\rm pred} = \frac{\langle \hat O \rangle^{\rm physics}_{\rm meas}}{\eta_{\hat O}} \ ,
\end{equation}
where the ``meas'' observable is the quantity extracted from the noisy device, and ``pred'' is the noiseless prediction.

\begin{figure}[ht!]
    \centering
\includegraphics[width=.45\linewidth]{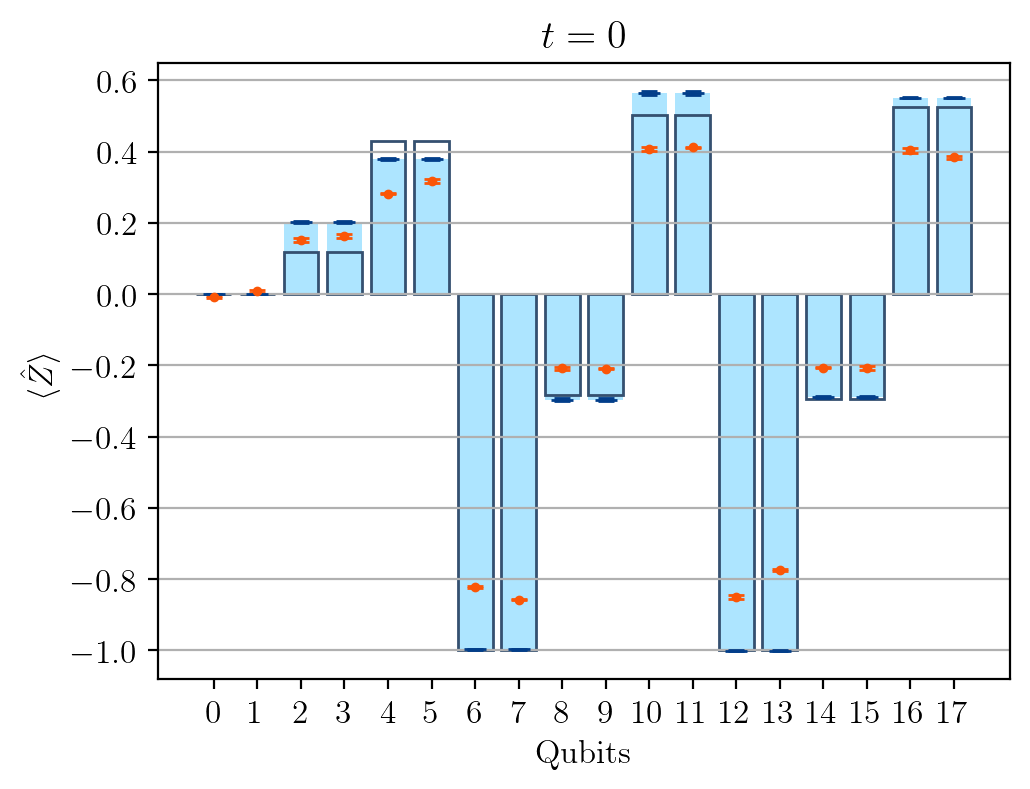}
\includegraphics[width=.45\linewidth]{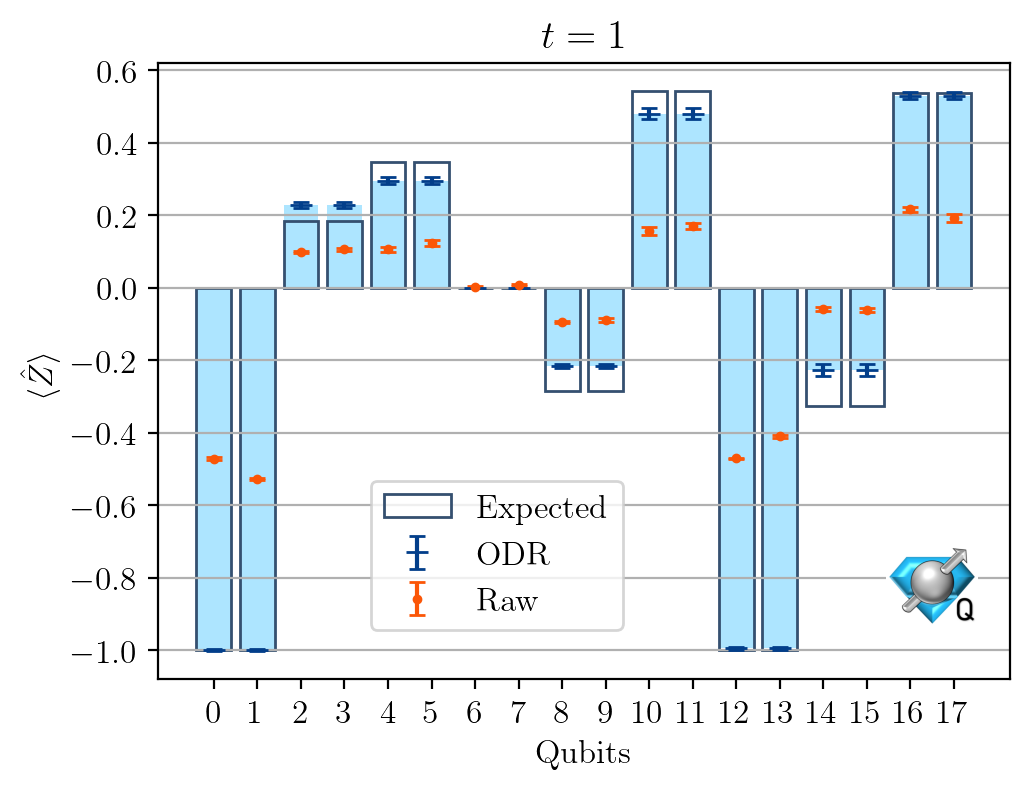}
\caption{
Results obtained from {\tt ibm\_pittsburgh} for $\langle \hat Z_j \rangle$ in the $L=3$ system.  
At $t=0$, the heavy quark is instantaneously moved from $x=0$ to $x=1$, then evolved forward in time to $t=1.0$ 
using one step of the second-order Trotterized time-evolution operator.
The SU(2) symmetry has been used to combine results for each red and green quark.
The left panel shows the expectation values of $\langle \hat Z_j \rangle$ before the heavy quark is moved; the right panel  shows the values
at $t=1$. The results are obtained using bootstrap resampling with the median and MAD.
The points with error bars correspond to the raw (orange circle) and mitigated (blue line) results obtained from {\tt ibm\_pittsburgh}, 
while the gray bar outlines are the expected results obtained using classical computing.
}
    \label{fig:IBMresultfig}
\end{figure}

The quantum circuits were transpiled to run on IBM hardware.
For each twirl of each type of circuit, 4096 measurements (shots) of $\langle \hat Z_j \rangle$ were performed in the computational basis. 
For each qubit, we obtain three results (one from the physics circuit, and two from the mitigation circuits), each for a total of 64 twirls.
For the mitigation circuits, $\langle \hat Z_j\rangle^{\rm mit}_{\rm pred}$ is expected to take three values, $\{-1,0,1\}$. 
The reason we chose to run two different mitigation circuits, is that with each of the mitigation circuits, some of the observables can take the value zero, incompatible with ODR.
With the two choices of $\theta_{\rm mit}$ selected, we have at least one non-zero value for $\langle \hat Z_j\rangle^{\rm mit}_{\rm pred}$.
Measurements with $|\langle \hat Z_j\rangle^{\rm mit}_{\rm meas}|<\lambda^{\rm mit}_{\rm cut}$ and $|\langle \hat Z_j\rangle^{\rm phys}_{\rm meas}|<\lambda^{\rm phys}_{\rm cut}$, are filtered out~\cite{Farrell:2024fit,Farrell:2025nkx}.
Additionally, the inherent SU(2) symmetry of the system, which requires that pairs of qubits within a staggered site to have the same value of $\langle \hat Z_j \rangle$, was used to combined results for the $r,g$ qubits.
The median and (re-scaled) median-average-deviation (MAD) were used as robust estimators of the predicted central value of $\langle \hat Z_j \rangle$ for each qubit, along with its $68\%$ confidence interval.
The value of $\lambda^{\rm mit}_{\rm cut}$ and $\lambda^{\rm phys}_{\rm cut}$ was set at $0.005$, data below which is considered unrecoverable for a circuit with $\sim 400$ two-qubit-gate depth. 
The results of this analysis are compiled in Table~\ref{tab:IBManalysis}, and displayed in Fig.~\ref{fig:IBMresultfig}.

The error-mitigated results obtained from {\tt ibm\_pittsburgh}
are in good agreement with expectations,
despite a transpiled two-qubit-gate circuit depth of $\sim 400$.  
The left panel of Fig.~\ref{fig:IBMresultfig} shows that there remain modest systematic errors that we have not identified, and that ODR does not remove all of the errors introduced by {\tt ibm\_pittsburgh} during circuit execution.
Some of these errors can be attributed to the time of execution of the circuits, where delay between device calibration and execution times can be 
significant.
The right panel displays the bootstrap distributions from 64 twirled samples for all the staggered sites from the raw data. While for some of the results their $\{r,g\}$-averaged values come from single peaked distributions, 
some come from non-trivial distributions, and in the worst case a double peaked distribution, making the use of robust statistics important.

\subsubsection{Energy Loss}
\label{sec:QSresultsdEdx}
\noindent
Determining
the energy loss during the motion of the heavy quark can be accomplished by measuring the difference between the total energy immediately before and 
after 
translating the heavy quark between adjacent spatial sites. 
Given that the time evolution circuit is predominantly energy-preserving, 
and any small systematic errors are negligible,
the post-translation time-evolution step is unnecessary.
As considered in Sec.~\ref{sec:L3Qmotion} and Sec.~\ref{sec:L3QmotionIM}, it is only the energy in the gauge field that changes discontinuously during this process.  Therefore, only this component of the Hamiltonian  needs to be evaluated.
At the level of circuits, there are well-established ways to determine the energy in a state.
We choose to use a set of low-depth entangling operations to transform modest sets of commuting operators contributing to the Hamiltonian into the computational basis (with simultaneous transformations).
These sets of contributions are then individually  evaluated probabilistically from measurements. 
As only the chromo-electric energy is required to be evaluated, 
before and after translation of the heavy quark, 
the same GHZ transformation used in the time evolution of the gauge term can be applied to transform $\sigma^\pm\sigma^\mp\sigma^\mp\sigma^\pm$ into the computational basis for simpler measurements.
This transformation is implemented 
in quantum simulations 
by applying $\hat{\cal G}^\dagger$ defined in Fig.~\ref{fig:hgauge_seg} to the final state wavefunction on the quantum computer,
followed by measurements in the computational basis.
This circuit is applicable for any pair of quarks and anti-quarks or quarks.\footnote{In addition to this method, we have also explored the Hadamard test to assess its viability as an alternate technique, see App.~\ref{app:Energycircs}. 
We find it to be inefficient.}

To determine the difference in energy of the system when the heavy quark is moved from $x=0$ to $x=1$,
the only terms that need to be evaluated before and after are
\begin{align}
\Delta \hat H_{g} = &  
\ \frac{g^2}{2} \sum_a \ 
\left[\ 
2 ( \hat{\mathcal{Q}}^{(a)}_{0} )^2
\ +\ 
4 \hat{\mathcal{Q}}^{(a)}_{0}\hat{\mathcal{Q}}^{(a)}_{1}
\ +\ 
2 \hat{\mathcal{Q}}^{(a)}_{0} \hat{\mathcal{Q}}^{(a)}_{2}
\ \right]
\ ,
\nonumber\\
= & 
\ \frac{g^2}{2} \left[ 
\frac{3}{4}(1-\hat Z_0\hat Z_1)
+ \frac{1}{4} (\hat Z_0-\hat Z_1)(\hat Z_2-\hat Z_3) + \frac{1}{8} (\hat Z_0-\hat Z_1)(\hat Z_4-\hat Z_5)
\ \right. 
\nonumber\\
 &
 \quad \quad \left.  
  + \  2 \left(\hat\sigma^+_0\hat\sigma^-_1\hat\sigma^-_2\hat\sigma^+_3 + {\rm h.c.} 
    \right)
 + \left(\hat\sigma^+_0\hat\sigma^-_1\hat\sigma^-_4\hat\sigma^+_5 + {\rm h.c.} 
    \right)    
\right]\ ,
\nonumber\\
= & \  \sum_{i=1}^3\ 
\hat {\cal O}_i^\dagger \ ( C_i \cdot \hat P_i ) \ \hat {\cal O}_i
\ ,
\label{eq:HdiffE}
\end{align}
The expectation values of the operators in Eq.~\eqref{eq:HdiffE} vanish after the translation of the heavy quark to $x=1$ because $\hat{\mathcal{Q}}^{(a)}_{0}|\psi\rangle=0$.
Equation~\eqref{eq:estimator} gives the operators evaluated on the final state, the list of Pauli operators,
 $\hat P_i$, and the list of their coefficients, $C_i$.
\begin{align}
\hat {\cal O}_1\ =\ \hat I 
\qquad & :
\hat P_1 = \left[
\hat I,\ 
\hat Z_0\hat Z_1,\  \hat Z_0\hat Z_2,\   \hat Z_0\hat Z_3,\   \hat Z_1\hat Z_2 , \  
\hat Z_1\hat Z_3,\    \hat Z_0\hat Z_4,\    \hat Z_0\hat Z_5,\    \hat Z_1\hat Z_4,\   
\hat Z_1\hat Z_5
\right] 
\nonumber\\
&
C_1 = 
\frac{g^2}{2} \frac{1}{8}\left[ 
+6,-6, +2, -2, -2, +2, +1, -1,  -1, +1
\right] 
\nonumber\\
\hat {\cal O}_2\ =\ \hat{\cal G}_{01;23}^\dagger & :  
\hat P_2 = \left[
        -\hat Z_0\hat Z_1\hat Z_3, 
        \ \hat Z_1\hat Z_2 \hat Z_3,   
        \ -\hat Z_0\hat Z_1\hat Z_2,   
        \ \hat Z_1, 
        \ -\hat Z_0\hat Z_1\hat Z_2\hat Z_3,   
        \ \hat Z_1\hat Z_2,  
        \ \hat Z_1\hat Z_3,  
         \ -\hat Z_0 \hat Z_1  
\right]
\nonumber\\
&
C_2 = \frac{g^2}{2} \frac{1}{4}
\left[+1, +1, +1, +1, +1, +1, +1, +1\right]
\nonumber\\
 \hat {\cal O}_3\ =\ \hat{\cal G}_{01;45}^\dagger & : 
\hat P_3 = \left[
        -\hat Z_0\hat Z_1\hat Z_5, 
        \ \hat Z_1\hat Z_4 \hat Z_5,   
        \ -\hat Z_0\hat Z_1\hat Z_4,   
        \ \hat Z_1, 
        \ -\hat Z_0\hat Z_1\hat Z_4\hat Z_5,   
        \ \hat Z_1\hat Z_4,  
        \ \hat Z_1\hat Z_5,  
         \ -\hat Z_0 \hat Z_1  
\right]
\nonumber\\
&
C_3 = 
\frac{g^2}{2} \frac{1}{8}\left[+1, +1, +1, +1, +1, +1, +1, +1
\right]
\ .
\label{eq:estimator}
\end{align}
The implementation in {\tt qiskit} involves three different circuit evaluations.
Taking the difference between the summed estimators
before and after the heavy quark translation
provides the energy loss from the translation.
The operator structure in Eqs.~\eqref{eq:HdiffE} and \eqref{eq:estimator} will differ for computing the energy loss from heavy-quark translations between different adjacent spatial sites on lattices of arbitrary length.

The energy transfer to the system due to the heavy-quark translating from $x=0$ to $x=1$ was determined by running three different physics circuits, along with 
corresponding
mitigation circuits,\footnote{The mitigation circuits for the operators in $\hat{P}_1$ and $\hat{P}_2$ have the non-Clifford rotation angles set to 0, while no mitigation circuits were run for $\hat{P}_3$. This is because, for generating the $\hat{P}_3$ mitigation circuits, setting non-Clifford rotation angles to 0 or $\pi/2$ both result in 0 for noiseless predictions of observables, incompatible with ODR.} on the quantum computer.  
As the last operator in the state preparation circuit does not change the measurements,
a shallower circuit, with a CZ depth of 109, is used to measure the diagonal Pauli strings defined in Eq.~\eqref{eq:estimator}.

In addition to the error mitigation methods applied in the charge distribution measurements, zero-noise extrapolation (ZNE)~\cite{ZNE1,ZNE2,PhysRevLett.120.210501,Klco:2018kyo,ZNE3} is employed through 
{\tt qiskit} Runtime options~\cite{qiskit}. 
Noise factors-$(1,2,3)$ are executed on the device, where the different noise factors indicate the increase in circuit depth due to additional CZ gates (e.g., factor 1 is the original circuit, and factor 3 has three times the CZ count of the original circuit).
In post-processing, and before performing ZNE of the raw data, 
we perform ODR for each noise level using the 32 sets of Pauli-twirled physical and mitigation circuits.
These distributions are then linearly extrapolated in the noise-factor to the point of zero noise, providing an error-mitigated estimate of the quantity of interest.
Computing the three contributions to the energy difference given in Eq.~\eqref{eq:estimator}, both with and without ZNE for the sake of comparison, gives,
\begin{align}
\langle \Delta \hat H_{g}  \rangle^{\rm IBM}_{t<0} & =
0.3715(16) - 0.7243(30) - 0.0314(04) \ = \ -0.3841(30)
\ \ ({\rm ODR})
\, 
\\
\langle \Delta \hat H_{g}  \rangle^{\rm IBM}_{t<0} & = 
0.3491(19) - 0.7538(73) - 0.0390(07) \ = \ -0.4437(72)
\ \ ({\rm ODR+ZNE})
\ ,
\label{eq:estIBMvalues}
\end{align}
which are to be compared with the exact values, 
$\langle \Delta \hat H_{g}  \rangle^{\rm exact}_{t<0} = 0.3314-0.7951-0.0421=-0.5058$, 
obtained from a noiseless simulation of the quantum circuits.
Combined with $\langle \Delta \hat H_{g}  \rangle^{\rm IBM}_{0<t<1}=0$, this result implies an energy loss of 
$\delta \langle \Delta \hat H_{g}  \rangle^{\rm IBM}=+0.4437(72)$
due to the translation of the heavy quark from $x=0$ to $x=1$.
Further details about this determination can be found in App.~\ref{app:IBMruns}.
It is important to note the importance of combining ODR and ZNE in obtaining a result from {\tt ibm\_pittsburgh}
that is close to the expected value.
This is, in part, due to the modest cancellations among contributions to the sum, that magnify the impact of the individual errors.

\section{Summary and Conclusions}
\label{sec:Summary}
\noindent
In this work, we have established a set of techniques that enable the real-time simulation of energy-loss and hadronization in non-Abelian lattice gauge theories in 1+1D.
As the action of the non-Abelian charge operators generally transform any given state, the methods used for such simulations in Abelian theories, including charges as discontinuities in Gauss's law, are inapplicable.  
To circumvent this problem, a non-dynamical  
heavy-quark field is included as an explicit degree of freedom.
With one light flavor of quark and anti-quark, and one flavor of heavy quark,
the 1+1D system with two colors can be mapped to a qubit register with six qubits per spatial site.
Motion of the heavy quark across the lattice is implemented by fermionic SWAP operators, with the condition that only complete translations between spatial sites are permitted, as anything different is incompatible 
with the 
time evolution of the light quarks.
This method introduces a further lattice-spacing artifact due to swapping in the presence of other heavy quarks in the lattice (that mimics matter), which can be mitigated at a subsequent time step.
Domain decomposition is introduced as a scalable way to 
prepare the (initial) ground state of a single heavy-quark positioned at one end of the lattice (and generalizes to all initial locations).
This permits the use to classical computing to provide an approximation (large) sections of the ground state, which can be ``stitched together'' on a quantum computer with localized boundary operators (due to confinement) found by variational methods, which we demonstrate for the (small) $L=3$ system.
The operator structures required for each step correspond to hadronic operators, both mesonic and baryonic, defined by position and form factors.  
The challenges that are naively faced, associated with color antisymmetry and entanglement, are naturally accommodated and rendered irrelevant by working with such hadronic operator structures.
For instance, a heavy quark can be created in a total color singlet lattice by acting with a 
heavy-baryon operator or heavy-meson operator
on the interacting vacuum state.   
This new understanding also allows for generalization to systems with larger number of colors,
or, as an example, with background charges in the adjoint representation.

Through extensive classical simulations of small systems, we have developed the necessary quantum circuits that can be scaled to arbitrarily large systems (made possible by truncations of the charge-charge operators enabled by confinement).
Using IBM's {\tt ibm\_pittsburgh},
these circuits (with a transpiled circuit depth $\sim 400$) have been used to perform simulations of single heavy-quark displaced by once spatial site and time evolved forward using one second-order Trotter step.
With a full complement of error-mitigation strategies, including Pauli-twirling, dynamic-decoupling, measurement twirling, ODR, and ZNE, results were obtained that are in good agreement with classical expectations.  

With all the elements in place to perform measurements of energy loss in non-Abelian systems in 1+1D, there are a number of near-term directions to pursue. 
One direction is to perform simulations on larger lattices, with the heavy quark initially located far from the boundaries.
This will permit a more robust implementation of domain decomposition for initial state preparation.
Further, it will allow for estimates of energy loss with reduced systematic uncertainties, and changing the Hamiltonian parameters to decrease the lattice spacing. 
Hadrons created by the motion of the heavy quark will have more time to propagate before encountering the boundaries and reflecting back to the heavy quark, providing an extended time interval to study the fundamental mechanisms without interference.
A second direction is to simulate energy loss and hadronization in the analogous SU(3) lattice gauge theory. 
Heavy quarks can now be included in existing frameworks of 1+1D SU(3) simulations~\cite{Ciavarella:2021nmj,Farrell:2022wyt,Farrell:2022vyh,Atas:2022dqm,Chernyshev:2025lil}. 
Such simulations with one flavor of light quarks require  9 qubits per spatial site ($rgb$ for the heavy and light quarks, and $\bar r \bar g \bar b$ for the light anti-quarks), and two-flavor simulations require 15 qubits per spatial site.
A third direction, which is less trivial and may require the development of additional techniques, is the extension to simulations in 2+1D and 3+1D.  
Significant progress is  being made in evolving quantum simulations of lattice field theories into 2+1D, with explorations in 3+1D in limited cases~\cite{Raychowdhury:2018osk,
Ciavarella:2021nmj,
Zache:2023dko,
Muller:2023nnk,
Kavaki:2024ijd,
Turro:2024pxu,
Gustafson:2023aai,
Kadam:2024ifg,
Fontana:2024rux,
Grabowska:2024emw,
Ciavarella:2024cyt,
Halimeh:2024bth,
Ciavarella:2024fzw,
Balaji:2025afl,
Illa:2025dou,
Illa:2025njz,
Jiang:2025ufg,
Yao:2025uxz}.
Typically, these involve simulating small volumes of the Kogut-Susskind Hamiltonian in the electric basis, with a significantly truncated gauge space, and without quantification of systematic errors.
However, full simulations with both quarks and gluons remain to be performed.  
Such simulations are a prerequisite for qualitative comparisons with experiment and observation, and for making robust predictions in regimes where experiment is not possible.

\begin{acknowledgments}
\noindent
We would like to thank Roland Farrell, Henry Froland, Nikita Zemlevskiy and Roland de Putter for helpful discussions. 
This work was supported, in part, 
by U.S. Department of Energy, Office of Science, Office of Nuclear Physics, InQubator for Quantum Simulation (IQuS)\footnote{\url{https://iqus.uw.edu}} under Award Number DOE (NP) Award DE-SC0020970 via the program on Quantum Horizons: QIS Research and Innovation for Nuclear Science,\footnote{\url{https://science.osti.gov/np/Research/Quantum-Information-Science}}
by the Quantum Science Center (QSC)\footnote{\url{https://www.qscience.org}} which is a National Quantum Information Science Research Center of the U.S.\ Department of Energy, and by PNNL’s Quantum Algorithms and Architecture for Domain Science (QuAADS) Laboratory Directed Research and Development (LDRD) Initiative. The Pacific Northwest National Laboratory is operated by Battelle for the U.S.\ Department of Energy under Contract DE-AC05-76RL01830.
It was also supported, in part, by the Department of Physics\footnote{\url{https://phys.washington.edu}}
and the College of Arts and Sciences\footnote{\url{https://www.artsci.washington.edu}} at the University of Washington. 
This research used resources of the Oak Ridge Leadership Computing Facility (OLCF), 
which is a DOE Office of Science User Facility supported under Contract DE-AC05-00OR22725.
We acknowledge the use of IBM Quantum services for this work. The views expressed are those of the authors, and do not reflect the official policy or position of IBM or the IBM Quantum team. This work was enabled, in part, by the use of advanced computational, storage and networking infrastructure provided by the Hyak supercomputer system at the University of Washington.\footnote{\url{https://itconnect.uw.edu/research/hpc}}
We have made extensive use of Wolfram {\tt Mathematica}~\cite{Mathematica},
{\tt python}~\cite{python3,Hunter:2007}, {\tt julia}~\cite{Julia-2017},
{\tt jupyter} notebooks~\cite{PER-GRA:2007} 
in the {\tt Conda} environment~\cite{anaconda},
and IBM's quantum programming environment {\tt qiskit}~\cite{qiskit}.
This research used resources of the National Energy Research Scientific Computing Center (NERSC), a Department of Energy Office of Science User Facility using NERSC award NP-ERCAP0032083.

\end{acknowledgments}

\bibliography{SU2dEdx}

\appendix

\vfill\eject

\section{Details on the Fermionic SWAP}
\addtocontents{toc}{\vspace{-1em}}
\label{app:fswap}
\noindent
In this appendix, 
we show how the value of the rotation in the FSWAP gate, Eq.~\eqref{eq:FSWAP}, cannot be arbitrary in these simulations.
To understand this limitation, it is helpful to consider a simple model system comprised of two sites $x=\{0,1\}$, defined in the ket (of one of the colors) $|\psi\rangle=|Q_0 Q_1\rangle$ in the occupation basis, with an initial condition of having the heavy quark residing at $x=0$, $|\psi\rangle_0=|\uparrow \downarrow \rangle$. 
In an attempt to move the heavy quark to $x = 1$ using continuous spatial translations, a FSWAP gate is applied, $\hat F(\theta)$, followed by time evolution, followed by a second FSWAP gate, $\hat F(\phi)$. 

Working in the reduced space with states $\{ |\uparrow\downarrow\rangle, |\downarrow\uparrow\rangle \}$, imagine that the evolution operator has the form 
\begin{equation}
\hat U = e^{i{\frac{\alpha}{2}}} e^{i  {\frac{\eta}{2}}\hat Z} 
\ ,
\end{equation}
where $\eta$ and $\alpha$ are phases resulting from the time evolution of the system between the applications of the two FSWAP gates.  
These phase shifts 
model the effect of the time evolution of the light quarks on the wavefunction including the heavy-quark.
The resulting state is 
\begin{align}
|\tilde\psi\rangle \ =\ 
\hat F(\phi) \cdot \hat U \cdot \hat F(\theta) |\psi\rangle_0
 = e^{i\alpha/2}\ e^{i\theta/2}\ e^{i\phi/2}\ &  \left[\ 
|\uparrow\downarrow\rangle 
\left(
\cos{\frac{\eta}{2}} \cos{\frac{\theta+\phi}{2}} + i \sin{\frac{\eta}{2}} \cos{\frac{\theta-\phi}{2}}
\right)
\right.\nonumber\\
& 
\left.
-i|\downarrow\uparrow\rangle 
\left(
\cos{\frac{\eta}{2}} \sin{\frac{\theta+\phi}{2}} - i \sin{\frac{\eta}{2}} \sin{\frac{\theta-\phi}{2}}
\right)
\ \right]
    \ ,
    \label{eq:QQmove}
\end{align}
In particular, when the system with a heavy-quark at $x=0$ is in its ground state, when moved instantaneously to $x=1$ it is in a combination of excited states.
In the case where 
$\theta\ne\phi\ne\pi$ but
$\theta+\phi=\pi$, it would have been ideal for the heavy-quark to have moved from the $x=0$ site to the $x=1$ site, but this is not the situation when $\eta\ne 0$.
Forming the location of the heavy quark from the expectation value, from matrix elements of the operator
\begin{equation}
\hat R = \frac{1}{4} \left(\  2\hat I +\hat Z_2-\hat Z_1\ \right)
    \label{eq:Rop}
\end{equation}
in $|\tilde\psi\rangle$ gives
\begin{equation}
\langle\tilde\psi| \ \hat R \ |\tilde\psi\rangle = 
\frac{1}{2}
\left(
1 - \cos^2{\frac{\eta}{2}} \cos ({\theta+\phi})
 - \sin^2{\frac{\eta}{2}} \cos ({\theta-\phi})
\right)
\ .
\label{eq:RinPsitilde}
\end{equation}
Attempting to move the heavy-quark from $x=0$ to $x=1$ via an intermediate $x=\frac{1}{2}$ location, $\theta=\phi=\pi/2$, leads to the location of the heavy quark being
$\langle \hat R \rangle = 1-\sin^2{\frac{\eta}{2}}$.  
In contrast, making one move to the adjacent lattice site, 
with $\theta=\pi$, $\phi=0$, 
or with $\theta=0$, $\phi=\pi$, 
gives $\langle \hat R \rangle = 1$.

\section{Details of the $L=2$ System}
\addtocontents{toc}{\vspace{-1em}}
\label{app:L2}
\noindent
Details of calculations in the $L=2$ system that are summarized in Sec.~\ref{sec:L2} are presented in this appendix.
While this system follows relatively straightforwardly from the $L=1$ system, 
there are aspects relevant for simulations in larger systems, 
as discussed in the main text.
Following the same path as in Sec.~\ref{sec:L1}, we present results for the eigenstates, energies, spin-observables, and the variational operators and wavefunctions in the 
$n_Q=\{0,1\}$ sectors.

The wavefunction for a state describing  an $L=2$ system
is formed from a linear combination of occupation basis states, 
$|\ Q_r\  Q_g\ q_r\  q_g\  \overline{q}_r\ \overline{q}_g
\ \ Q_r\  Q_g\ q_r\  q_g\  \overline{q}_r\ \overline{q}_g\  \rangle$,
with staggered site labels $n=\{0,1,\ldots,5\}$.
The kinetic energy operator receives contributions from the light quarks, 
\begin{align}
\hat H_k = &
-\frac{1}{2}
\sum_{n=1,4}\sum_{c=0,1}
\left(\ 
\hat \sigma^+_{2n+c} \hat Z_{2n+1+c} \hat \sigma^-_{2n+2+c} 
\ +\ {\rm h.c.}
\right)
\nonumber\\
&
-\frac{1}{2}
\sum_{n=2}\sum_{c=0,1}
\left(\ 
\hat \sigma^+_{2n+c} \hat Z_{2n+1+c} \hat Z_{2n+2+c} \hat Z_{2n+3+c} \hat \sigma^-_{2n+4+c} 
\ +\ {\rm h.c.}
\right)
\ \ ,
\end{align}
where the second term comes for hopping between anti-quarks at site $x=0$ and quarks at $x=1$.
The mass term, using the same arguments as before, takes the form
\begin{equation}
\hat H_m = 
\frac{m_q}{2}
\sum\limits_{n=1,2,4,5}\sum_{c=0,1}
(-)^{\eta_n}\  \hat Z_{2n+c}
\ \ ,
\end{equation}
where $\eta_{1,4}=0$ and $\eta_{2,5} = 1$.
The gauge-field contribution to the Hamiltonian is,
\begin{align}
\hat H_g & = 
\frac{g^2}{2}
\sum\limits_{\substack{a=1,2,3}}
\left[\ 
\left(\hat{\mathcal{Q}}_0^{(a)} +  \hat{\mathcal{Q}}_1^{(a)}\right)^2
\ +\ 
\left(\hat{\mathcal{Q}}_0^{(a)} + \hat{\mathcal{Q}}_1^{(a)}+ \hat{\mathcal{Q}}_2^{(a)}\right)^2
\ +\ 
\left( \hat{\mathcal{Q}}_0^{(a)} + \hat{\mathcal{Q}}_1^{(a)} + \hat{\mathcal{Q}}_2^{(a)} + \hat{\mathcal{Q}}_3^{(a)} + \hat{\mathcal{Q}}_4^{(a)}\right)^2
\ \right]
\ \ , 
\nonumber\\
\hat H_\lambda & = 
\frac{\lambda^2}{2}
\sum\limits_{a=1,2,3}
\left( \hat{\mathcal{Q}}_0^{(a)} + \hat{\mathcal{Q}}_1^{(a)} + \hat{\mathcal{Q}}_2^{(a)} + \hat{\mathcal{Q}}_3^{(a)} + \hat{\mathcal{Q}}_4^{(a)} + \hat{\mathcal{Q}}_5^{(a)} \right)^2
\ \ ,
\end{align}
with the charge operators defined in Eq.~\eqref{eq:Qadefs},
which can be replaced with the more compact expression (see App.~\ref{app:explicitham})
\begin{equation}
\hat H_g = 
\frac{g^2}{2}
\sum\limits_{\substack{a=1,2,3}}
\left[\ 
\left( \hat{\mathcal{Q}}_0^{(a)} + \hat{\mathcal{Q}}_1^{(a)}\right)^2
\ +\ 
\left( \hat{\mathcal{Q}}_0^{(a)} + \hat{\mathcal{Q}}_1^{(a)} + \hat{\mathcal{Q}}_2^{(a)}\right)^2
\ +\ 
\left( \hat{\mathcal{Q}}_5^{(a)}\right)^2
\ \right]
\ \ ,
\end{equation}
with $\lambda\rightarrow \infty$ to enforce the color-neutrality constraint.

We use the Lanczos algorithm~\cite{Lanczos:1950zz} to generate the ground-state energies and wavefunctions.
Working with $g=1.0$ and $m_q=0.1$ in the $n_Q=0$ sector,
our starting state is the SC vacuum is
\begin{equation}
|SC; 2, 0\rangle = 
| \downarrow\downarrow\downarrow\downarrow\ \uparrow\uparrow\ 
\downarrow\downarrow\downarrow\downarrow\ \uparrow\uparrow\ \rangle
\ \ .
\label{eq:psiSCL2Q0}
\end{equation}
Iterating the Lanczos algorithm to convergence with the system Hamiltonian gives
\begin{align}
|I; 2, 0\rangle  = & \ 
0.47806 | \downarrow\downarrow\downarrow\downarrow\ \uparrow\uparrow\ 
\downarrow\downarrow\downarrow\downarrow\ \uparrow\uparrow\ \rangle
\nonumber\\
& +  
0.27134 | \downarrow\downarrow\downarrow\downarrow\ \uparrow\uparrow\ 
\downarrow\downarrow\downarrow\uparrow\ \uparrow\downarrow\ \rangle
\ +\ 
0.27134 | \downarrow\downarrow\downarrow\uparrow\ \uparrow\downarrow\ 
\downarrow\downarrow\downarrow\downarrow\ \uparrow\uparrow\ \rangle
\nonumber\\
& -
0.27134 
| \downarrow\downarrow\downarrow\downarrow\ \uparrow\uparrow\ 
\downarrow\downarrow\uparrow\downarrow\ \downarrow\uparrow\ \rangle
\ -\ 
0.27134 
| \downarrow\downarrow\uparrow\downarrow\ \downarrow\uparrow\ 
\downarrow\downarrow\downarrow\downarrow\ \uparrow\uparrow\ \rangle
\nonumber\\
& + 
0.22182 
| \downarrow\downarrow\uparrow\uparrow\ \downarrow\downarrow\ 
\downarrow\downarrow\downarrow\downarrow\ \uparrow\uparrow\ \rangle
\ +\ 
0.22182 
| \downarrow\downarrow\downarrow\downarrow\ \uparrow\uparrow\ 
\downarrow\downarrow\uparrow\uparrow\ \downarrow\downarrow\ \rangle
\ +\ \cdots
\ \ ,
\end{align}
which exhibits the structure of mesons and baryon-antibaryon-pair excitations above the SC vacuum.
The ground-state energy of 
$E_0=-1.5179$, 
with corresponding energy density of 
$E_0/L\equiv \epsilon_0=-0.7589$.
The matrix elements of the components of the Hamiltonian density are
\begin{equation}
\langle \hat H_k \rangle/L \ =\ 
-1.04757
\ ,\ 
\langle \hat H_m \rangle/L \ =\ 
+0.13346
\ ,\ 
\langle \hat H_g \rangle/L = 
+0.15512
\ ,\ 
\langle \hat H_{\rm ext.}\rangle/L \ =\ 0
\ \ ,
\end{equation}
and the resulting spin alignments are 
\begin{equation}
\langle \hat Z_j \rangle =
\{  -1, -1, -0.38569, -0.38569, 0.279695, 0.279695, -1, -1,
-0.279695, -0.279695, 0.38569, 0.38569 \}
\ \ .
\end{equation}

There are three meson operators (two acting on a single spatial site, $\hat O_{M0}^{(0)}$ and $\hat O_{M0}^{(1)}$, and one between sites, $\hat O_{M1}^{(0,1)}$), and three baryon operators with analogous actions ($\hat O_{B0}^{(0)}$, $\hat O_{B0}^{(1)}$, and $\hat O_{B1}^{(0,1)}$) that can be used to prepare the interacting ground state with sub-percent precision,
\begin{align}
    \hat O_{M0}^{(0)} & \equiv
    \hat I^2 \left(
\hat X \hat Z \hat Y \hat I 
- \hat Y\hat Z \hat X \hat I
+ \hat I \hat X \hat Z \hat Y
- \hat I \hat Y \hat Z \hat X \right) \hat I^6 \ ,
    \nonumber\\
\hat O_{M0}^{(1)} & \equiv
\hat I^8 \left(
    \hat X\hat Z \hat Y \hat I 
    - \hat Y\hat Z \hat X \hat I
    + \hat I \hat X\hat Z \hat Y  
    - \hat I\hat Y\hat Z \hat X \right) \ ,
        \nonumber\\
    \hat O_{M1}^{(0,1)} & \equiv
    -\hat I^4
    \left(
    \hat X\hat Z^3 \hat Y \hat I
    -\hat Y\hat Z^3 \hat X \hat I
+
    \hat I\hat X   \hat Z^3 \hat Y
    -\hat I \hat Y  \hat Z^3 \hat X
\right)  \hat I^2 \ ,
        \nonumber\\
    \hat O_{B0}^{(0)} & \equiv 4 i \hat I^2\left(
    \hat\sigma^+\hat\sigma^+\hat\sigma^-\hat\sigma^-
    -
    \hat\sigma^-\hat\sigma^-\hat\sigma^+\hat\sigma^+ \right)
    \hat I^6 \ ,
        \nonumber\\
    \hat O_{B0}^{(1)}  & \equiv 4 i \hat I^8\left(
    \hat\sigma^+\hat\sigma^+\hat\sigma^-\hat\sigma^-
    -
    \hat\sigma^-\hat\sigma^-\hat\sigma^+\hat\sigma^+ \right) \ ,
        \nonumber\\
    \hat O_{B1}^{(0,1)}  & \equiv -4 i \hat I^4\left(
     \hat\sigma^+\hat\sigma^+ \hat{I^2}\hat\sigma^-\hat\sigma^-
    -
    \hat\sigma^-\hat\sigma^-\hat{I^2} \hat\sigma^+\hat\sigma^+ \right)\hat I^2
\ .
\label{eq:OPcomsL2}
\end{align}
To identify the optimal operator ordering and subsequent layer-by-layer parameter optimizations, the meson operators are applied first
to the SC vacuum in Eq.~\eqref{eq:psiSCL2Q0},
followed by the baryon operators, to give Eq.~\eqref{eq:L2UNITQ0},
\begin{equation}
\hat U = 
e^{-i \theta_5 \hat O_{B1}^{(0,1)}}
e^{-i \theta_4 \hat O_{M1}^{(0,1)}}
e^{-i \theta_3 (\hat O_{B0}^{(0)}+\hat O_{B0}^{(1)})}
e^{-i \theta_2 (\hat O_{M0}^{(0)}+\hat O_{M0}^{(1)})}
e^{-i \theta_1 \hat O_{M1}^{(0,1)}}
\ .
\label{eq:L2UNITQ0app}
\end{equation}
The form of this variational ansatz is motivated by the symmetries of the system - the translational invariance (modulo OBCs) and color neutrality.
Table~\ref{tab:L2VQEmanuelQ0} shows the results of tuning the parameters defining the wavefunction ansatz in Eq.~\eqref{eq:L2UNITQ0} to the exact $L=2$, $n_Q=0$ wavefunction.
\begin{table}[!ht]
\renewcommand{\arraystretch}{1.1}
\begin{tabularx}{\textwidth}{| Y || Y | Y | Y | Y | Y |  Y | Y | Y |}
 \hline
 {\rm Quantity}
 & $\hat I$
 & $\hat O_{M1}^{(0,1)}$
 & $\hat O_{M0}^{(0)}+\hat O_{M0}^{(1)}$
 & $\hat O_{B0}^{(0)}+\hat O_{B0}^{(1)}$
 & $\hat O_{M1}^{(0,1)}$
 & $\hat O_{B1}^{(0,1)}$
 \\
\hline\hline 
 $\infiL$
& 0.3857 
& 0.3479 & 0.0202 & 0.0133 & 0.0020  & 0.0007\\
\hline
\hline
$\theta_i$ 
& - 
& 0.1907 & & & &\\
 \hline
& - 
& 0.1291 & 0.2967 & & &\\
 \hline
& - 
& 0.1291 & 0.2543 & 0.0469 & &\\
 \hline
& - 
& 0.2264 & 0.2802 & 0.0588 & -0.1498 &\\
 \hline
& - 
& 0.2316 & 0.2790 & 0.0637 & -0.1691 &  0.0289 \\
 \hline
 \hline
\end{tabularx}
\caption{
Results from  fitting the variational wavefunction  in Eq.~\eqref{eq:L2UNITQ0} with $m_q=0.1$ and $g=1.0$ through five layers, optimized at each level.  
Shown are the infidelity density $\infiL$,  defined in Eq.~\eqref{eq:infidelitydens}, and fit angles $\theta_i$, for the $n_Q=0$ state with $L=2$.
}
\label{tab:L2VQEmanuelQ0}
\end{table}
The expectation values $\langle \hat Z_j \rangle$ are given in Table~\ref{tab:L2Q0Z} and displayed in Fig.~\ref{fig:L2nQ01Z}.
They show the convergence with the addition of each layer of the state preparation in Eq.~\eqref{eq:L2UNITQ0}, and subsequent re-optimization.
\begin{table}[!ht]
\renewcommand{\arraystretch}{1.1}
\begin{tabularx}{\textwidth}{|c || Y || Y | Y | Y | Y | Y |  Y | Y | Y |  Y | Y | Y |}
 \hline
 state
 & 0
 & 1
 & 2
 & 3
 & 4
 & 5
 & 6
 & 7
 & 8
 & 9
 & 10
 & 11
 \\
\hline\hline 
$|SC; 2, 0\rangle$ 
&  -1 & -1 & -1 & -1 & 1 & 1 & -1 & -1 & -1 & -1 & 1 & 1 \\
layer-1 
&  -1 & -1 & -1 & -1 & 0.7220 & 0.7220 & -1 & -1 & -0.7220 & -0.7220
   & 1 & 1 \\
layer-2 &  -1 & -1 & -0.4127 & -0.4127 & 0.2810 & 0.2810 & -1 & -1 & -0.2810
   & -0.2810 & 0.4127 & 0.4127 \\
layer-3 &  -1 & -1 & -0.4048 & -0.4048 & 0.2731 & 0.2731 & -1 & -1 & -0.2731 &
   -0.2731 & 0.4048 & 0.4048 \\
layer-4 &  -1 & -1 & -0.3936 & -0.3936 & 0.2841 & 0.2841 & -1 & -1 & -0.2841
   & -0.2841 & 0.3936 & 0.3936 \\
layer-5 &  -1 & -1 & -0.3890 & -0.3890 & 0.2842 & 0.2842 & -1 & -1 & -0.2842
   & -0.2842 & 0.3890 & 0.3890 \\
$|I; 2, 0\rangle$ &  -1 & -1 & -0.3856 & -0.3856 & 0.2796 & 0.2796 & -1 & -1 & -0.2796
   & -0.2796 & 0.3856 & 0.3856 \\
 \hline
 \hline
\end{tabularx}
\caption{
The $\langle \hat Z_j \rangle$ as a function of the fermion site $j$, starting from the SC vacuum, through the classically optimized layers, 
and comparing to the exact values,
for the $n_Q=0$ state with $L=2$, $m_q=0.1$ and $g=1.0$.
These results are displayed in Fig.~\ref{fig:L2nQ01Z}.
}
\label{tab:L2Q0Z}
\end{table}

In the $n_Q=1$ sector with the heavy quark located at $x=0$, 
the SC vacuum is
\begin{equation}
|SC; 2, 1\rangle = 
\frac{1}{\sqrt{2}}
\biggl[\ 
| \uparrow\downarrow\downarrow\uparrow\ \uparrow\uparrow\ 
\downarrow\downarrow\downarrow\downarrow\ \uparrow\uparrow\ \rangle
- 
| \downarrow\uparrow\uparrow\downarrow\ \uparrow\uparrow\ 
\downarrow\downarrow\downarrow\downarrow\ \uparrow\uparrow\ \rangle
\biggr]
\ \ ,
\label{eq:nq2L2}
\end{equation}
and is no longer spatially symmetric because of the heavy quark at $x=0$. 
However,  as with the $n_Q=0$ sector, 
to identify the optimal operator ordering and subsequent layer-by-layer parameter optimizations, 
we first implement the meson operators followed by the baryon operators.
The meson operators are sequentially implemented with distance from the heavy quark, 
followed by two baryon operators, and then followed by the closest meson operator,
\begin{equation}
\hat U  = 
e^{-i \theta_6 \hat O_{M0}^{(0)}}\ 
e^{-i \theta_5 \hat O_{B1}^{(0,1)}}\ 
e^{-i \theta_4 \hat O_{B0}^{(1)}}\ 
e^{-i \theta_3 \hat O_{M0}^{(1)}}\ 
e^{-i \theta_2 \hat O_{M1}^{(0,1)}}\ 
e^{-i \theta_1 \hat O_{M0}^{(0)}}
\ ,
\label{eq:L2UNITQ1app}
\end{equation}
applied to $|SC; 2, 1\rangle$, reproducing Eq.~\eqref{eq:L2UNITQ1} in the main text. 
The operator sequence has the mesons implemented sequentially away from the heavy quark, one at the boundary, baryon-antibaryon pairs sequentially are applied moving closer to the heavy quark, with a meson operator at the heavy quark.
The fit angles and infidelities are given in Table~\ref{tab:L2VQEmanuelQ1},
\begin{table}[!ht]
\renewcommand{\arraystretch}{1.2}
\begin{tabularx}{\textwidth}{| Y || Y | Y | Y | Y | Y |  Y | Y | Y | Y |}
 \hline
 {\rm Quantity}
 & $\hat I$
 & $\hat O_{M0}^{(0)}$
 & $\hat  O_{M1}^{(0,1)}$
 & $\hat O_{M0}^{(1)}$
 & $\hat O_{B0}^{(1)}$
 & $\hat O_{B1}^{(0,1)}$
 & $\hat O_{M0}^{(0)}$
 \\
\hline\hline 
 $\infiL$
& 0.336982 
& 0.2965 & 0.1855 & 0.02256 & 0.02048 & 0.01959 & 0.003619\\
\hline
\hline
$\theta_i$ 
& - 
& 0.2309 & & & & &\\
& - 
& 0.2096 & 0.2473 & & & &\\
& - 
& 0.2231 & 0.2152 & 0.2563 & & &\\
& - 
& 0.2231 & 0.2149 & 0.2318 & 0.03233 & &\\
& - 
& 0.2223 & 0.2025 & 0.2283 & 0.03208 &  0.02261 &\\
& - 
& 0.3862 & 0.2358 & 0.2282 & 0.03233 &  0.02613 & -0.1837\\
 \hline
 \hline
\end{tabularx}
\caption{
Results from  fitting the variational wavefunction  in Eq.~\eqref{eq:L2UNITQ1}
with $m_q=0.1$ and $g=1.0$ through six layers, optimized at each level.  
Shown are the infidelity density, $\infiL$, 
defined in Eq.~\eqref{eq:infidelitydens},
and fit angles, $\theta_i$, for the $n_Q=1$ state with $L=2$.
}
\label{tab:L2VQEmanuelQ1}
\end{table}
and the expectation values $\langle \hat Z_j \rangle$ are given in Table~\ref{tab:L2Q1Z} and displayed in Fig.~\ref{fig:L2nQ01Z}.  
They show the pattern of convergence with the addition of each layer of state preparation
in Eq.~\eqref{eq:L2UNITQ1}, and subsequent re-optimizations.
\begin{table}[!ht]
\renewcommand{\arraystretch}{1.2}
\begin{tabularx}{\textwidth}{|c || Y | Y | Y | Y | Y | Y |  Y | Y | Y |  Y | Y | Y |}
 \hline
 State
 & 0
 & 1
 & 2
 & 3
 & 4
 & 5
 & 6
 & 7
 & 8
 & 9
 & 10
 & 11
 \\
\hline\hline 
$|SC; 2, 1\rangle$
&  0 & 0 & 0 & 0 & 1 & 1 & -1 & -1 & -1 & -1 & 1 & 1 \\
layer-1 
&  0 & 0 & 0.1971 & 0.1971 & 0.8028 & 0.8028 & -1 & -1 & -1 & -1 & 1
   & 1. \\
layer-2 &  0 & 0 & 0.1660 & 0.1660 & 0.4148 & 0.4148 & -1 & -1 & -0.5809 &
   -0.5809 & 1 & 1 \\
layer-3 &  0 & 0 & 0.1839 & 0.1839 & 0.4941 & 0.4941 & -1 & -1 & -0.2752 &
   -0.2752 & 0.5971 & 0.5971 \\
layer-4 &  0 & 0 & 0.1844 & 0.1844 & 0.5049 & 0.5049 & -1 & -1 & -0.2843 &
   -0.2843 & 0.5948 & 0.5948 \\
layer-5 &  0 & 0 & 0.1854 & 0.1854 & 0.5061 & 0.5061 & -1 & -1 & -0.2796 &
   -0.2796 & 0.5880 & 0.5880 \\
   layer-6 & 0 & 0 & 0.1533 & 0.1533 & 0.5034 & 0.5034 & -1 & -1 & -0.2581 &
   -0.2581 & 0.6013 & 0.6013 \\ \hline
$|I; 2, 1\rangle$ &  0 & 0 & 0.1554 & 0.1554 & 0.4989 & 0.4989 & -1 & -1 & -0.2498 & -0.2498 & 0.5955 & 0.5955 \\
 \hline
 \hline
\end{tabularx}
\caption{
The $\langle \hat Z_j \rangle$ as a function of fermion site $j$, 
from the SC vacuum, $|SC; 2, 1\rangle$,  
through the classically optimized layers, 
through to the exact values from $|I; 2, 1\rangle$
for the $n_Q=0$ state with $L=2$, $m_q=0.1$ and $g=1.0$.
These results are displayed in Fig.~\ref{fig:L2nQ01Z}.
}
\label{tab:L2Q1Z}
\end{table}
The energy of the ground state in this sector is 
$E_1=-1.162$, giving a hadron mass of 
$\Lambda_Q=E_1 - E_0 =0.3557$.

\section{Details of the $L=3$ System}
\addtocontents{toc}{\vspace{-1em}}
\label{app:L3}
\noindent
Further details related to the $L=3$ results presented in Sec.~\ref{sec:L3SP} are presented in this appendix.
Specifically, we
present results for the eigenstates, energies, spin-observables, color-entanglement, and variational operators and wavefunctions in the $n_Q=0,1$ sectors.

Working with $g=1.0$ and $m_q=0.1$ (for the sake of comparison), 
in the light-sector ($n_Q=0$),
the ground-state wavefunction is determined using the Lanczos algorithm
applied to the SC vacuum, 
$|SC; 3, 0\rangle$, requiring more than 50 iterations to converge,
\begin{align}
|I; 3, 0\rangle = & \
0.345173 | \downarrow\downarrow\downarrow\downarrow\ \uparrow\uparrow\ 
\downarrow\downarrow\downarrow\downarrow\ \uparrow\uparrow\ 
\downarrow\downarrow\downarrow\downarrow\ \uparrow\uparrow \rangle
\nonumber\\
& \ +\ 
0.19079 | \downarrow\downarrow\downarrow\uparrow\ \uparrow\downarrow\ 
\downarrow\downarrow\downarrow\downarrow\ \uparrow\uparrow\
\downarrow\downarrow\downarrow\downarrow\ \uparrow\uparrow \rangle
\ -\ 
0.19079 | \downarrow\downarrow\downarrow\downarrow\ \uparrow\uparrow\ 
\downarrow\downarrow\downarrow\downarrow\ \uparrow\uparrow\ 
\downarrow\downarrow\uparrow\downarrow\ \downarrow\uparrow \rangle
\nonumber\\
& \ +\ 
0.19079 
| \downarrow\downarrow\downarrow\downarrow\ \uparrow\uparrow\ 
\downarrow\downarrow\downarrow\downarrow\ \uparrow\uparrow\
\downarrow\downarrow\downarrow\uparrow\ \uparrow\downarrow \rangle
\ -\ 
0.19079 
| \downarrow\downarrow\uparrow\downarrow\ \downarrow\uparrow\ 
\downarrow\downarrow\downarrow\downarrow\ \uparrow\uparrow\
\downarrow\downarrow\downarrow\downarrow\ \uparrow\uparrow \rangle
\ +\ \cdots
\ \ .
\label{eq:L3exact}
\end{align}
It has an energy of $E_0=-2.3994$
(with a corresponding energy density of $\epsilon_0=-0.7998$),
and matrix elements of the components of the Hamiltonian density,
\begin{equation}
\langle \hat H_k \rangle/L \ =\ 
-1.0846
\ ,\ 
\langle \hat H_m \rangle/L \ =\ 
+0.1264
\ ,\ 
\langle \hat H_g \rangle/L = 
+0.1583
\ ,\ 
\langle \hat H_{\rm ext.}\rangle \ =\ 0
\ \ .
\label{eq:HdesL3}
\end{equation}
The $\langle \hat Z_j \rangle$,
exhibiting the expected reflection symmetry,
are 
\begin{align}
\langle \hat Z_j \rangle = &  \ 
\big\{-1, -1, -0.4411, -0.4411, 0.2877, 0.2877,
\nonumber  \\ 
& \ \ -1, -1,
-0.3744, -0.3744, 0.3744, 0.3744,
\nonumber \\
& \ \   -1, -1, -0.2877,
-0.2877, 0.4411, 0.4411\  
\big\}
\ \ .
\label{eq:ZnL3n0}
\end{align}

In the $n_Q=1$ sector with the heavy quark located at $x=0$, 
the SC wavefunction is
\begin{equation}
|I; 3, 1, x=0\rangle  = 
\frac{1}{\sqrt{2}}
\biggl[
| \uparrow\downarrow\downarrow\uparrow\ \uparrow\uparrow\ 
\downarrow\downarrow\downarrow\downarrow\ \uparrow\uparrow\
\downarrow\downarrow\downarrow\downarrow\ \uparrow\uparrow \rangle
\ -\  
| \downarrow\uparrow\uparrow\downarrow\ \uparrow\uparrow\ 
\downarrow\downarrow\downarrow\downarrow\ \uparrow\uparrow\
\downarrow\downarrow\downarrow\downarrow\ \uparrow\uparrow \rangle
\biggl]
\ .
\label{eq:SCL3n1x0}
\end{equation}
and the ``exact'' wavefunction, determined by the Lanczos algorithm, is 
\begin{align}
|I; 3, 1, x=0\rangle = & \
0.273228| \uparrow\downarrow\downarrow\uparrow\ \uparrow\uparrow\ 
\downarrow\downarrow\downarrow\downarrow\ \uparrow\uparrow\
\downarrow\downarrow\downarrow\downarrow\ \uparrow\uparrow \rangle
\ -\  
0.273228| \downarrow\uparrow\uparrow\downarrow\ \uparrow\uparrow\ 
\downarrow\downarrow\downarrow\downarrow\ \uparrow\uparrow\
\downarrow\downarrow\downarrow\downarrow\ \uparrow\uparrow \rangle
\nonumber\\
&
\ -\ 
0.161826 | \uparrow\downarrow\downarrow\uparrow\ \uparrow\downarrow\ 
\downarrow\downarrow\downarrow\uparrow\ \uparrow\uparrow\
\downarrow\downarrow\downarrow\downarrow\ \uparrow\uparrow\ \rangle
\ -\ 
0.161826 | \downarrow\uparrow\uparrow\downarrow\ \downarrow\uparrow\ 
\downarrow\downarrow\uparrow\downarrow\ \uparrow\uparrow\
\downarrow\downarrow\downarrow\downarrow\ \uparrow\uparrow\ \rangle
\ +\ \cdots
\ \ ,
\label{eq:L3exactn1}
\end{align}
with an energy of $E_1=-2.0806$.
This gives rise to a hadron mass of $\Lambda_Q = 0.3188$. 
The resulting spin alignments are 
\begin{align}
\langle \hat Z_j \rangle  = &  \
\big\{0, \ \ \ 0,\ \  0.1169, \ \ \ 0.1169, 0.4204, 0.4204, 
\nonumber\\
& \ -1, -1, -0.2705, 
-0.2705, 0.4939, 0.4939, 
\nonumber \\
& \ -1, -1, -0.2878, -0.2878,  0.5270, 0.5270
\big\}
\ \ .
\end{align}
\begin{table}[!ht]
\renewcommand{\arraystretch}{1.2}
\begin{tabularx}{\textwidth}{| Y || Y | Y | Y | Y | Y |  Y | Y | Y |Y | Y |  Y | Y | Y |}
 \hline
 {\rm Quantity}
 & $\hat I$
 & $\hat O_{M0}^{(0)} $
 & $\hat O_{M0}^{(2)} $
 & $\hat O_{M1}^{(0,1)} $
 & $\hat O_{B0}^{(2)} $
 & $\hat O_{M0}^{(1)}$
 & $\hat O_{B0}^{(1)}$
 & $\hat O_{B1}^{(0,1)}$
 & $\hat O_{M0}^{(0)} $
 & $\hat O_{M1}^{(1,2)} $
 & $\hat O_{M2}^{(1,2)}$
 \\
\hline\hline 
 $\infiL$
& 0.2835 
& & 0.2258 
& & 0.1557 
& 0.0889 & 0.0881  & 0.0875
& 0.0765 & 0.0171 & 0.0043
 \\
\hline
\hline
 ${\cal M}_2$ 
 & 0 
 & & 1.519(0)
 & & 1.993(0)
 & 3.04(2)
 & 3.06(3)
 & 3.07(3)
 & 3.21(3)
 & 4.07(4)
 & 4.16(3)
\\
\hline
\hline
 $\theta_i$ 
& - & 0.2270 & 0.2687 & & & &&&&&\\
 & - & 0.2024 & 0.2363 & 0.2644 & 0.0328 &&&&&&\\
 & - & 0.2141 & 0.2411 & 0.2369 & 0.0358 & 0.2280&&&&& \\
 & - & 0.2137 & 0.2375 & 0.2360 & 0.0380 & 0.2092&0.0268&&&&\\
 & - & 0.2165 & 0.2386 & 0.2203 & 0.0386 & 0.2072& 0.0261& 0.0247 &&&\\
 & - & 0.3706 & 0.2401 & 0.2591 & 0.0370 & 0.2102& 0.0235& 0.0278 & -0.1878 &&\\
 & - & 0.3753 & 0.2139 & 0.2675 & 0.0325 & 0.1831& 0.0187& 0.0401 & -0.1997 & 0.2002&\\
 & - & 0.3802 & 0.2200 & 0.2642 & 0.0270 & 0.1820& 0.0196& 0.0407 & -0.2000 & 0.2314& -0.0995\\
 \hline
 \hline
\end{tabularx}
\caption{
Results from  fitting the variational wavefunction in Eq.~\eqref{eq:L3U}
with $m_q=0.1$ and $g=1.0$ through eight layers, optimized at each level.  
DDec of the $L=3$ system is used to determine the sequence of operators from the
$L=1$ and $L=2$ systems, along with their initial angles in optimization.
Shown are the infidelity density, $\infiL$, 
defined in Eq.~\eqref{eq:infidelitydens},
the stabilizer Renyi entropy ${\cal M}_2$
and fit angles $\theta_i$, for the $n_Q=1$ state with $L=3$.
}
\label{tab:L3DoDeQ1}
\end{table}
Table~\ref{tab:L3DoDeQ1} shows the results of a layer-by-layer optimization of  parameters in the variational wavefunction in Eq.~\eqref{eq:L3U}.
Table~\ref{tab:L3Q1Z} provides the numerical values of 
$\langle \hat Z_j\rangle$ obtained 
from optimizing each layer preparing 
$|I; 3, 1, x=0\rangle$ state with $m_q=0.1$ and $g=1.0$,
which are displayed in Fig.~\ref{fig:L3nQ1Z}.
\begin{table}[!ht]
\renewcommand{\arraystretch}{1.2}
\begin{tabularx}{\textwidth}{| c || Y | Y | Y | Y || Y | Y |  Y | Y || Y |  Y | Y | Y |}
 \hline
 State
 & 2
 & 3
 & 4
 & 5
 & 8
 & 9
 & 10
 & 11
 & 14
 & 15
 & 16
 & 17
 \\
\hline\hline 
$|SC; 3, 1\rangle$
 & 0 & 0 & 1 & 1  & -1 & -1 & 1 & 1  & -1 & -1 & 1 & 1
   \\
layer-1 &  
 0.1935 & 0.1935 & 0.8064 & 0.8064  & -1 & -1 & 1 & 1  & $-0.4727$ & $-0.4727$ & 0.4727 & 0.4727 
   \\
layer-2 &  
 0.1593 & 0.1593 & 0.3711 & 0.3711  & -0.5305 &
   -0.5305 & 1 & 1  & -0.4961 & -0.4961 & 0.4961 & 0.4961
   \\
layer-3  & 0.1705 & 0.1705 & 0.4447 & 0.4447  & -0.3033 &
   -0.3033 & 0.6880 & 0.6880 &  -0.4444 & -0.4444 & 0.4444 &
   0.4444 
   \\
layer-4  &   0.1728 & 0.1728 & 0.4501 & 0.4501  & -0.3069 &
   -0.3069 & 0.6839 & 0.6839  & -0.4533 & -0.4533 &
   0.4533 & 0.4533 \\
layer-5  & 0.1763 & 0.1763 & 0.4437 & 0.4437 & -0.2980 &
   -0.2980 & 0.6779 & 0.6779 &  -0.4558 & -0.4558 &
   0.4558 & 0.4558 \\
layer-6  &  0.1264 & 0.1264 & 0.4557 & 0.4557 &  -0.2747 &
   -0.2747 & 0.6926 & 0.6926 &  -0.4536 & -0.4536 &
   0.4536 & 0.4536 \\
layer-7  &  0.1115 & 0.1115 & 0.4253 & 0.4253 &  -0.3049 &
   -0.3049 & 0.5661 & 0.5661 &  -0.3581 & -0.3581 &
   0.5600 & 0.5600 \\
layer-8  &   0.1177 & 0.1177 & 0.4310 & 0.4310 &  -0.2834 &
   -0.2834 & 0.5035 & 0.5035 &  -0.2929 & -0.2929 & 0.5240
   & 0.5240 \\ \hline
$|I; 3, 1\rangle$
   & 0.1169 & 0.1169 & 0.4204 & 0.4204 &  -0.2705 &
   -0.2705 & 0.4940 & 0.4940 &  -0.2878 & -0.2878 &
   0.5270 & 0.5270 \\
\hline
\hline
\end{tabularx}
\caption{
The $\langle \hat Z_j \rangle$ as a function of fermion site number $j$, starting from the $L=3$ SC ground state, 
$|SC; 3, 1, x=0\rangle$, 
through the classically optimized layers,  
and comparing to the exact values $|I; 3, 1, x=0\rangle$ with $m_q=0.1$ and $g=1.0$.
The values for the heavy-quark sites are
$\langle \hat Z_j \rangle=0$ for $j=\{0,1\}$, and
$\langle \hat Z_j \rangle=-1$ for $j=\{6,7,12,13\}$.
These results are displayed in Fig.~\ref{fig:L3nQ1Z}.
}
\label{tab:L3Q1Z}
\end{table}

Table~\ref{tab:L3Q1Iijtau4} displays the values of the mutual information and 4-tangles among the heavy quark and light quarks and anti-quarks for $L=3$.
\begin{table}[!ht]
\renewcommand{\arraystretch}{1.0}
\begin{tabularx}{\textwidth}{| c || Y | Y | Y ||}
 \hline
Quantity
& $x=0$
& $x=1$
& $x=2$
 \\
\hline\hline 
$I^{(0,x)}_q$ & 1.5182 & 0.0367 & $1.59\times 10^{-4}$\\
$I^{(0,x)}_{\bar{q}}$ & 0.0448 & $2.52\times 10^{-3}$ & $9.20\times 10^{-5}$\\
\hline
$\tau^{(0,x)}_{4q}$ & 0.618 & $8.03 \times 10^{-3}$ & $3.07\times 10^{-5}$\\
$\tau^{(0,x)}_{4\bar{q}}$ & $8.95 \times 10^{-3}$ & $4.17 \times 10^{-4}$ 
& $1.30\times 10^{-5}$ \\
\hline
\hline
\end{tabularx}
\caption{ The mutual information and 4-tangles among the heavy quark 
at $x=0$
and light quarks and anti-quarks for $L=3$ with $m_q=0.1$ and $g=1.0$.}
\label{tab:L3Q1Iijtau4}
\end{table}
%

\section{Details of the Heavy-Quark Motion for $L=3$}
\begin{figure}[ht!]
    \centering
    \includegraphics[width=0.8\linewidth]{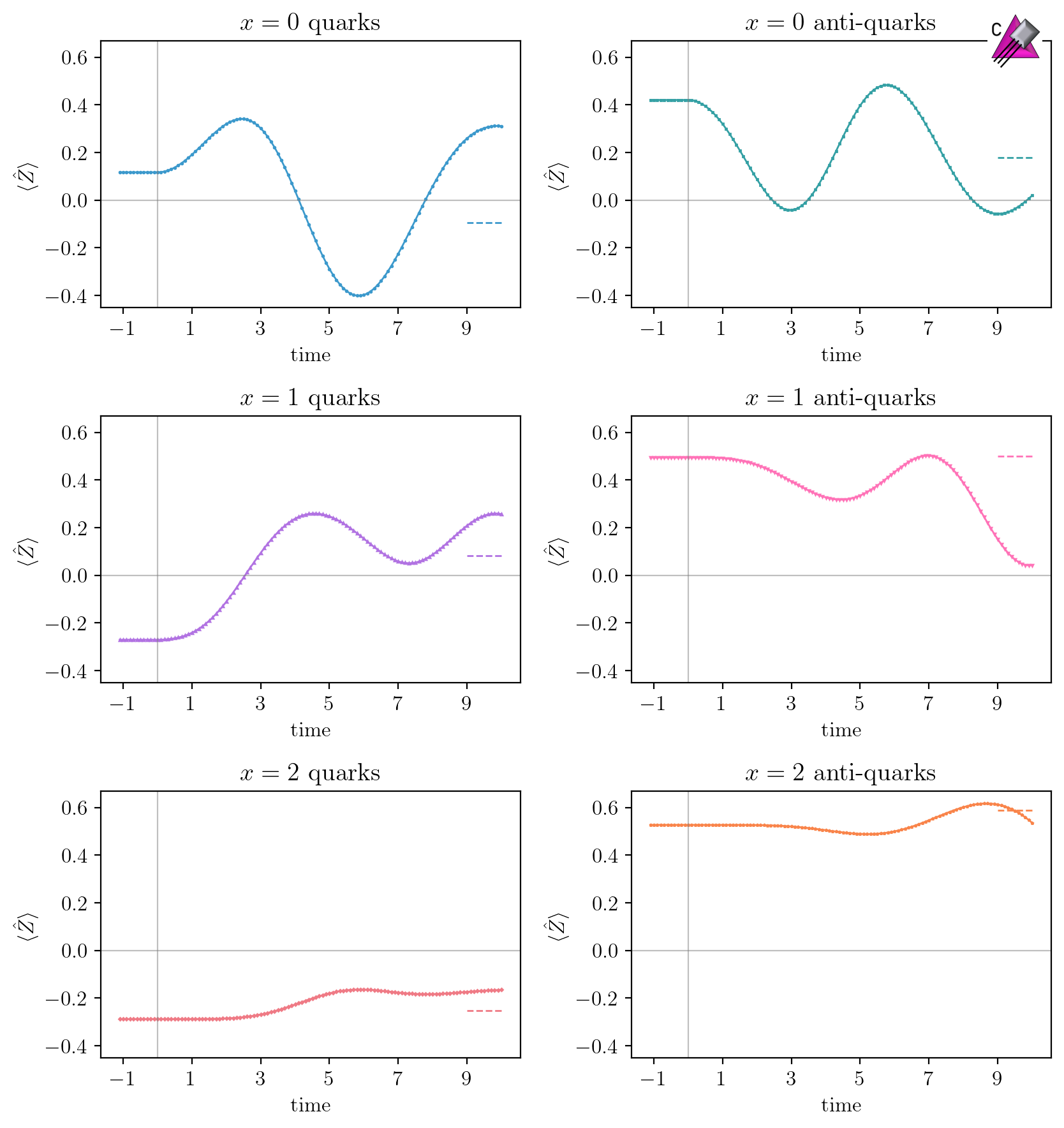}
\caption{The individual expectation values of the $\hat Z_j$ as a function of time in the $L=3$ system.  
At $t=0$, the heavy quark is instantaneously moved form $x=0$ to $x=1$.
}
    \label{fig:L3Ziind}
\end{figure}

\addtocontents{toc}{\vspace{-1em}}
\label{app:L3motion}
\noindent
Figure.~\ref{fig:L3Hi} of the main text displays the time evolution of the 
$\langle \hat Z_j \rangle$ subsequent to moving the heavy quark from $x=0$ to $x=1$.
It is helpful to display the time dependencies individually, as shown in 
Fig.~\ref{fig:L3Ziind}.
\begin{table}[!ht]
\renewcommand{\arraystretch}{1.0}
\begin{tabularx}{\textwidth}{| c || Y | Y | Y ||}
 \hline
Quantity
& $x=0$
& $x=1$
& $x=2$
 \\
\hline\hline 
$\tau^{(1,x)}_{4q}$ & 0.0209 & $0.3704$ & $2.800\times 10^{-3}$\\
$\tau^{(1,x)}_{4\bar{q}}$ & $2.924 \times 10^{-3}$ 
& $1.238 \times 10^{-2}$ 
& $8.016\times 10^{-4}$ \\
\hline
\hline
\end{tabularx}
\caption{ The mutual information and 4-tangles among the heavy quark 
at $x=1$
and light quarks and anti-quarks for $L=3$ with $m_q=0.1$ and $g=1.0$.}
\label{tab:L3Q1tau4x1}
\end{table}
%

\section{Circuit Construction}
\addtocontents{toc}{\vspace{-1em}}
\label{appen:cir}
\noindent 
This appendix is a supplement to
Sec.~\ref{sec:ApTpC} and gives further details 
about the construction of the quantum circuits used
in our simulations.

\begin{figure}[!ht]
    \centering
    \includegraphics[width=0.75\textwidth]{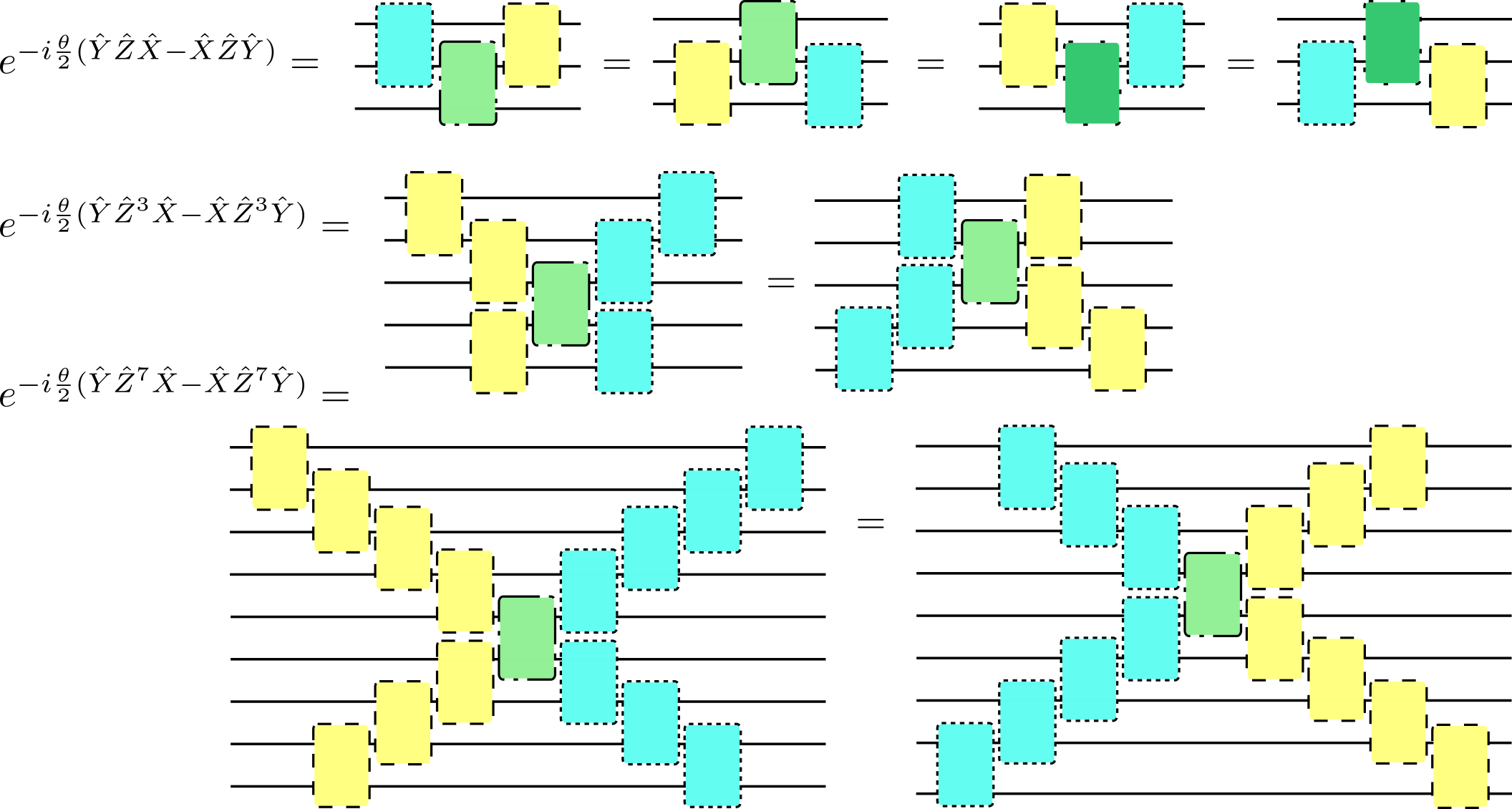}
    \caption{Different circuits implementing $e^{-i \frac{\theta}{2} (\hat Y \hat Z^n \hat X - \hat X \hat Z^n\hat Y )}$, the building blocks for implementing the meson operators defined in Eq.~\eqref{eq:mesOps}.
    The R-box composite elements in the circuit are defined in Fig.~\ref{fig:rbox}.
    }
    \label{fig:rboxbig}
\end{figure}

\begin{figure}[!ht]
    \centering
    \includegraphics[width=0.65\textwidth]{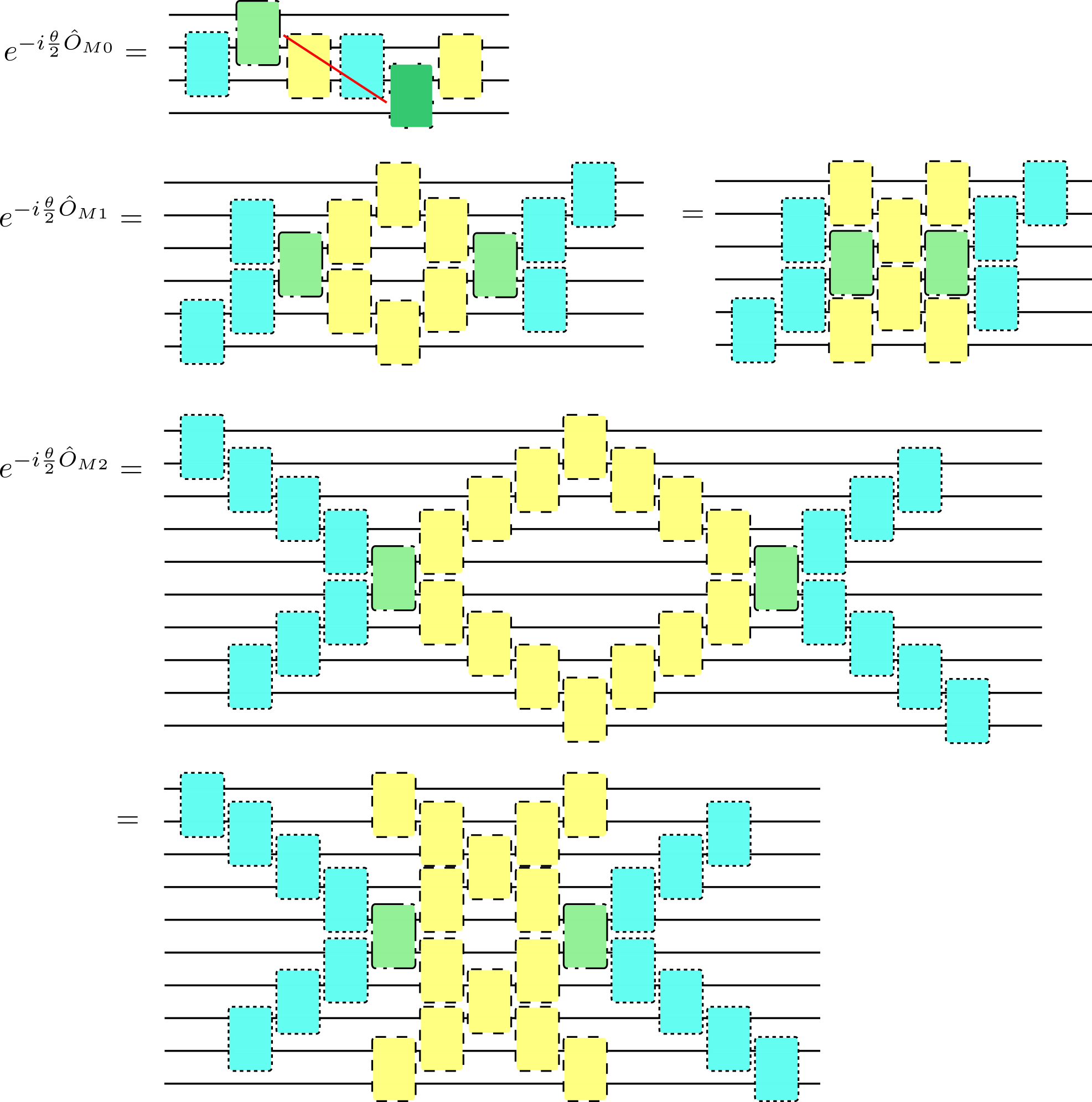}
    \caption{Circuits that implement unitary transformations
    from the meson operators.  
    The red line in first circuit denotes a cancellation between block operators.
    The CNOT-depth of these circuits are: 8, 28, 60, respectively. The R-box composite elements in the circuit are defined in  Fig.~\ref{fig:rbox}.
    }
    \label{fig:ohn}
\end{figure}

In the operator pool for state preparation, there are two types of operators, meson operators and baryon operators, as explained in Sec.~\ref{sec:L3SP} and Sec.~\ref{sec:StatePrepCir}. 
Meson operators are in the form of $e^{-i \frac{\theta}{2} \left((\hat Y \hat Z^n \hat X - \hat X \hat Z^n\hat Y )\hat I+\hat I(\hat Y \hat Z^n \hat X - \hat X \hat Z^n\hat Y )\right)}$.
Since $(\hat Y \hat Z^n \hat X - \hat X \hat Z^n\hat Y )\hat I$ commutes with $\hat I(\hat Y \hat Z^n \hat X - \hat X \hat Z^n\hat Y )$, they can be implemented exactly by subsequently applying the segments shown in Fig~\ref{fig:rboxbig}.
Figure~\ref{fig:rboxbig} also
shows examples of how unitary operators of the form
$e^{-i \frac{\theta}{2} (\hat Y \hat Z^n \hat X - \hat X \hat Z^n\hat Y )}$
can be implemented in multiple ways with different ``dressings'' of $R_{+}^{XY}$ or $R_{-}^{XY}$. 
This flexibility allows us to select a
variation to use in 
subsequent applications, 
and further reduce the depth of the circuit by taking advantage of the SU(2) symmetry.
Figure~\ref{fig:ohn} implements such products, along with circuit compression from identified cancellations. In particular, the variations chosen form two ``V''-shaped (or ``$\Lambda$''-shaped) operator structures in the middle that can be replaced with an ``X''-shape that reduces the circuit depth. 

\begin{figure}[htpb]
    \centering 
    \includegraphics[width = 0.6\textwidth]{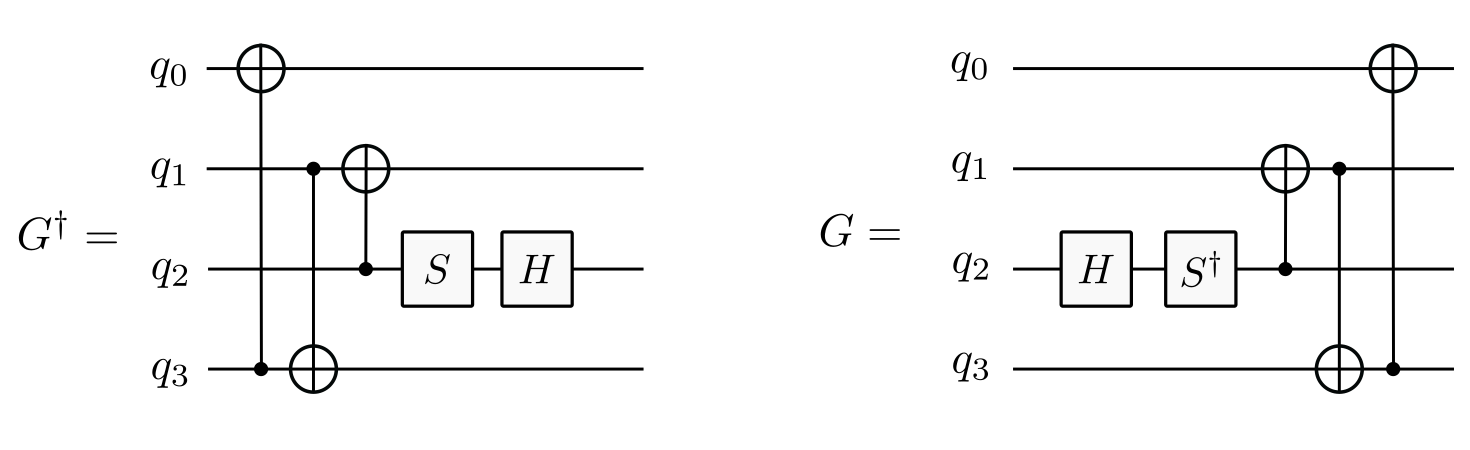}
    \caption{Circuits that furnish the GHZ transformation for baryon operators in state preparation.
    }
    \label{fig:ghz_state_prep}
\end{figure}

For baryon operators, GHZ transformations are used to transform the operator into the computational basis
\begin{align}
\label{eqn:stateprepGHZ}
\hat O_{B0} & \equiv 4i(\hat\sigma^+\hat\sigma^+\hat\sigma^-\hat\sigma^- - {\rm h.c.}) \nonumber \\
& = \frac{1}{2}(-\hat{X}\hat{X}\hat{Y}\hat{X}+\hat{X}\hat{Y}\hat{X}\hat{X}+\hat{Y}\hat{Y}\hat{X}\hat{Y}+\hat{Y}\hat{X}\hat{X}\hat{X}-\hat{X}\hat{X}\hat{X}\hat{Y}-\hat{X}\hat{Y}\hat{Y}\hat{Y}+\hat{Y}\hat{Y}\hat{Y}\hat{X}-\hat{Y}\hat{X}\hat{Y}\hat{Y}) \nonumber \\
& = \frac{1}{2} G(-\hat{I}\hat{I}\hat{Z}\hat{I}+\hat{I}\hat{Z}\hat{Z}\hat{I}-\hat{Z}\hat{Z}\hat{Z}\hat{I}+\hat{Z}\hat{Z}\hat{Z}\hat{Z}-\hat{I}\hat{Z}\hat{Z}\hat{Z}+\hat{I}\hat{I}\hat{Z}\hat{Z}-\hat{Z}\hat{I}\hat{Z}\hat{Z}+\hat{Z}\hat{I}\hat{Z}\hat{I})G^{\dagger} \ ,
\end{align}
where $G$ is defined in Fig.~\ref{fig:ghz_state_prep}. The operators in the diagonal basis commute, so they can be arranged in a fashion that minimizes the number of CNOT gates required.
In Eq.~\eqref{eqn:stateprepGHZ}, the diagonal operators are listed in the order in which they are applied in the quantum circuit.

\begin{figure}[!ht]
    \centering
    \includegraphics[width=0.75\textwidth]{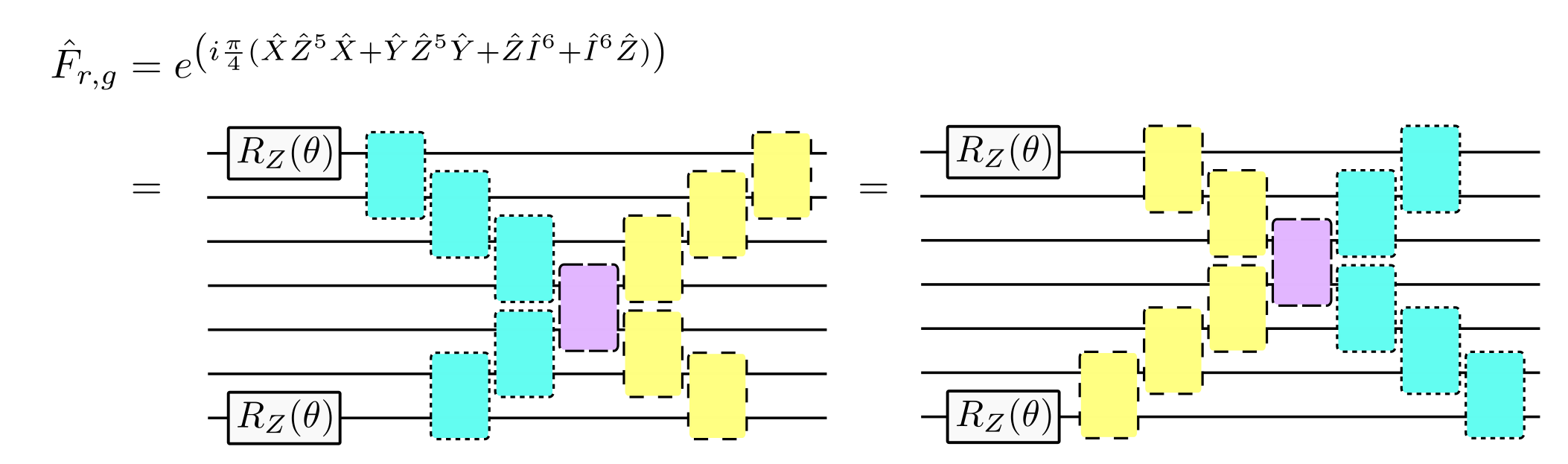}
    \caption{Circuits to implement the spatial translation of one color of SU(2) heavy quark between adjacent spatial lattice sites using an FSWAP circuit. 
    The $\theta$s appearing in circuit elements take the value of $-\pi/2$.
    }
    \label{fig:fswap_seg}
\end{figure}

The FSWAP gate for one color is shown in Fig.~\ref{fig:fswap_seg}, which is then combined in similar ways as for the meson operators in state preparation to perform the full Fermionic swap.

\begin{figure}[!ht]
    \centering
    \includegraphics[width=0.65\textwidth]{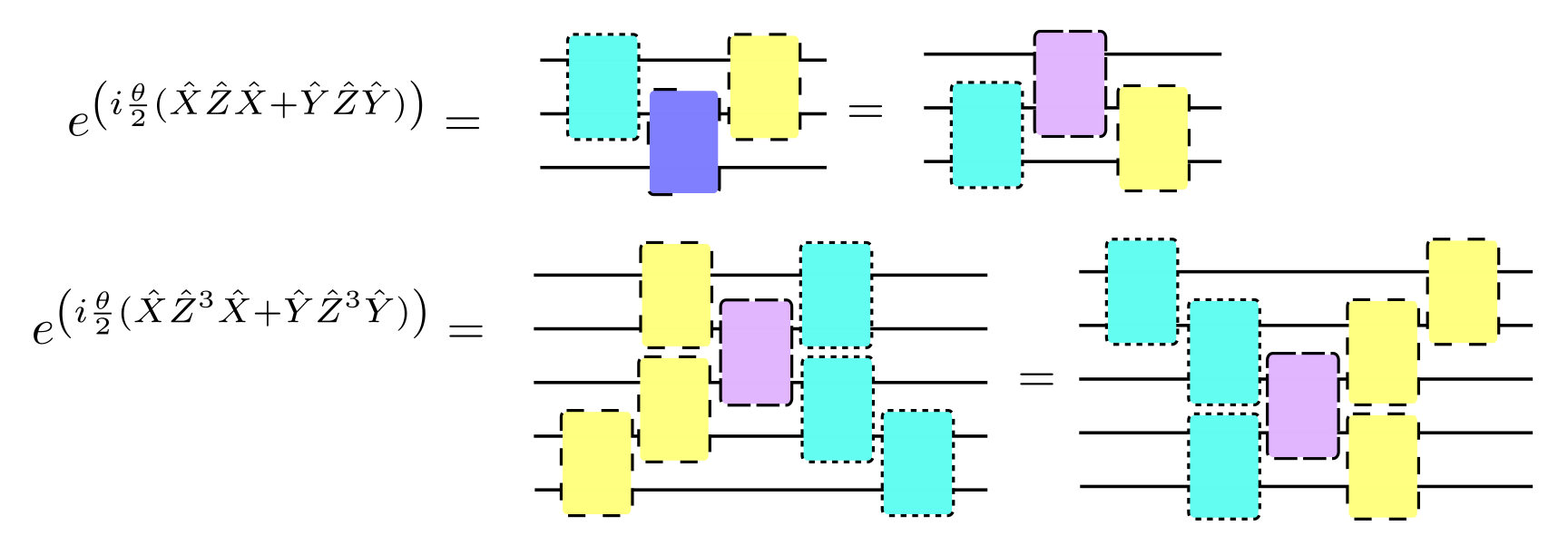}
    \caption{
    Elements of the quantum circuits used to implement the kinetic-energy operator.
    The different circuit elements are given in Fig.~\ref{fig:rbox}.}
    \label{fig:hkin}
\end{figure}

The single-color components used to construct the kinetic energy term in the Hamiltonian
are shown in Fig.~\ref{fig:hkin}. 
%

\section{Additional Circuit Resource Analysis}
\addtocontents{toc}{\vspace{-1em}}
\label{appen:resources}
\noindent
This appendix presents additional analysis of quantum circuit resources, including the CNOT depth of circuit segments, CNOT distributions over the lattice, and analysis of Trotter errors. 
The depth of each component in state preparation and time evolution is shown in Table~\ref{tab:time_evo_resource}. Note that evolution of the mass term has only single qubit rotations, and are not included in the table. 
\begin{table}[htpb]
    \centering
    \renewcommand{\arraystretch}{1.1}
    \begin{tabular}{|c || c | c | c | }
        \hline
        State prep. & CZ All-to-All & CZ Transpiled   \\
        \hline \hline
        $\hat O_{M0}$ & 6 & 6  \\
        $\hat O_{M1}$ & 14 & 14 \\
        $\hat O_{M2}$ & 26 & 27 \\
        $\hat O_{B0}$ & 14 & 35 \\
        $\hat O_{B1}$ & 14 & 43 \\
         \hline \hline
    \end{tabular}
    \hspace{.2in}
    \begin{tabular}{|c || c | c | c |}
        \hline
        Time ev. & CZ All-to-All & CZ Transpiled  \\
        \hline  \hline 
        $\hat{H}_{k0}$ & 6 & 11\\
        $\hat{H}_{k1}$ & 14  & 19\\
        $\hat{H}_{g}$ (total) & 78 & 202\\ \hline
        $\hat{\mathcal{Q}}_n\hat{\mathcal{Q}}_{n+1}$ & 16 & 41 \\
        $\hat{\mathcal{Q}}_n\hat{\mathcal{Q}}_{n+2}$ & 16 & 43\\
        $\hat{\mathcal{Q}}_n\hat{\mathcal{Q}}_{n+3}$ & 16 & 49\\
         \hline \hline
    \end{tabular}
    \caption{
    The two-qubit-gate depth
    of each component in the state preparation (left table) and time evolution (right table). The ``CZ Transpiled'' column correspond to the depth after transpiling onto devices with heavy-hex connectivity. This number could vary between different layouts and transpilation optimization.
    }
\label{tab:time_evo_resource}
\end{table}

\begin{figure}[!ht]
    \centering
    \includegraphics[width=0.9\linewidth]{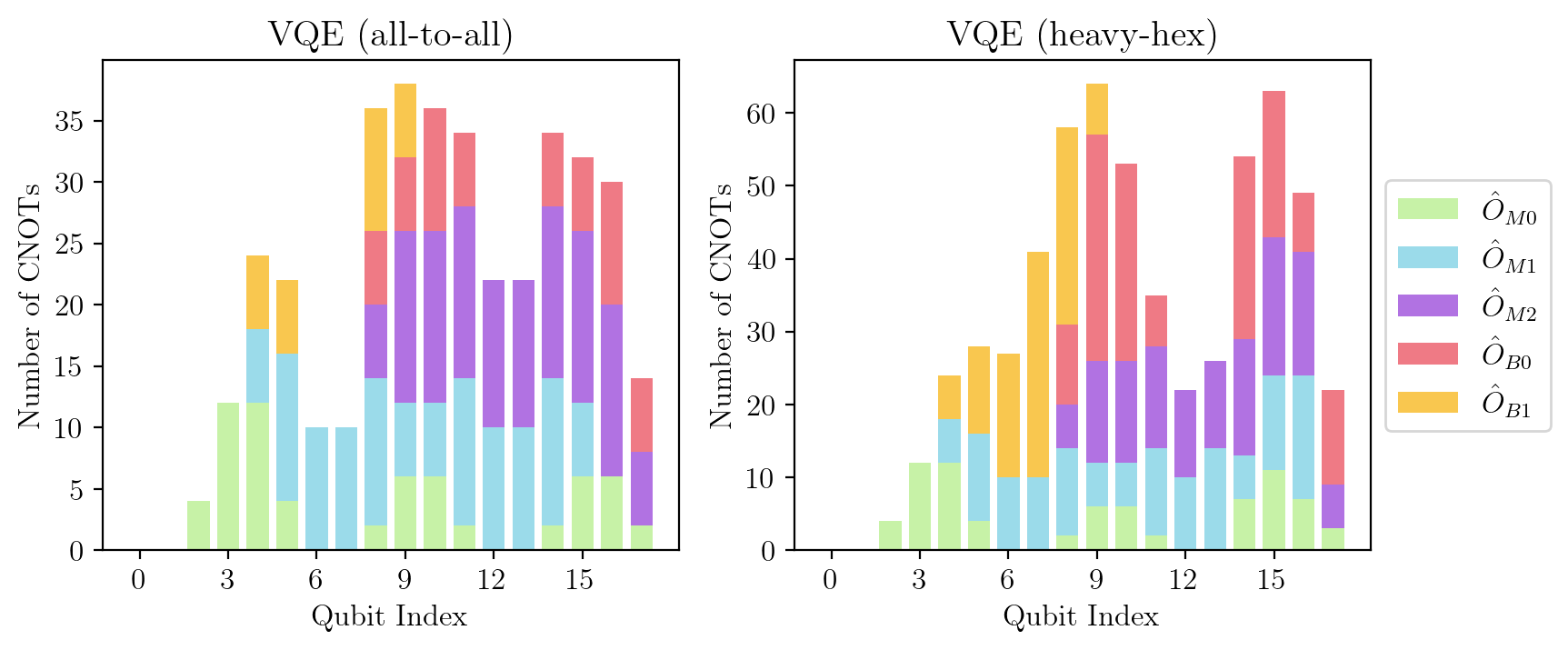}
    \caption{The CNOT depth distribution of components in the state preparation. 
    The left panel corresponds to  all-to-all connectivity,
    while the right panel is after transpiling to {\tt ibm\_pittsburgh}. 
    The number of CNOTs on each qubit is smaller than the total CNOT depth because some operators can't be executed in parallel.}
    \label{fig:vqe_cnot_dist}
\end{figure}

\begin{figure}[!ht]
    \centering
    \includegraphics[width=0.9\linewidth]{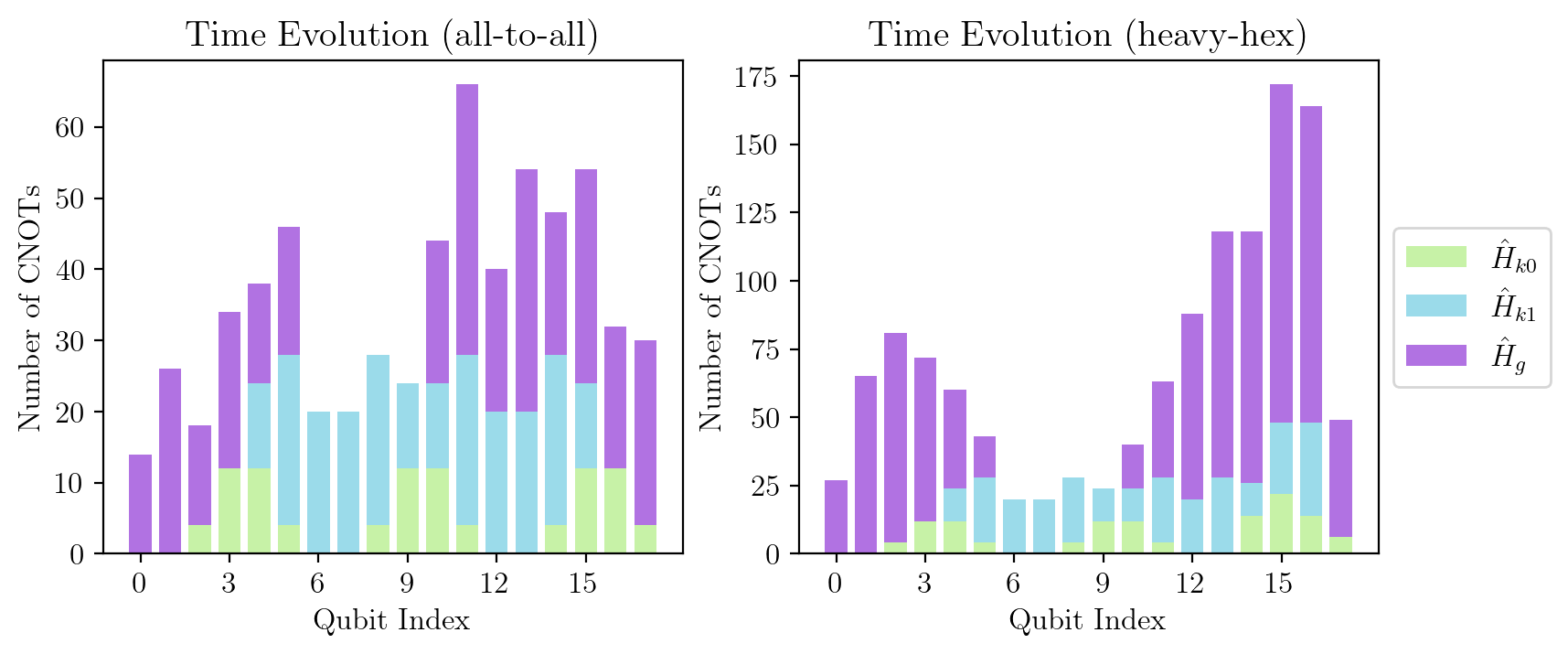}
    \caption{The CNOT depth distribution of components in the time evolution. 
    The left panel corresponds to  all-to-all connectivity, while the right panel is after transpiling to {\tt ibm\_pittsburgh}. 
    Because of the use of second-order Trotterized time evolution, the gate-counts associated with $\hat{H}_{k0}$ and $\hat{H}_{k1}$ count are doubled.
    The number of CNOTs on each qubit is smaller than the total CNOT depth because some operators can't be executed in parallel.    }
    \label{fig:time_evo_cnot_dist}
\end{figure}

\begin{figure}[htpb]
    \centering
    \includegraphics[width=0.9\linewidth]{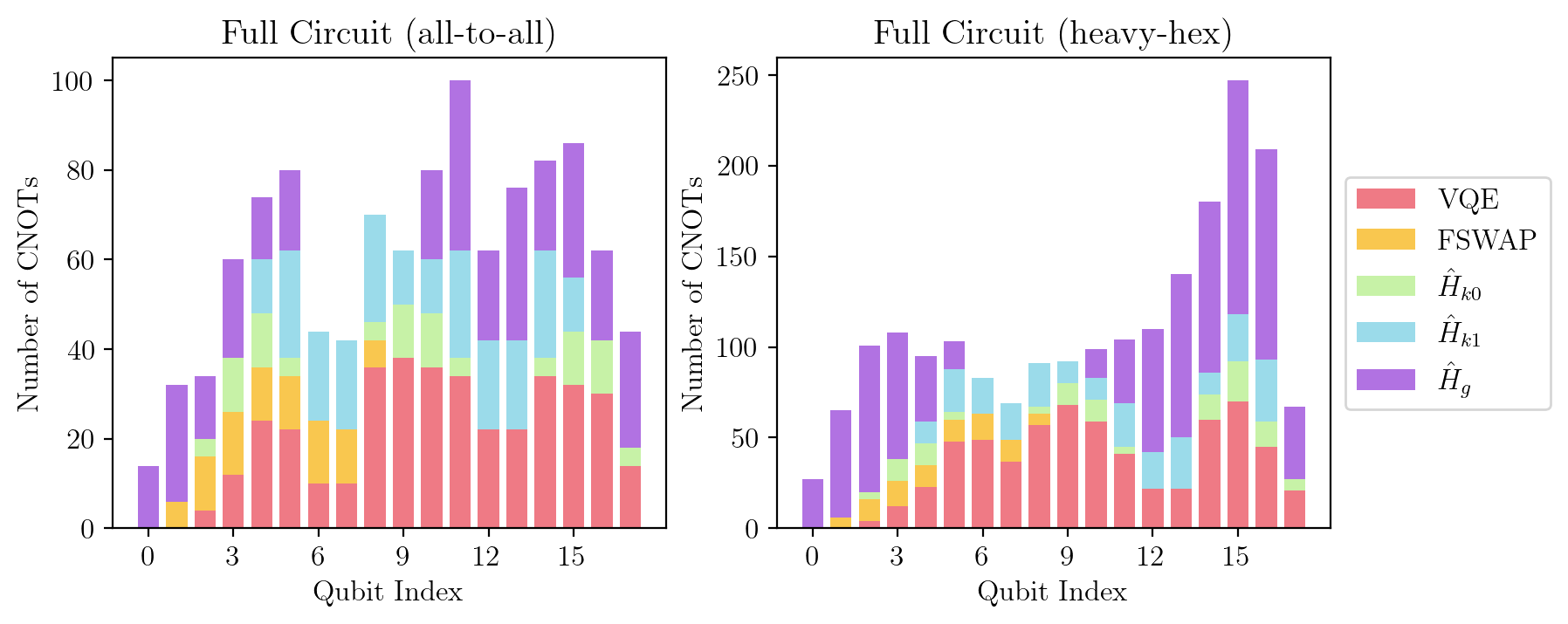}
    \caption{CNOT depth distribution of components in the complete process. 
    The left panel corresponds to  all-to-all connectivity, while the right panel is after transpiling to {\tt ibm\_pittsburgh}.
    }
    \label{fig:full_cnot_dist}
\end{figure}

CNOT depth distributions over qubits for state preparation (Fig.~\ref{fig:vqe_cnot_dist}), time evolution (Fig.~\ref{fig:time_evo_cnot_dist}) and for the full process (Fig.~\ref{fig:full_cnot_dist})  are analyzed, both for all-to-all connectivity and for the layout we use on {\tt ibm\_pittsburgh}
(with heavy-hex connectivity).
This allows for better understanding of the contributions to depth from each component over sectors of the lattice.
Upon transpilation to nearest-neighbor connectivity, the CNOT depth of $\hat O_{B1}$ and $e^{-it\hat H_g}$ increase significantly, while the other terms remain nearly unchanged, as they are already constructed for nearest-neighbor connectivity. 

\begin{figure}[htpb]
    \centering
    \includegraphics[width=0.85\linewidth]{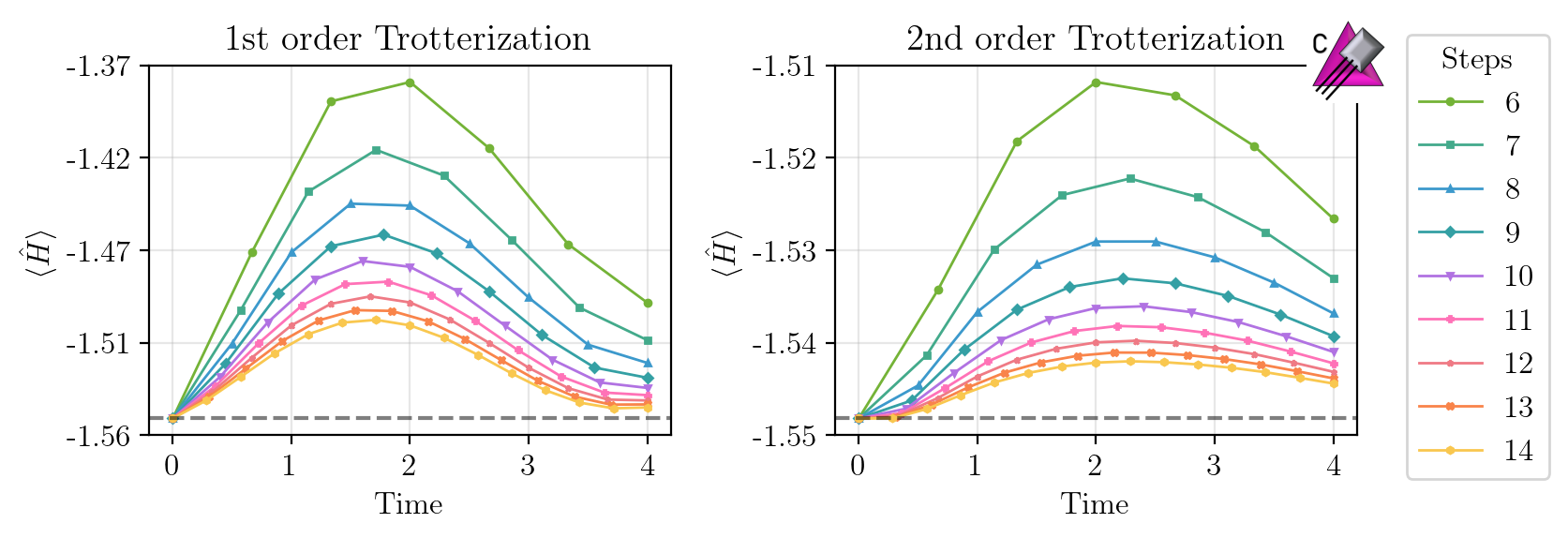}
    \caption{
    Energy conservation in the presence of Trotter errors.
    The left panel shows $\langle \hat H \rangle$ as a function of time for first-order Trotterization, 
    while the right panel shows the same for second-order Trotterization,
    using a number of Trotter steps between 6 to 19.
    }
    \label{fig:energy_scans}
\end{figure}

Violations of energy conservation
caused by Trotter errors over various step sizes and number of steps 
are shown in Fig.~\ref{fig:energy_scans} for both first- and second-order Trotterization. Second-order Trotterization preserves energy for longer times with comparable circuit depth.

\section{The Hadamard Test for Computing Energies}
\begin{figure}[htpb!]
    \centering
       \begin{tikzpicture}
        \node[scale=1.1] {
            \begin{quantikz}[row sep={0.9cm,between origins}]
    \lstick{$|\psi\rangle$} &          &  & \gate{\hat U(t)}     &  & \\
    \lstick{$|0\rangle$}    & \gate{H} & \gate{S}  &  \ctrl{-1} & \gate{H} &  \meter{}
           \end{quantikz}
        };
    \end{tikzpicture}
    \caption{The circuit implementing the evolution operator via a 
    Hadamard test, 
    from which part or all of the total energy of the system can be extracted.}
  \label{eq:HadamardTestHcirc}
\end{figure}
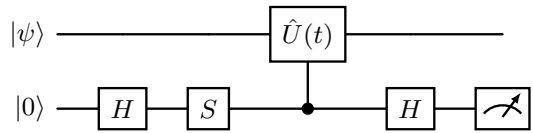

\addtocontents{toc}{\vspace{-1em}}
\label{app:Energycircs}
\noindent 
While we have measured energy losses using the method discussed in the main text, with results presented in Sec.~\ref{sec:QSresultsdEdx},
we have also assessed the implementation of the Hadamard test for determining the energy in the system
using classical simulations.
We have found it to be effective but inefficient,
requiring a suite of measurements and  subsequent extrapolation.
Including an ancilla with the register containing a state $|\psi\rangle$,
and acting with the controlled-evolution operator $C\hat U(t)$ 
(controlled by the ancilla qubit),
\begin{align}
    &|\phi\rangle =  |0\rangle |\psi\rangle
    \ ,\ 
     \hat S\hat H\otimes\hat I|\phi\rangle\ =\ 
     \frac{1}{\sqrt{2}} \left( |0\rangle + i|1\rangle\right)|\psi\rangle
     \ ,
     \nonumber\\
&(\hat H \otimes \hat I) \cdot C\hat U(t) \cdot (\hat S\hat H \otimes \hat I)|\phi\rangle\
     =
          \frac{1}{2}
     \left[
     |0\rangle \left(\hat I+i\hat U \right)|\psi\rangle
     \ +\ 
          |1\rangle \left(\hat I-i\hat U\right)|\psi\rangle
     \right]
     \ ,
\end{align}
implemented with the circuit shown in Fig.~\ref{eq:HadamardTestHcirc}.

The probabilities of the ancilla qubit measurements are 
\begin{equation}
{\rm Prob}(0) = \frac{1}{4} \langle \psi | 2 \hat I - i \left(\hat U^\dagger - \hat U\right) |\psi\rangle
\ ,\ 
{\rm Prob}(1) =  \frac{1}{4} \langle \psi | 2 \hat I + i \left(\hat U^\dagger - \hat U\right) |\psi\rangle
\ \ .
\end{equation}
Using the exact evolution operator $\hat U (t) = e^{-i t \hat H}$
leads to
\begin{equation}
{\rm Prob}(0) - {\rm Prob}(1) = t  \langle \psi | \hat H |\psi\rangle\ +\ {\cal O}(t^3)
\ \ .
\label{eq:fromprobstoH}
\end{equation}
By evaluating this quantity over a range of short times, 
$\langle\hat H\rangle$ can be determined.

For a Trotterized evolution operator, the extracted energy corresponds to that of an effective Hamiltonian, modified by the finite time step.  
A leading-order Trotterization will give rise to a correction  to Eq.~\eqref{eq:fromprobstoH} that is quadratic in time, giving a relative linear correction.
To mitigate this systematic error, a series of simulations should be performed over a range of time steps, with a polynomial extrapolation of the results to estimate the energy for a vanishing time step size.

\begin{figure}[htpb]
    \centering
    \includegraphics[width=0.4\linewidth]{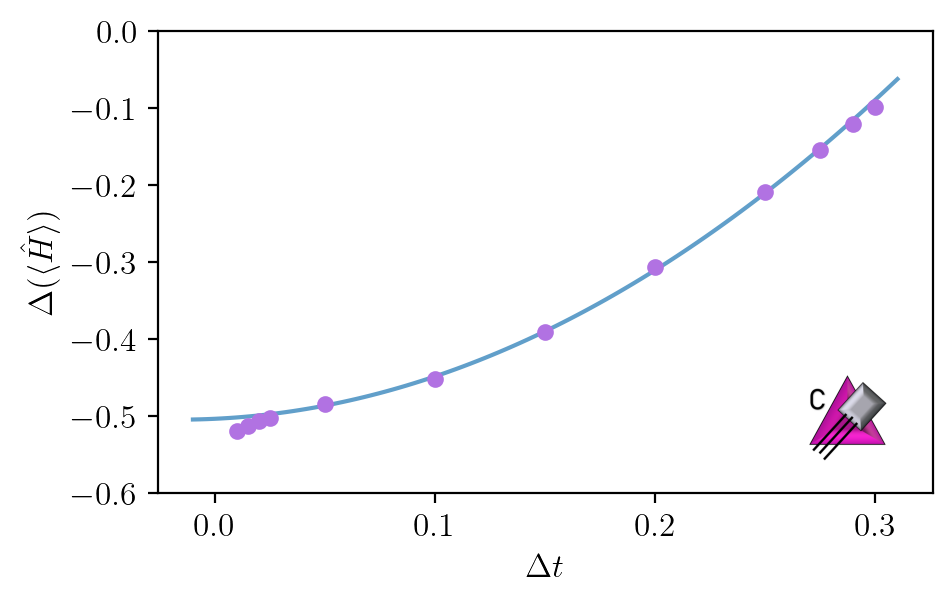}
    \caption{Results obtained from the  {\tt qiskit} simulator for the energy injected into the light quarks by moving the heavy quark from $x=0$ to $x=1$ in the $L=3$ system (without other matter present) (purple points), 
    with Hamiltonian parameters $m_q=0.1$ and $g=1.0$. 
    The blue curve corresponds to a (representative) quadratic fit to the results from $\Delta t= 0.05-0.275$.
    }
    \label{fig:L3x0x1SIM}
\end{figure}

Using the {\tt qiskit} simulator to measure the energy difference induced by the $x=0$ to $x=1$ displacement of the heavy quark without matter present,
we have implemented this technique over a range of time steps, obtaining the result shown in Fig.~\ref{fig:L3x0x1SIM}.
The results for small $\Delta t$ are subject to precision issues, and are not included in the fit. 
From these (classical) simulations, the central value of the 
extrapolated energy is found to be 
$\delta \langle \Delta \hat H_{g}  \rangle=-0.50(2)$ 
which is to be compared with the exact result of $0.5058$,
and the result that we have obtained from {\tt ibm\_pittsburgh} using the other method 
$\delta \langle \Delta \hat H_{g}  \rangle^{\rm IBM}=-0.5301(48)$. 
The quoted error is estimated from the range of values determined from varying the order of fit polynomials and selection of fit data.
While encouraged by the results and the fit displayed in Fig.~\ref{fig:L3x0x1SIM}, 
this makes clear that multiple sets of measurements are required, 
and an understanding of their systematic errors are needed for the method to provide reliable results.
Further, the required extrapolation adds additional errors that lead to a result that is inferior to that from the method used in this work, presented in Sec.~\ref{sec:QSresultsdEdx}.

\section{Explicit form of the Hamiltonian}
\addtocontents{toc}{\vspace{-1em}}
\label{app:explicitham}
\noindent 
The Hamiltonian for a single flavor of light and heavy quarks, with an arbitrary number of colors $N_c$ and sites $L$, is given by:
\begin{align}
    \hat{H}_{k} = & \ \frac{1}{2}\sum_{x=0}^{L-1}\sum_{c=0}^{N_c-1} \left[ \sigma_{i(3x+1,c)}^+ \left ( \bigotimes_{j=1}^{N_c-1}[-Z_{i(3x+1,c)+j} ] \right ) \sigma_{i(3x+1,c) + N_c}^- +\rm{h.c.} \right] \nonumber\\
    & \ + \frac{1}{2}\sum_{x=0}^{L-2}\sum_{c=0}^{N_c-1} \left[ \sigma_{i(3x+2,c)}^+ \left ( \bigotimes_{j=1}^{2 N_c-1}[-Z_{i(3x+2,c)+j} ] \right ) \sigma_{i(3x+2,c) + 2 N_c}^- +\rm{h.c.} \right] \ , \\
    \hat{H}_m  = & \ \frac{1}{2} \sum_{x=0}^{L-1} \sum_{c=0}^{N_c-1} \left(m_Q \left[  Z_{i(3x,c)} + 1 \right]+m_q \left[  Z_{i(3x+1,c)} + 1 \right]+m_q \left[  -Z_{i(3x+2,c)} + 1 \right]\right) \ , \\
    \hat{H}_g = & \ \frac{g^2}{2} \sum_{n=0}^{3L-2} \left( 2L-\lceil \tfrac{2n+1}{3} \rceil \right) \hat{\mathcal{Q}}^{(a)}_n \hat{\mathcal{Q}}^{(a)}_n + g^2 \sum_{n=0}^{3L-3} \sum_{m=n+1}^{3L-2} \left( 2L-\lceil \tfrac{2m+1}{3} \rceil \right) \hat{\mathcal{Q}}^{(a)}_n \hat{\mathcal{Q}}^{(a)}_m \ , \\
    \hat{H}_\lambda = & \ \frac{\lambda^2}{2} \sum_{n=0}^{3L-2} \hat{\mathcal{Q}}^{(a)}_n \hat{\mathcal{Q}}^{(a)}_n + \lambda^2 \sum_{n=0}^{3L-3} \sum_{m=n+1}^{3L-2} \hat{\mathcal{Q}}^{(a)}_n \hat{\mathcal{Q}}^{(a)}_m \ ,
\end{align}
with $i(n,c) = (N_c n+ c)$, and the products of the charges defined as~\cite{Farrell:2022wyt}
\begin{align}
    4 \hat{\mathcal{Q}}_{n}^{(a)} \, \hat{\mathcal{Q}}_{n}^{(a)} =& \ \frac{N_c^2-1}{2} - \left (1+\frac{1}{N_c} \right )\sum_{c=0}^{N_c-2}\sum_{c' = c+1}^{N_c-1} Z_{i(n,c)} Z_{i(n,c')} \ ,  \nonumber \\[4pt]
    8 \hat{\mathcal{Q}}_{n}^{(a)} \, \hat{\mathcal{Q}}_{m}^{(a)} =& \ 4 \sum_{c=0}^{N_c-2}\sum_{c'=c+1}^{N_c-1} \left[ \sigma^+_{i(n,c)} \ \left(\otimes Z \right)_{(n,c,c')} \ \sigma^-_{i(n,c')} \sigma^-_{i(m,c)} \ \left(\otimes Z \right)_{(m,c,c')} \ \sigma^+_{i(m,c')} + {\rm h.c.}
    \right] \  \nonumber \\
    &+ \sum_{c=0}^{N_c-1} \sum_{c'=0}^{N_c-1}\left (\delta_{cc'} - \frac{1}{N_c} \right)Z_{i(n,c)} Z_{i(m,c')}\ , \nonumber \\
    \left(\otimes Z \right)_{(n,c,c')}  \equiv & \ \bigotimes_{k=1}^{c'-c-1} Z_{i(n,c)+k}
    \ .
    \label{eq:QnfQmfpN}
\end{align}
As mentioned in Refs.~\cite{Farrell:2023fgd,Farrell:2024fit}, in the $Q=0$ sector, 
there is a freedom in the way Gauss' law can be imposed.
Instead of starting with ${\bf E}^{(a)}=0$ at one end of the lattice and working toward the other side of the lattice,
the contributions can be determined by starting from ${\bf E}^{(a)}=0$ at both ends, and working toward the middle.
In our case, it takes the following form,
\begin{align}
    \hat{H}_g = \sum_{b=0}^{L-2}E_b^2+\sum_{b=L}^{2L-1}E_b^2 =\ & \frac{g^2}{2} \sum _{n=0}^{\lfloor \tfrac{3L-1}{2} \rfloor } \left[\left(L- \lceil \tfrac{2n+1}{3}\rceil \right) \hat{\mathcal{Q}}^{(a)}_n \hat{\mathcal{Q}}^{(a)}_n + \left(L-\lceil \tfrac{2n-1}{3}\rceil \right) \hat{\mathcal{Q}}^{(a)}_{3L-n-1}\hat{\mathcal{Q}}^{(a)}_{3L-n-1} \right] \nonumber\\
    & + g^2 \sum _{n=0}^{\lfloor \tfrac{3 L}{2}\rfloor } \left[\sum_{m=n+1}^{\lfloor \tfrac{3 L-2}{2} \rfloor } \left(L-\lceil \tfrac{2m+1}{3} \rceil \right) \hat{\mathcal{Q}}^{(a)}_n \hat{\mathcal{Q}}^{(a)}_m + \sum_{m=1}^{n-1} \left(L+1-\lceil \tfrac{2n}{3}\rceil \right) \hat{\mathcal{Q}}^{(a)}_{3L-n} \hat{\mathcal{Q}}^{(a)}_{3L-m}\right]
\end{align}
\begin{align}
    \hat{H}_g = \sum_{b=0}^{L-1}E_b^2+\sum_{b=L+1}^{2L-1}E_b^2 =\ & \frac{g^2}{2} \sum _{n=0}^{\lfloor \tfrac{3L}{2} \rfloor } \left[\left(L + 1 - \lceil \tfrac{2n+1}{3}\rceil \right) \hat{\mathcal{Q}}^{(a)}_n \hat{\mathcal{Q}}^{(a)}_n + (1-\delta_{n,0})\left(L-\lceil \tfrac{2n}{3}\rceil \right) \hat{\mathcal{Q}}^{(a)}_{3L-n}\hat{\mathcal{Q}}^{(a)}_{3L-n} \right] \nonumber\\
    & + g^2 \sum _{n=0}^{\lfloor \tfrac{3 L}{2}\rfloor } \left[\sum_{m=n+1}^{\lfloor \tfrac{3 L}{2} \rfloor } \left(L-\lceil \tfrac{2m-2}{3} \rceil \right) \hat{\mathcal{Q}}^{(a)}_n \hat{\mathcal{Q}}^{(a)}_m + \sum_{m=1}^{n-1} \left(L-\lceil \tfrac{2n}{3}\rceil \right) \hat{\mathcal{Q}}^{(a)}_{3L-n} \hat{\mathcal{Q}}^{(a)}_{3L-m}\right]
\end{align}
%

\section{Results from Simulations using IBMs Quantum Computers}
\label{app:IBMruns}
\noindent 
The results obtained from simulations performed using {\tt ibm\_pittsburgh} 
discussed in Sec.~\ref{sec:QSresults}
and displayed in Fig.~\ref{fig:IBMresultfig}
are shown in Table~\ref{tab:IBManalysis}. 
\begin{table}[!ht]
\renewcommand{\arraystretch}{1.1}
\begin{tabularx}{\textwidth}{|c|| Y | Y | Y | Y | Y | Y | Y | Y | Y |}
 \hline
\diagbox{Set up}{Qubit} & 0,1 & 2,3 & 4,5 & 6,7 & 8,9 & 10,11 & 12,13 & 14,15 & 16,17 \\
\hline\hline
\multirow{2}{*}{Raw} & -0.008(3) & 0.152(5) & 0.282(2) & -0.822(2) & -0.208(4) & 0.408(5) & -0.850(7) & -0.206(3) & 0.404(6) \\
 & 0.008(4) & 0.163(7) & 0.318(6) & -0.859(1) & -0.210(1) & 0.413(2) & -0.776(3) & -0.208(5) & 0.385(4) \\
\hline
ODR & 0.000(0) & 0.203(4) & 0.380(4) & -0.997(2) & -0.297(4) & 0.564(5) & -1.002(2) & -0.289(3) & 0.552(3) \\
\hline
Expected & 0.000 & 0.118 & 0.431 & -1.000 & -0.282 & 0.503 & -1.000 & -0.294 & 0.524 \\
\hline
\end{tabularx}
\\
\begin{tabularx}{\textwidth}{|c|| Y | Y | Y | Y | Y | Y | Y | Y | Y |}
 \hline
\diagbox{Set up}{Qubit} & 0,1 & 2,3 & 4,5 & 6,7 & 8,9 & 10,11 & 12,13 & 14,15 & 16,17 \\
\hline\hline
\multirow{2}{*}{Raw} & -0.472(4) & 0.099(3) & 0.106(7) & 0.002(4) & -0.095(3) & 0.156(11) & -0.470(1) & -0.059(5) & 0.216(7) \\
 & -0.527(3) & 0.106(5) & 0.123(8) & 0.006(3) & -0.089(5) & 0.170(8) & -0.410(4) & -0.062(6) & 0.193(11) \\
\hline
ODR & -1.000(3) & 0.228(7) & 0.295(9) & 0.000(0) & -0.216(5) & 0.480(15) & -0.994(4) & -0.227(16) & 0.530(10) \\
\hline
Expected & -1.000 & 0.183 & 0.348 & 0.000 & -0.285 & 0.541 & -1.000 & -0.325 & 0.537 \\
\hline
\end{tabularx}
\caption{
The values of $\langle \hat Z_j \rangle$ measured using  {\tt ibm\_pittsburgh} for the $L=3$ systems using Hamiltonian parameters $m_q=0.1$ and $g=1.0$.  An initial state was prepared with a heavy quark residing on the  $x=0$ spatial site (upper table). The heavy quark
is then moved to the $x=1$ spatial site, then the system is
time evolved forward to $t=1.0$ using one second-order Trotter step (lower table), as discussed in the main text. 
As discussed in Sec.~\ref{sec:chargeD}, the applied cuts are
$\lambda^{\rm mit}_{\rm cut} = \lambda^{\rm phys}_{\rm cut} = 0.005$ for the ODR points. 
}
\label{tab:IBManalysis}
\end{table}

\begin{figure}[ht!]
    \centering
\includegraphics[width=\linewidth]{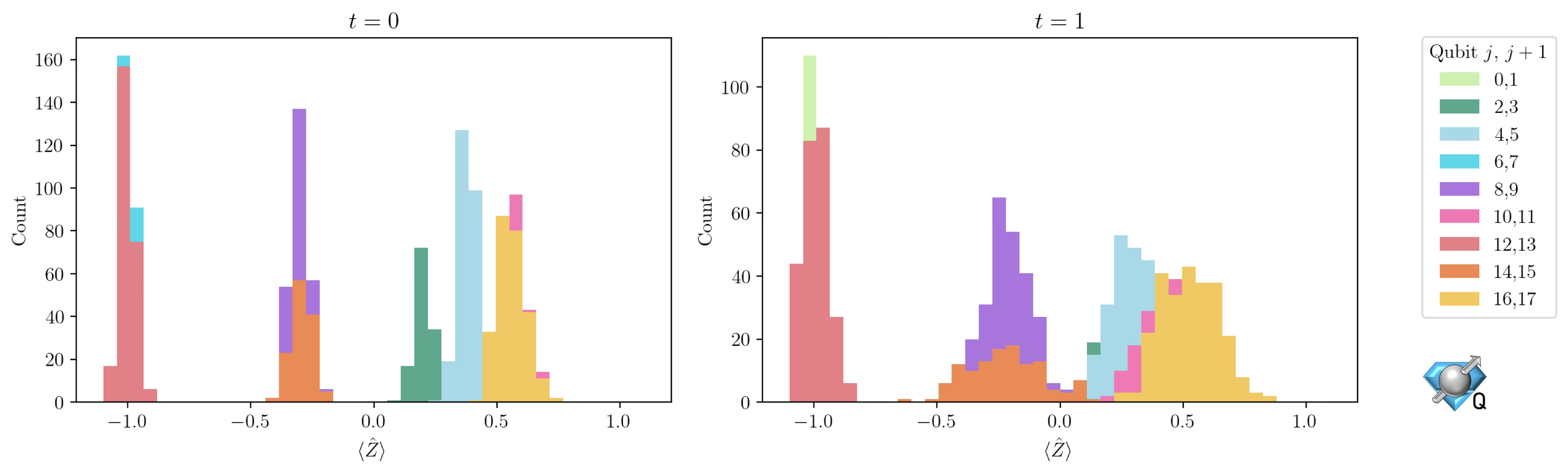}
\caption{
Results obtained from {\tt ibm\_pittsburgh} for $\langle \hat Z_j \rangle$ in the $L=3$ system.  
At $t=0$, the heavy quark is instantaneously moved from $x=0$ to $x=1$, then evolved forward in time to $t=1.0$ 
using one step of the second-order Trotterized time-evolution operator.
The SU(2) symmetry has been used to combine results for each red and green quark.
The left panel shows the distribution of each $\langle \hat Z_j \rangle$ before the heavy quark is moved; the right panel  shows the distribution after time evolution.
}
    \label{fig:IBMresultdist}
\end{figure}

Fig.~\ref{fig:IBMresultdist} shows histograms of results for $\langle \hat Z_j \rangle$ for each pair of qubits. Each color corresponds to a different staggered site.

The evaluation of the energy of the system at times immediately before and after translating the heavy quark by one spatial lattice site
is accomplished by running physics and mitigation circuits (described in the text) 
with different levels of noise for ZNE on IBM's quantum computers.
For each of the three ensemble measurements, matrix elements of operators diagonal in the computational basis are evaluated in post-processing directly from measurements, as described in the main text. 
\begin{figure}[!ht]
    \centering
    \includegraphics[width=0.32\linewidth]{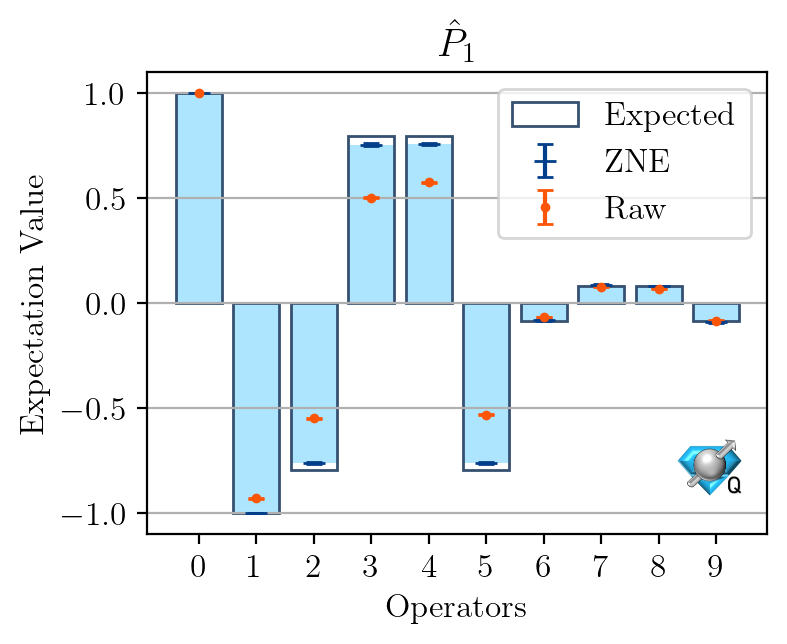}
    \includegraphics[width=0.32\linewidth]{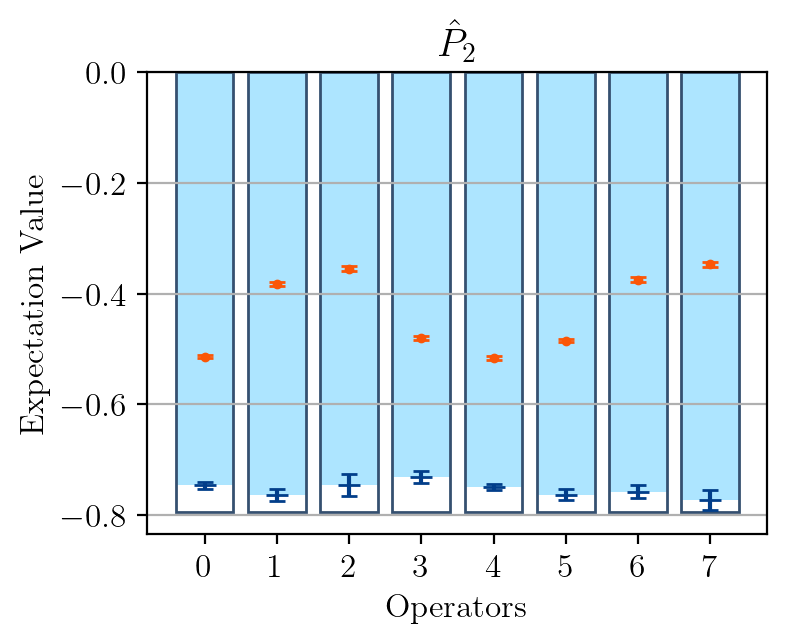}
    \includegraphics[width=0.325\linewidth]{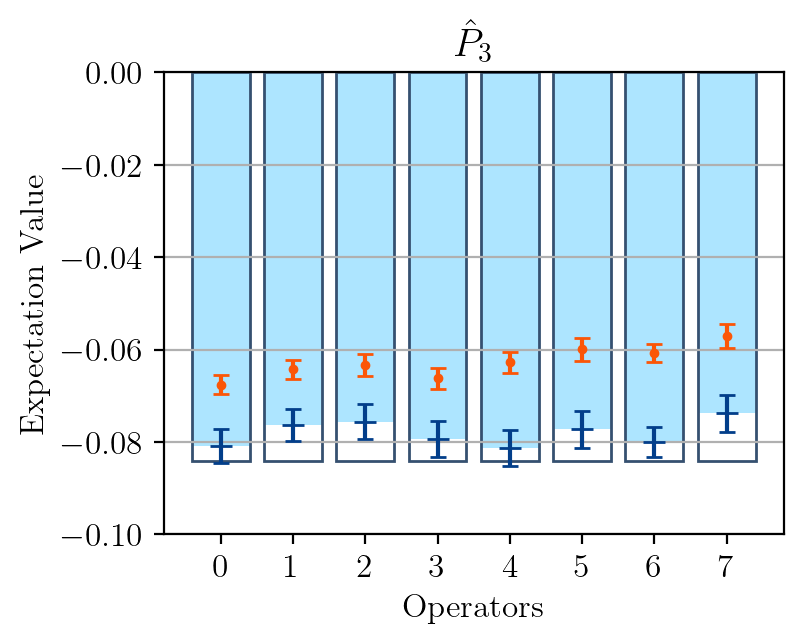}
    \caption{Results for matrix elements of Pauli strings from the three quantum circuits used to evaluate the change in energy due to the heavy quark being moved from $x=0$ to $x=1$.
    The left panel shows the raw and mitigate results, along with expected noiseless values from classical computation, corresponding to $\hat {P}_1 $ given in Eq.~\eqref{eq:estimator}, while the middle and right panels correspond to $\hat {P}_2$ and $\hat {P}_3$,   respectively.
    The points with error bars correspond to the raw and mitigated results obtained from {\tt ibm\_pittsburgh}, 
while the gray bar outlines are the expected results obtained using classical computing.
 The numerical values of the displayed results are given in Tables~\ref{tab:MEest1}, \ref{tab:MEest2} and \ref{tab:MEest3}.
 }
    \label{fig:est123}
\end{figure}
\begin{figure}[!ht]
    \centering
    \includegraphics[width=0.95\linewidth]{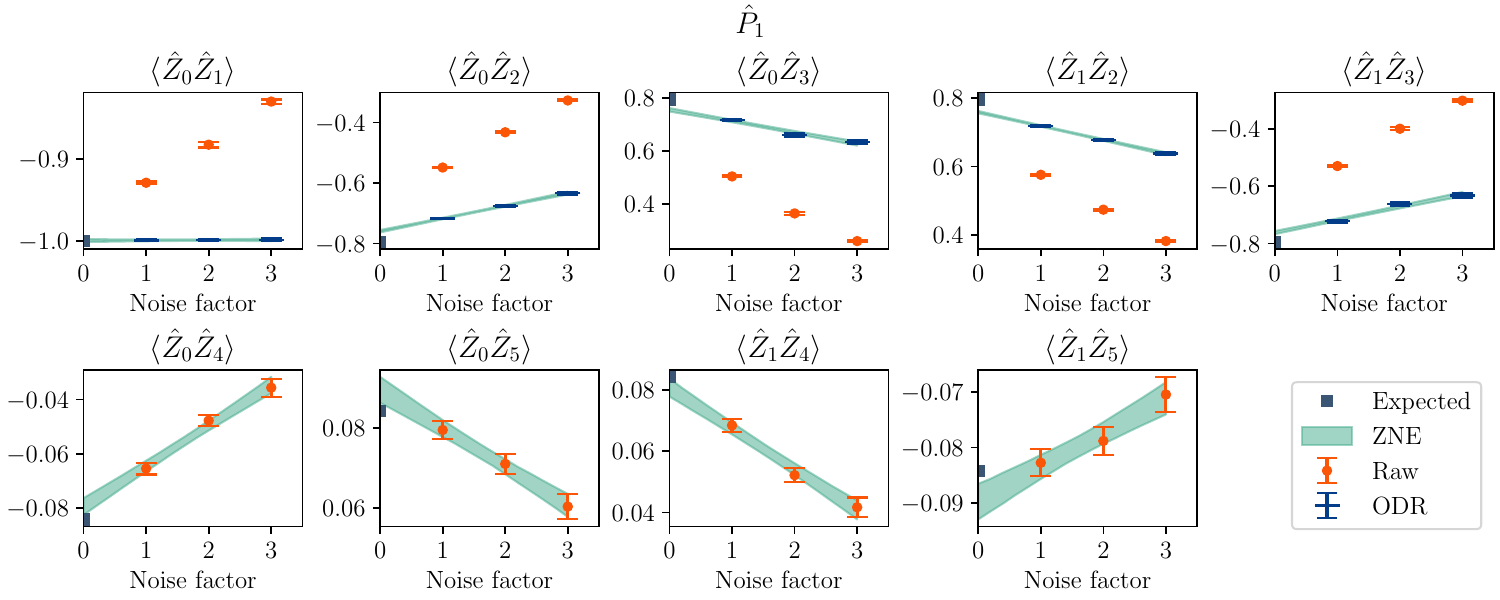}
    \includegraphics[width=0.76\linewidth]{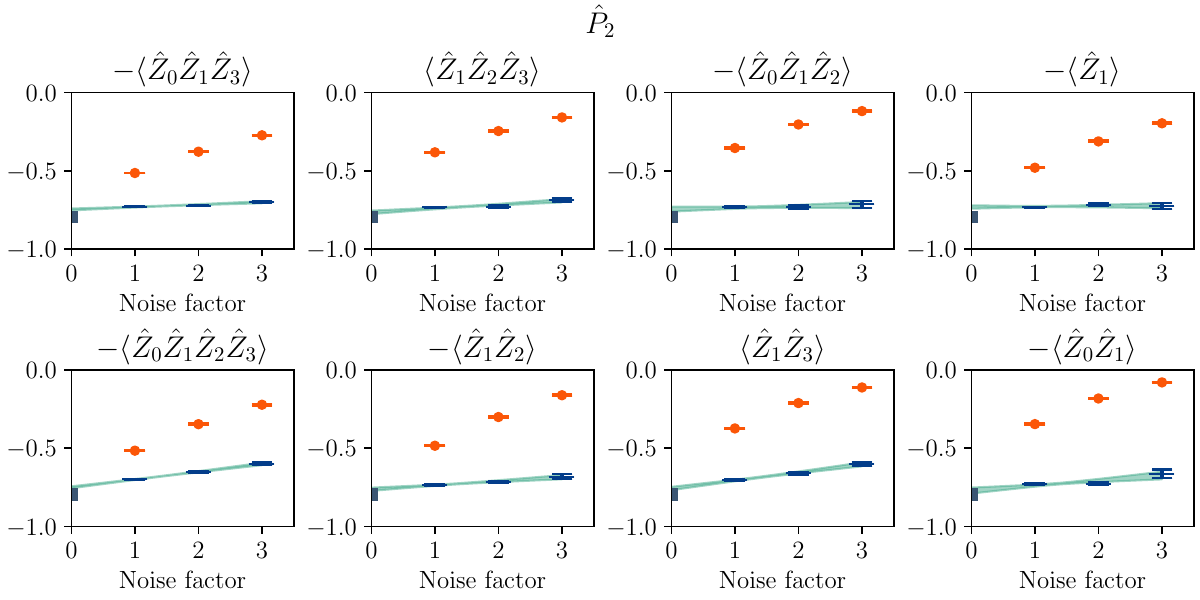}
    \includegraphics[width=0.76\linewidth]{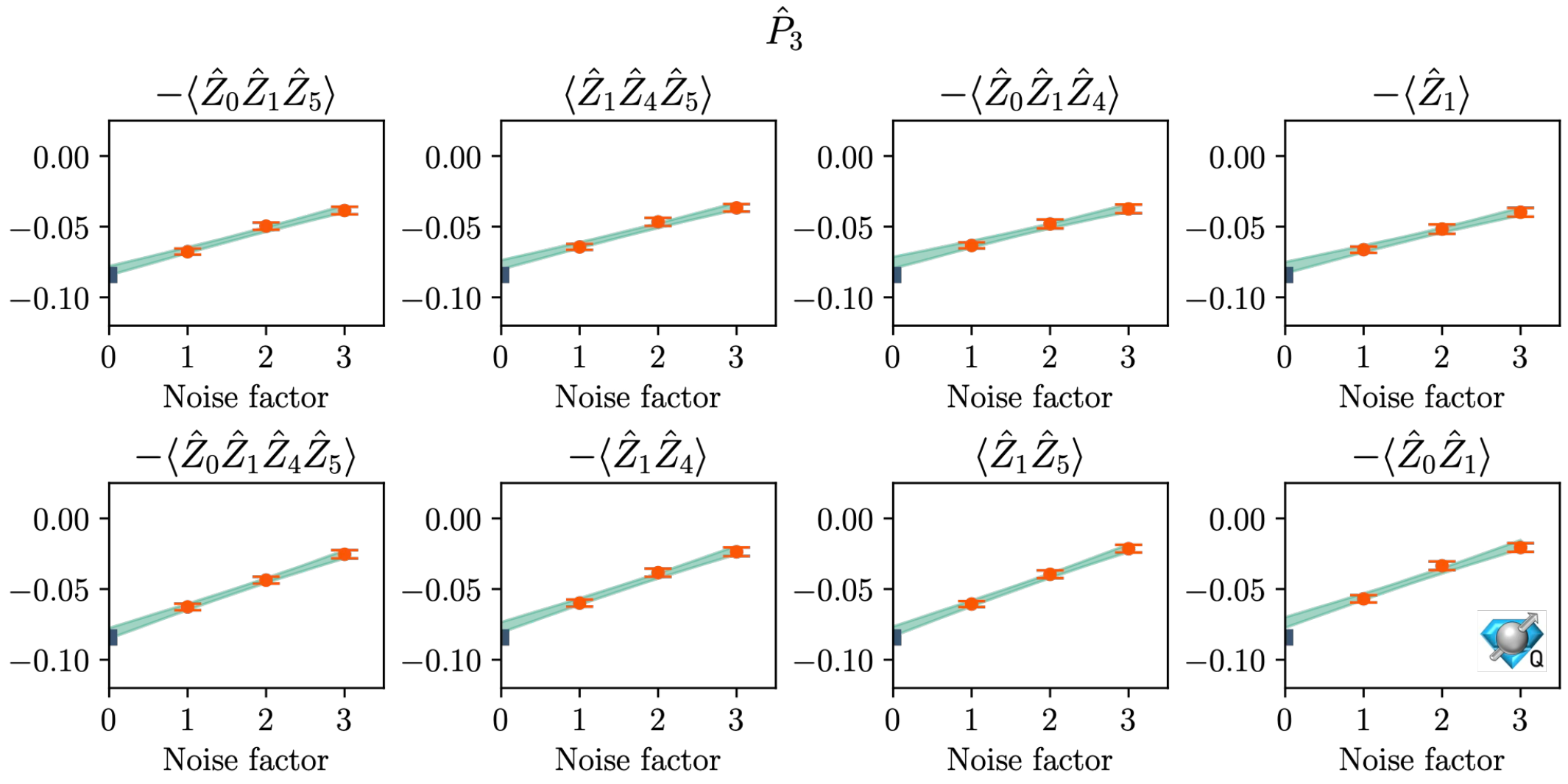}
    \caption{ZNE results for the matrix elements of Pauli strings from the three quantum circuits used to evaluate the change in energy due to the heavy quark being moved from $x=0$ to $x=1$.
    In each panel ($\hat {\cal O}_1$ top, $\hat {\cal O}_2$ middle, $\hat {\cal O}_3$ bottom), each subpanel shows the raw and ODR-mitigated results (when available) for each Pauli string, along with expected noiseless values from classical computation, and the linear ZNE band fit.
    }
    \label{fig:est123_zne}
\end{figure}
The final results for the matrix elements of the Pauli strings given in Eq.~\eqref{eq:estimator}
obtained from the three different circuits run on {\tt ibm\_pittsburgh} are displayed in Fig.~\ref{fig:est123} and 
given in Tables~\ref{tab:MEest1}, \ref{tab:MEest2} and \ref{tab:MEest3}. 
The ZNE fits are displayed in Fig.~\ref{fig:est123_zne}.
\begin{table}[!ht]
\renewcommand{\arraystretch}{1.1}
\begin{tabularx}{\textwidth}{|c || Y | Y | Y |Y| Y| }
 \hline
\multicolumn{1 }{|c||}{ }
&  \multicolumn{3}{ c |}{Matrix Elements $\hat P_1 $ from {\tt ibm\_pittsburgh}} & \multicolumn{2}{c |}{Classical }\\
 \hline
Operator & Raw & ODR & ODR $+$ ZNE & Expected & Exact \\
\hline\hline 
$\hat I$ & 1.0000(0) & 1.0000(0) & 1.0000(0) & 1.0 & 1.0\\
$\hat Z_0 \hat Z_1$ & -0.9288(15) & -0.9991(11) & -0.9994(20) & -1.0 & -1.0\\
$\hat Z_0 \hat Z_2$ & -0.5488(25) & -0.7174(28) & -0.7597(53) & -0.7951  & -0.7862\\
$\hat Z_0 \hat Z_3$ & 0.5039(38) & 0.7185(45) & 0.7561(85) & 0.7951 & 0.7862\\
$\hat Z_1 \hat Z_2$ & 0.5761(22) & 0.7189(24) & 0.7581(49) & 0.7951 & 0.7862\\
$\hat Z_1 \hat Z_3$ & -0.5302(37) & -0.7224(45) & -0.7613(81) & -0.7951 & -0.7862\\
$\hat Z_0 \hat Z_4$ & -0.0653(22) & - & -0.0792(37) & -0.0842 & -0.0946\\
$\hat Z_0 \hat Z_5$ & 0.0794(25) & - & 0.0894(40) & 0.0842 & 0.0946\\
$\hat Z_1 \hat Z_4$ & 0.0683(22) & - & 0.0807(36) & 0.0842 & 0.0946\\
$\hat Z_1 \hat Z_5$ & -0.0826(25) & - & -0.0895(39) & -0.0842 & -0.0946\\
 \hline
 \hline
\end{tabularx}
\caption{
Matrix elements of diagonal Pauli strings in the computational basis of the initial state ($t<0$) prepared on  {\tt ibm\_pittsburgh}, 
corresponding to $\hat {\cal O}_1 $ 
given in Eq.~\eqref{eq:estimator}.
These results are displayed in Fig.~\ref{fig:est123}.
The expected results show the value from a noiseless circuit applied to the ground state prepared on the quantum computer, while the value in parentheses corresponds to the result from the exact ground state.
Results for the last few operators were not mitigated via ODR because setting non-Clifford rotation angles to 0 or $\pi/2$ both result in 0 for noiseless predictions of these observables, incompatible with ODR.}
\label{tab:MEest1}
\end{table}
\begin{table}[!ht]
\renewcommand{\arraystretch}{1.1}
\begin{tabularx}{\textwidth}{|c || Y | Y | Y |Y| Y| }
 \hline
\multicolumn{1 }{|c||}{ } 
&  \multicolumn{3}{ c |}{Matrix Elements $\hat P_2 $ from {\tt ibm\_pittsburgh}} & \multicolumn{2}{c |}{ Classical}\\
 \hline
Operator & Raw & ODR &  ODR $+$ ZNE & Expected & Exact  \\
\hline\hline 
$-\hat Z_0\hat Z_1\hat Z_3$ & -0.5141(29) & -0.7286(27) & -0.746(06) & -0.7951 & -0.7862 \\
$\hat Z_1\hat Z_2 \hat Z_3$ & -0.3820(41) & -0.7334(46) & -0.764(11) & -0.7951 & -0.7862 \\
$-\hat Z_0\hat Z_1\hat Z_2$ & -0.3545(48) & -0.7328(50) & -0.746(20) & -0.7951 & -0.7862 \\
$\hat Z_1$ & -0.4804(36) & -0.7323(37) & -0.731(10) & -0.7951 & -0.7862 \\
$-\hat Z_0\hat Z_1\hat Z_2\hat Z_3$ & -0.5160(33) & -0.6996(27) & -0.749(06) & -0.7951 & -0.7862 \\
$\hat Z_1\hat Z_2$ & -0.4846(33) & -0.7344(40) & -0.763(09) & -0.7951 & -0.7862 \\
$\hat Z_1\hat Z_3$ & -0.3742(47) & -0.7039(44) & -0.758(12) & -0.7951 & -0.7862 \\
$-\hat Z_0 \hat Z_1$ & -0.3466(44) & -0.7283(56) & -0.772(18) & -0.7951 & -0.7862 \\
\hline
\hline
\end{tabularx}
\caption{Matrix elements of diagonal Pauli strings in the computational basis of the initial state ($t<0$) prepared on  {\tt ibm\_pittsburgh}, corresponding to $\hat {\cal O}_2$ given in Eq.~\eqref{eq:estimator}.
These results are displayed in Fig.~\ref{fig:est123}. The expected results show the value from a noiseless circuit applied to the ground state prepared on the quantum computer, while the value in parentheses corresponds to the result from the exact ground state.
}
\label{tab:MEest2}
\end{table}
\begin{table}[!ht]
\renewcommand{\arraystretch}{1.1}
\begin{tabularx}{\textwidth}{|c || Y | Y | Y | Y |}
 \hline
\multicolumn{1}{|c||}{} & \multicolumn{2}{c|}{Matrix Elements $\hat P_3$ from {\tt ibm\_pittsburgh}} & \multicolumn{2}{c|}{Classical}\\
 \hline
Operator & Raw & ZNE & Expected & Exact \\
\hline\hline
$-\hat Z_0\hat Z_1\hat Z_5$ & -0.0676(22) & -0.0808(37) & -0.0842 & -0.0946 \\
$\hat Z_1\hat Z_4 \hat Z_5$ & -0.0643(21) & -0.0763(35) & -0.0842 & -0.0946 \\
$-\hat Z_0\hat Z_1\hat Z_4$ & -0.0633(24) & -0.0756(37) & -0.0842 & -0.0946 \\
$\hat Z_1$ & -0.0662(23) & -0.0794(39) & -0.0842 & -0.0946 \\
$-\hat Z_0\hat Z_1\hat Z_4\hat Z_5$ & -0.0628(22) & -0.0813(39) & -0.0842 & -0.0946 \\
$\hat Z_1\hat Z_4$ & -0.0600(25) & -0.0773(40) & -0.0842 & -0.0946 \\
$\hat Z_1\hat Z_5$ & -0.0607(20) & -0.0800(32) & -0.0842 & -0.0946 \\
$-\hat Z_0 \hat Z_1$ & -0.0570(26) & -0.0738(40) & -0.0842 & -0.0946 \\
 \hline
 \hline
\end{tabularx}
\caption{
Matrix elements of diagonal Pauli strings in the computational basis of the initial state ($t<0$) prepared on  {\tt ibm\_pittsburgh}, 
corresponding to $\hat {\cal O}_3$ 
given in Eq.~\eqref{eq:estimator}.
These results are displayed in Fig.~\ref{fig:est123}.
The expected results show the value from a noiseless circuit applied to the ground state prepared on the quantum computer, while the value in parentheses corresponds to the result from the exact ground state. ODR mitigation circuits were not run for $\hat{P}_3$ because setting non-Clifford rotation angles to 0 or $\pi/2$ both result in 0 for the noiseless predictions of observables, 
incompatible with implementing ODR.
}
\label{tab:MEest3}
\end{table}
%


\end{document}